\def\ISDRAFT{0} 
\def\ISLONG{1} 

\pdfoutput=1
\newcommand{\DRAFT}[1]{\ifodd\ISDRAFT#1\fi} 
\newcommand{\LONG}[1]{\ifnum\ISLONG=1#1\fi} 
\newcommand{\SHORT}[1]{\ifnum\ISLONG=0#1\fi} 

\newcommand{\lampersand}{\LONG{&}}

\documentclass[preprint,12pt]{elsarticle}
\journal{SecReT}

\usepackage[utf8x]{inputenc}
\usepackage{hyperref}
\usepackage{listings}
\usepackage{eurosym}
\hypersetup{pdfborder ={0 0 0}, colorlinks=true}
\usepackage[english]{babel}
\usepackage{comment}
\usepackage[\SHORT{cmex10}]{amsmath}
\SHORT{\interdisplaylinepenalty=2500}
\usepackage{amssymb,amsthm}
\usepackage{stmaryrd}
\usepackage{empheq}
\usepackage{mathtools}
\usepackage[algoruled,linesnumbered]{algorithm2e} 
\usepackage{longtable}
\usepackage{xcolor}
    \usepackage{microtype}
\usepackage{graphicx} 
\usepackage{stfloats}
\usepackage{wrapfig}
\usepackage{fmtcount} 

\newcounter{tableitemcntr}[table]
\renewcommand{\thetableitemcntr}{\arabic{tableitemcntr}}
\newcommand{\tableitem}{%
        \refstepcounter{tableitemcntr}%
        \thetableitemcntr. 
}

\newtheorem{theorem}{Theorem}
\newtheorem{prop}{Proposition}
\newtheorem{lemma}{Lemma}
\newtheorem{cor}{Corollary}

\theoremstyle{definition}
\newtheorem{example}{Example}
\newtheorem{df}{Definition}[section]

\DeclareMathOperator*{\orut}{\ensuremath{root}}
\DeclareMathOperator*{\otaag}{\ensuremath{tag}}

\DeclareMathOperator*{\opbin}{\ensuremath{bin}}
\DeclareMathOperator*{\opaci}{\ensuremath{\cdot}}
\DeclareMathOperator*{\openc}{\ensuremath{enc}}
\DeclareMathOperator*{\opaenc}{\ensuremath{aenc}}
\DeclareMathOperator*{\opsig}{\ensuremath{sig}}
\DeclareMathOperator*{\oppriv}{\ensuremath{priv}}
\DeclareMathOperator*{\oppair}{\ensuremath{pair}}

\DeclareMathOperator*{\oBASEsubterms}{\ensuremath{Sub}}
\DeclareMathOperator*{\oPREFIXsubterms}{\ensuremath{Q}}

\DeclareMathOperator*{\omeasure}{\ensuremath{measure}}

\DeclareMathOperator*{\osubtermsII}{\ensuremath{Sub}}
\DeclareMathOperator*{\ovars}{\ensuremath{Vars}}
\DeclareMathOperator*{\oder}{\ensuremath{Der}}

\DeclareMathOperator*{\odom}{\ensuremath{dom}}
\DeclareMathOperator*{\oDAGsize}{\ensuremath{size_{DAG}}} 
\DeclareMathOperator*{\oelems}{\ensuremath{elems}}

\DeclareMathOperator*{\opairing}{\ensuremath{\pi}}

\newcommand*{\rut}[1]{\ensuremath{\orut\left(#1\right)}}
\newcommand*{\taag}[1]{\ensuremath{\otaag\left(#1\right)}}

\newcommand*{\bin}[2]{\ensuremath{\opbin\left(#1,#2\right)}}
\newcommand*{\aci}[1]{\ensuremath{\opaci\left(#1\right)}}

\newcommand*{\pair}[2]{\ensuremath{\oppair\left(#1,#2\right)}}
\newcommand*{\enc}[2]{\ensuremath{\openc\left(#1,#2\right)}}
\newcommand*{\aenc}[2]{\ensuremath{\opaenc\left(#1,#2\right)}}
\newcommand*{\sig}[2]{\ensuremath{\opsig\left(#1,#2\right)}}
\newcommand*{\priv}[1]{\ensuremath{\oppriv\left(#1\right)}}

\newcommand*{\sub}[1]{\ensuremath{\osubterms\left(#1\right)}}
\newcommand*{\subII}[1]{\ensuremath{\osubtermsII\left(#1\right)}}
\newcommand*{\vars}[1]{\ensuremath{\ovars\left(#1\right)}}
\newcommand*{\der}[1]{\ensuremath{\oder\left(#1\right)}}
\newcommand*{\derdy}[1]{\ensuremath{\oder_{DY}\left(#1\right)}}
\newcommand*{\xder}[2]{\ensuremath{\oder_{#1}\left(#2\right)}}
\newcommand*{\dom}[1]{\ensuremath{\odom\left(#1\right)}}

\newcommand*{\DAGsize}[1]{\ensuremath{\oDAGsize\left(#1\right)}}
\newcommand*{\elems}[1]{\ensuremath{\oelems\left(#1\right)}}

\newcommand*{\card}[1]{\ensuremath{\lvert#1\rvert}}

\newcommand*{\subo}[1]{\ensuremath{\oPREFIXsubterms\hspace{-1pt}\mathring{\oBASEsubterms} \left(#1\right)}}

\newcommand*{\measure}[1]{\ensuremath{\omeasure\left(#1\right)}}

\newcommand*{\pairing}[1]{\ensuremath{\opairing(#1)}}

\newcommand*{\bool}[1]{\ensuremath{2^{#1}}}

\newcommand*{\snd}[2]{\ensuremath{!_{#1} #2}}
\newcommand*{\rcv}[2]{\ensuremath{?_{#1} #2}}
\newcommand*{\channel}[2]{\ensuremath{{#1} \rightharpoonup {#2}}}

\newcommand*{\Universe}{\ensuremath{\mathcal{T}}}
\newcommand*{\UniverseG}{\ensuremath{\mathcal{T}_g}}
\newcommand*{\ConstrSys}[1]{\ensuremath{\mathcal{#1}}}
\newcommand*{\UniAt}[1]{\ensuremath{\mathcal{#1}}}
\newcommand*{\UniVar}[1]{\ensuremath{\mathcal{#1}}}

\newcommand*{\tuple}[1]{\ensuremath{\left\langle #1 \right\rangle}}
\newcommand*{\set}[1]{\ensuremath{\left\lbrace #1 \right\rbrace}}
\newcommand*{\lst}[1]{\ensuremath{\left\lbrace #1 \right\rbrace}}
\newcommand*{\norm}[1]{\ensuremath{\left\ulcorner #1 \right\urcorner}}
\newcommand*{\call}[1]{\ensuremath{\mathcal{#1}}}

\newcommand*{\dlta}[1]{\ensuremath{\delta\left(#1\right)}}

\newcommand*{\ahreal}[1]{\ensuremath{H^{\ConstrSys{S},\sigma}\left(#1\right)}}
\newcommand*{\ah}[1]{\ensuremath{H\left(#1\right)}}

\DeclareMathOperator*{\pahs}{\ensuremath{\pairing{\ah{\sigma}}}}
\newcommand*{\pah}[1]{\ensuremath{\pairing{\ah{#1}}}}

\newcommand{\as}{\UniVar{A}\xspace}

\newcommand{\css}{\ConstrSys{S}\xspace}

\newcommand*{\edges}[1]{\mathbb{E}\left(#1\right)}

\newcommand{\br}{\\} 

\begin{document}
\begin{frontmatter}
\title{Satisfiability of General Intruder Constraints with and without a Set Constructor\tnoteref{thanks}}
\tnotetext[tnotetext]{The work presented in this paper was partially supported by the FP7-ICT-2007-1 Project no. 216471, ``AVANTSSAR: Automated Validation of Trust and Security of Service-oriented Architectures'' (\url{http://www.avantssar.eu})
and  FP7-ICT Project no. 256980, ``NESSoS:  Network of Excellence on Engineering Secure Future Internet Software Services and Systems'' (\url{http://www.nessos-project.eu}).
}

\author[loria]{Tigran Avanesov}
\ead{Tigran.Avanesov@loria.fr}
\author[sabatier]{Yannick Chevalier}
\ead{ychevali@irit.fr}
\author[loria]{Micha\"el Rusinowitch}
\ead{Michael.Rusinowitch@loria.fr}
\author[loria]{Mathieu Turuani}
\ead{Mathieu.Turuani@loria.fr}

\address[loria]{ Loria, Inria Nancy - Grand Est, Campus Scientifique --- BP 239, 54506 Vand\oe uvre-l\`es-Nancy, France }
\address[sabatier]{ IRIT - Université Paul Sabatier, 118 route de Narbonne, 31020 Toulouse Cedex, France }


    \begin{abstract}
    Many decision problems on security protocols can be reduced to solving so-called 
    intruder constraints in Dolev Yao model. Most constraint solving procedures for  protocol security   
    rely on two properties of constraint systems  called \emph{monotonicity} and  \emph{variable-origination}. 
    In this work we relax these restrictions by 
    giving  a decision procedure for solving general intruder constraints (that do not have these properties) 
    that stays in NP. Our result extends a first work by L. Mazaré in several directions: we allow non-atomic keys, 
    and an  associative, commutative and idempotent symbol (for modeling sets). We also discuss several   new applications of the results. 
    \end{abstract}


\end{frontmatter}
\section{Introduction}

Detecting flaws in security protocol specifications  
under the perfect cryptography assumption in Dolev-Yao intruder model 
is an approach  that has been extensively investigated 
in recent years \cite{MillenShmatikov2001,ofmc2005, Turuani06, Cr2008Scyther}.
In particular  symbolic constraint solving has proved to be a very successful approach in the area. 
It amounts to express the possibility of mounting  an attack, e.g. the derivation of a secret, 
as a list of steps where for each step some message has to be derived from the current intruder knowledge. 
These steps correspond in general to the progression of the protocol execution, 
up to the last one which is the secret derivation. 

Enriching standard Dolev-Yao intruder model 
with different equational theories  \cite{BasinMV05,ChevRus-Combin2009}  
like exclusive OR, modular exponentiation, Abelian groups, etc. \cite{CDL05-survey, ChevalierKRT05xor,DLLT-IC07} 
helps to find flaws that could not be detected considering free symbols only.
A particularly useful theory is the  theory of an \emph{ACI} operator (that is associative commutative and idempotent) 
since it  allows one to express sets in cryptographic protocols.

Up to one exception~\cite{Mazare05,MazareThese}, all  proposed algorithms rely on two strong assumptions about the 
constraints to be processed: knowledge monotonicity and variable origination. 
Constraints satisfying this hypothesis 
are  called \emph{well-formed constraints} in the literature 
and they are not restrictive as these conditions hold
when handling standard security problems with a single Dolev-Yao intruder.  
However, we will see that in some situations it can be quite 
useful to relax these hypotheses and consider  \emph{general constraints}, 
that is constraints without the restrictions above. 
General  constraints naturally occur when considering security problems 
involving several  non-communicating Dolev-Yao intruders (see \S~\ref{ssec:severalintruders}).
Remark that if intruders can communicate during protocol execution, the model becomes attack-equivalent to 
one with a unique Dolev-Yao intruder \cite{Syverson00dolev-yaois}.

\subsection{Contributions of the paper} 
First, we will show 
that as for the standard case, in this more general framework it is still possible to derive an NP 
decision procedure for detecting attacks on a bounded number of protocol  sessions (Sections~\ref{sec:dy},~\ref{sect:complexity}). 
Second, our result extends previous ones  by allowing non-atomic keys and 
the usage of an associative commutative idempotent operator (Sections~\ref{sect:resolutionofcs},~\ref{sect:complexity}) 
that can be used for instance to model sets of nodes in XML document (see \S~\ref{ssec:xmlinj}). 
Third, we will remark that the satisfiability procedure we obtain for   \emph{general constraints} is a non trivial extension 
of the one for  \emph{well-formed constraints} by showing that this procedure 
cannot be extended to handle operators with subterm convergent theories since satisfiability gets undecidable in this case (\ref{sec:undec}). 
On the other hand  it is known that satisfiability remains  decidable for the standard  case of \emph{well-formed constraints}
with the same operator properties~\cite{Baudet05}. %
Finally we will sketch the potential applications of our results (Section~\ref{sec:motiv}).

\subsection{Related works}
The decision procedure for satisfiability of well-formed constraint systems  can be used to decide the insecurity of cryptographic protocols with a 
bounded number of sessions \cite{RusinowitchT03}. 
In this domain, several works deviated from the perfect cryptography assumption and started to consider algebraic properties of functional symbols. 
For example properties of XOR operator and exponentiation were considered in 
\cite{ChevalierKRT03exp,ChevalierKRT05xor,Comon-Lundh03intruderdeductions,Shmatikov04decidableanalysis} 
and together with homomorphic symbol in \cite{THESE-delaune06}. 
Some algebraic properties (like associative and commutative symbol) make the insecurity problem undecidable \cite{Bursuc07associative-commutativededucibility}.

All the works mentioned above consider systems of constraints with two restrictions %
namely knowledge monotonicity (the left-hand side of a constraint representing the current knowledge of the intruder is included into the left-hand side of the next one) and variable origination (variable appears first in the right-hand side of some constraint): this limitation is not impeding  the solution of  usual protocol insecurity problems  
since the constraints generated with an active Dolev-Yao intruder are of the required type. 
An attempt to swerve from well-formed constraints was made by Mazar\'e \cite{Mazare05}. He considered ``quasi well-formed'' constraint systems by partially relaxing the knowledge monotonicity. 
Later, in his thesis \cite{MazareThese}, he raised a similar decidability problem, but now for general constraint systems. 
He succeeded to find a decision procedure for satisfiability of general constraint systems with the restriction that keys used for encryption are atomic.
However to our knowledge no extension of Dolev-Yao
deduction system to non-atomic key or to algebraic properties has been shown decidable
for general constraint systems. 
Moreover, satisfiability of well-formed constraints with ACI theory was not considered before.

\section{Motivating examples}
\label{sec:motiv}
\subsection{Protocol analysis with several intruders}
\label{ssec:severalintruders}
In the domain of  security protocol analysis 
Dolev-Yao model is widely used  in spite of its limitations. 
We propose here  to consider instead of a  powerful Dolev-Yao intruder that controls 
the whole network,   several  non communicating Dolev-Yao intruders 
with smaller controlled domains. We give  below an  application of this model. 

\begin{figure}
\includegraphics[width=\columnwidth]{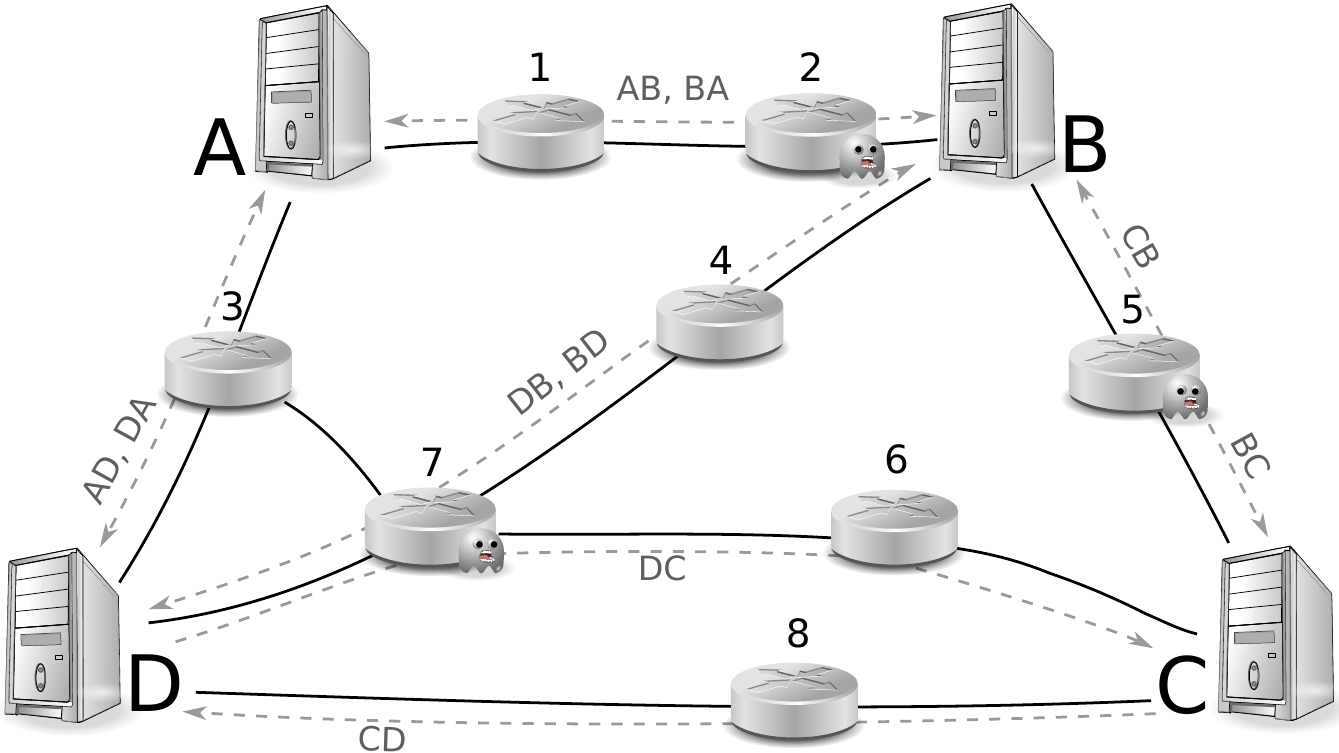}	
\caption{Untrusted routers}\label{fig:ex2}
\end{figure}

Suppose several agents ($A,B\dots$, see \figurename{}~\ref{fig:ex2}) execute a message exchange protocol 
(every agent has a finite list of actions in a send/receive format that is known to everybody).
Due to their (long distance) layout  they have to transmit data through routers ($1,2,3 \dots$).
The routing tables of all honest routers/agents are static (messages follow always the same path).
Some routers ($2,5,7$) may be compromised: an intruder managed to install a device controlling input and output of the router
or implanted there his malicious code.
A message circulated via such an untrusted channel (e.g. $DB$) is consumed by the corresponding compromised 
device (\emph{local intruder}) ($7$) thereby increasing his knowledge.
Moreover, a local intruder can forge and emit to an endpoint ($C,B,D$) of any channel he controls ($BD, DB, DC$) 
any message he can build using the content of his memory and some available transformations specified by a deduction system.
Because of the network topology malicious routers have no means to communicate (there is no links between them, neither direct nor via other routers),
but at some point the intruder can gather the knowledge of all the compromised routers (by physically collecting devices or reading their memory).

In this framework the \emph{security problem} is to know whether it is possible  to initially give instructions to  
compromised routers (e.g. by reprogramming malicious devices)
to force such an execution that honest agents (that strictly follow their list of actions)  will reveal some secret data 
to the intruder (i.e. intruder can build this data from the gathered at the end knowledge of all local intruders).

\subsubsection{Formalizing the coordinated attack problem}
\label{sssec:formalizmotiv}

To formalize the problem  we introduce some notations and definitions that are more detailed in Subsection 3.1. 
\paragraph{Messages}
We consider first-order terms built from a set of function symbols 
(such as encryption, pairing, etc.), a set of constants $\UniAt{A}$ (representing elementary pieces of data: texts, 
public keys, names of agents, etc. also called \emph{atoms}) and a set of variables $\UniVar{X}$.  
Let $\Universe$  be the  set of all possible terms. %
For a term $t$  we write $\vars{t}$ the  set of all variables in $t$ (see Def. \ref{df:vars}).	
A term $t$ is a \emph{ground term}, if $\vars{t} = \emptyset$. The set of  ground terms is denoted by  $\UniverseG$.
We assume that terms are interpreted modulo an equational theory  
and that we can compute for every term $t$ a unique normal form denoted by $\norm{t}$ modulo 
this equational theory. (We will focus later on the special case where we 
have a  function symbol $\cdot$ and the theory is generated by  
the  commutativity, associativity and idempotency of$\cdot$). 
A term $t$ is \emph{normalized} if $t=\norm{t}$. 
Two terms $p$ and $q$ are \emph{equivalent}, if $\norm{p}=\norm{q}$. 
Given a  set of terms $T$ we define $\norm{T}=\set{\norm{t}: t\in T}$.
More details are given in Section \ref{sect:resolutionofcs}.

We define a substitution $\sigma =\{x_1 \mapsto t_1, \dots, x_k \mapsto t_k \}$  
(where $x_i \in \UniVar{X}$ and $t_i \in \Universe$) to be the mapping $\sigma : \Universe \rightarrow \Universe$, 
such that $t\sigma$ is a term obtained by replacing, for all $i$,  each occurrence of variable $x_i$  by the corresponding term $t_i$.
The set of variables $\set{x_1, \dots, x_k}$ is called the \emph{domain} of $\sigma$ and denoted by $\dom{\sigma}$.
If $T \subseteq \Universe$, then by definition $T\sigma = \set{t\sigma : t\in T}$.
A substitution $\sigma$ is \emph{ground} if for any $i \in \set{1,\dots,k}$, $t_i$ is ground.
We will say that the substitution $\sigma$ is normalized, if for all $x\in\dom{\sigma}$, $x\sigma$ is normalized.

\paragraph{Agents}
We will call communicating parties \emph{agents}. Every agent is identified by its name. We denote a set of agent names as $A$.

\paragraph{Channels}
Any two agents $a$ and $b$ communicate through a channel denoted as $\channel{a}{b}$.  
We will suppose, that channels are directed. The set of all channels is denoted as $\mathbb{C}$.
A channel supports a queue of messages: for example, 
if $a$ sends sequentially two messages  to $b$ (via channel  $\channel{a}{b}$), then $b$ cannot process the second message before  the first one; the sent messages are stored in queue to be processed in order of arrival. 

\paragraph{Agents behavior}
We define a protocol session $PS = \set{\tuple{a_i,l_i}}_{i=1,\dots,k}$ as a finite set of pairs of an agent name and 
a finite list of actions to be executed by this agent\footnote{For simplicity,  we suppose that for a protocol session, one agent cannot have more than one list of actions to execute, but this restriction can be relaxed.}.  
We also suppose that $\vars{l_i}\cap \vars{l_j} = \emptyset$, for all $i\neq j$ (where $\vars{\cdot}$ is naturally extended on lists of actions).

Every action is of receiving type $\rcv{f}{r}$ or sending type $\snd{t}{s}$, where 
\begin{itemize}
 \item $f$ is an agent name, whom a message is to be received from;
 \item $r$ is a term (a template for the message) expected to be received from $f$;
 \item $t$ is an agent name, whom the message is expected to be sent to;
 \item $s$ is a term (a template for the message) to be sent to $t$.
\end{itemize}

Let us consider any agent $a\in A$ participating in the protocol session $PS$ and let $\tuple{a, \set{\rho_i}_{i=1,\dots,k}} \in PS$. 

Case 1. If $\rho_1 = \rcv{f_1}{r_1}$ then the first action  agent $a$  can do, is 
to accept  a message $m$, admittedly from agent $f_1$ on channel $\channel{f_1}{a}$, matching the pattern $r_1$, i.e. such that $\norm{r_1\sigma}=\norm{m}$ 
for some substitution $\sigma$. Agent is blocked (does not execute any other actions) by awaiting a message. 
If $a$ receives a message that does not match the expected pattern, then $a$ terminates his participation in $PS$. %
Note that no notification is sent to the sender, thus a sender continues his execution%
\footnote{A way to model another behavior, is to explicitly provide for every sending a succedent receive of an acknowledge message 
and for every receive a succedent send of an acknowledge message.}.
Once $a$ has received message $m$  matching the pattern $r_1$ with substitution $\sigma$, he
instantiate $\vars{r_1}$ with $\sigma$ and execute his remaining actions using these values, i.e. $a$ moves to a state 
where the list of actions to be executed is  $\set{\rho_i\sigma}_{i=2,\dots,k}$, with 
\[
\rho\sigma=
 \begin{cases}
 	\rcv{f}{(r\sigma)}, & \mbox{if } \rho = \rcv{f}{r};\\
  	\snd{t}{(s\sigma)}, & \mbox{if } \rho = \snd{t}{s}.\\
 \end{cases}
\] 
We will say that an action $\rho$ is ground, if $\rho=\rcv{f}{r}$ and $r$ is a ground term; or $\rho=\snd{t}{s}$ and $s$ is ground.

Case 2. If $\rho_1$ is  $\snd{t_1}{s_1}$ then the first action of agent $a$ is  sending  message $s_1$ to agent $t_1$ 
(i.e. putting it to channel $\channel{a}{t_1}$) 
and then, moving to a state where $\set{\rho_i}_{i=2,\dots,k}$ has to be executed.

We suppose that agents cannot have a sending pattern that contains variables not instantiated before, i.e.
for any $\tuple{a,\rho_1.\cdots.\rho_{k_a}}\in PS$
if $\rho_i=\snd{t}{s}$ then 
for any variable $x\in \vars{s}$ there exists $j<i$ such that $\rho_j=\rcv{f}{r}$ and $x\in\vars{r}$.

\paragraph{Intruder model}

We assume that some communication channels are controlled  by $N$ \emph{local} intruders $\set{I_i}_{i=1,\dots,N}$ 
and there is no channel  controlled by more than one intruders.
We introduce an \emph{intruders layout} represented by a function $\iota: \mathbb{C} \mapsto \mathbb{I}\cup\set{\varnothing}$
mapping every channel to the local intruder 
that controls it if there is one,  to $\varnothing$ otherwise.

Every intruder $I$ is given some initial knowledge $K^0_I$ that is a set of ground terms. 
Once an agent sends a message via a channel controlled by intruder, the intruder reads it and blocks it.
Reading the message means extending intruder's current knowledge with this message.
An intruder controlling a  channel can generate a message from his knowledge using 
deduction rules  and send it to its endpoint.

We now specify the intruder capabilities:
\begin{df}
A \emph{rule} is a tuple of terms written as $s_1,\dots,s_k \rightarrow s$, where $s_1,\dots,s_k, s$ are terms. 
A \emph{deduction system} $\UniVar{D}$ is a set of rules.
\end{df}

From now to the end of this section  rules are assumed to belong to  a fixed deduction system $\UniVar{D}$.

\begin{df}
A \emph{ground instance} of a rule $d  = s_1,\dots,s_k \rightarrow s$ is a rule $l = l_1,\dots, l_k \rightarrow r$ where $l_1,\dots, l_k, r$ are ground terms and there exists $\sigma$ --- ground substitution, such that $l_i = s_i\sigma, $ for  $i=1,\dots,k$ and $r=s\sigma$. 
We  call a ground instance of a rule a \emph{ground rule}.%
\end{df}

Given two sets of ground terms $E$, $F$ and a rule $l \rightarrow r$, we write $E \rightarrow_{l\rightarrow r} F$ iff $F=E\cup \set{r}$ and $l \subseteq E$, where $l$ is a set of terms. We write $E \rightarrow F$ 
iff there exists rule $l \rightarrow r$ such that $E \rightarrow_{l\rightarrow r} F$.

\begin{df}
 A \emph{derivation} $D$ of length $n \geq 0$ is a sequence of finite sets of ground terms $E_0, E_1,\dots, E_n$ such that $E_0 \rightarrow E_1 \rightarrow \cdots \rightarrow E_n$, where $E_i = E_{i-1} \cup \set{t_i}, \forall i = \set{1,\dots,n}$.
 A term $t$ is \emph{derivable} from a set of terms $E$  iff there exists a derivation $D  = \,\, E_0,\dots,E_n$ such that 
$E_0 = E$ and $t \in E_n$.
A set of terms $T$ is derivable from $E$, iff every   $t \in T$ is derivable from $E$.
 We denote $\der{E}$ set of terms derivable from $E$.
\end{df}

Local intruder $I$ can send a message $m$, if $m\in\der{K_I}$, where $K_I$ is a current knowledge of intruder $I$.

\paragraph{Protocol session execution}
Now,%
we \LONG{can}\SHORT{could} present a course of a protocol execution.
We  first introduce a notion of \emph{symbolic} execution, where data exchanged among the agents and intruders are not instantiated and represented as (possibly non-ground) terms. This execution is constrained by some conditions.
Whenever these conditions are satisfied by an appropriate ground instantiation of variables, 
we obtain a \emph{concrete} execution, or simply an \emph{execution}.
These conditions 
are defined by \emph{constraint systems}: 
\begin{df}
Let $E$ be a set of terms and $t$ be a term, we define the couple $(E,t)$ denoted  $E \rhd t$ to be  a \emph{constraint}. 
A \emph{constraint system} is a set $$\ConstrSys{S} = \set{E_i \rhd t_i}_{i=1, \dots, n}$$
where $n$ is an integer and  $E_i \rhd t_i$ is  a  constraint  for  $i \in \{1,\dots,n\}$. 
\end{df}

We  extend the definition of $\vars{\cdot}$  to constraint system $\ConstrSys{S}$ in a natural way. 
We  say that $\ConstrSys{S}$ is normalized, if every term  in $\ConstrSys{S}$ is normalized. 
By $\norm{\ConstrSys{S}}$ we will denote a constraint system $ \set{\norm{E_i} \rhd \norm{t_i}}_{i=1, \dots, n}$.

\begin{df}
A ground substitution $\sigma$ is a \emph{model} of constraint $E \rhd t$ (or $\sigma$ satisfies this constraint), if $\norm{t\sigma} \in \der{\norm{E \sigma}}$.
A ground substitution $\sigma$ is a \emph{model} of a constraint system $\ConstrSys{S}$, if it satisfies all the constraints of $\ConstrSys{S}$ 
and $\dom{\sigma} = \vars{\ConstrSys{S}}$. 
\end{df}

\begin{df}\label{def:stateOfExec}
 A configuration $\Pi$ of a protocol session  is a quadruple 
$\tuple{PS, \mathcal{K}, \mathcal{Q}, \mathcal{S}}$,
where $\mathcal{K}=\set{\tuple{I_i, K_{i}}}_{i=1,\dots,N}$ represents current knowledges of intruders,
and $\mathcal{Q}=\set{\tuple{c, m_c}}_{c\in \mathbb{C}}$ is a configuration of channels: for every channel $c$ queue of messages $m_c$ is given.
\end{df}

\emph{Transitions} on configurations are defined in Table\footnote{$\uplus$ represents the union of two disjoint sets: $A \uplus B = A \cup B$ iff $A\cap B = \emptyset$.}~\ref{tab:conftransitions} and will be explained later. 
Transitions are written in form $\Pi_1 \xrightarrow{cond} \Pi_2$
and state that configuration $\Pi_1$ can evolve to a new configuration $\Pi_2$ if condition $cond$ is satisfied.

\begin{table*}
\centering
\begin{tabular}{|l l|}
 \hline
 \tableitem\label{tr:isends}&
    $\tuple{\set{\tuple{a, (\rcv{f}{r}).l_a}} \uplus PS, \set{\tuple{I,K}}\uplus\mathcal{K}, \mathcal{Q}, \mathcal{S}} \xrightarrow{\iota(\channel{f}{a}) = I}$ \LONG{ \\  \lampersand }
      $\tuple{\set{\tuple{a, l_a}} \cup PS, \set{\tuple{I,K}}\cup\mathcal{K}, \mathcal{Q}, \mathcal{S}\cup\set{K \rhd r}} $
\\
 \tableitem\label{tr:iintercepts}& 
    $\tuple{\set{\tuple{a, (\snd{t}{s}).l_a}} \uplus PS, \set{\tuple{I,K}}\uplus\mathcal{K}, \mathcal{Q}, \mathcal{S}} \xrightarrow{\iota(\channel{a}{t}) = I}$ \LONG{\\ \lampersand} 
    $\tuple{\set{\tuple{a, l_a}} \cup PS, \set{\tuple{I,K\cup{s}}}\cup\mathcal{K}, \mathcal{Q}, \mathcal{S}}$ 
    \\
 \tableitem\label{tr:asends}& 
    $\tuple{\set{\tuple{a, (\snd{t}{s}).l_a}} \uplus PS, \mathcal{K}, \set{\tuple{\channel{a}{t}, m_{\channel{a}{t}}}}\uplus\mathcal{Q}, \mathcal{S}} \xrightarrow{\iota(\channel{a}{t}) = \varnothing}$ \LONG{\\ \lampersand} 
    $ \tuple{\set{\tuple{a, l_a}} \cup PS, \mathcal{K}, \set{\tuple{\channel{a}{t}, m_{\channel{a}{t}}.s}}\cup\mathcal{Q}, \mathcal{S}}$
    \\
 \tableitem\label{tr:areads}&
$\tuple{\set{\tuple{a, (\rcv{f}{r}).l_a}} \uplus PS, \mathcal{K}, \set{\tuple{\channel{f}{a}, s.m_{\channel{f}{a}}}}\uplus\mathcal{Q}, \mathcal{S}} 
	 \xrightarrow{\iota(\channel{f}{a}) = \varnothing }$  \\ &
     $\tuple{\set{\tuple{a, l_a}} \cup PS, \mathcal{K}, \set{\tuple{\channel{f}{a}, m_{\channel{f}{a}}}}\cup\mathcal{Q}, \mathcal{S}\cup\set{\set{\enc{s}{k}} \rhd \enc{r}{k}}}$
    \\
 \hline
\end{tabular}
\caption{Configuration transitions}%
\label{tab:conftransitions}
\end{table*}

\begin{df} \label{def:symexecution}
A \emph{symbolic execution} $E^S_{PS}$ of protocol session $PS$ (with intruders layout  $\iota$) is 
a sequence of configurations obtained by application of transitions to the initial configuration  $\tuple{PS, \set{\tuple{I,K^0_I}}_{I\in\mathbb{I}}, \set{\tuple{c,\emptyset}}_{c\in\mathcal{C}}, \emptyset}$.
\end{df}

For a substitution $\sigma$ and a configuration $\Pi = \br \tuple{\set{\tuple{a_i, l_i}}_{i=1,\dots,k}, \set{\tuple{I_i, K_{i}}}_{i=1,\dots,N}, \set{\tuple{c, m_c}}_{c\in \mathbb{C}}, \set{E_i \rhd t_i}_{i=1, \dots, n}}$  we define  $\Pi\sigma$ as 
$\tuple{\set{\tuple{a_i, l_i\sigma}}_{i=1,\dots,k}, \set{\tuple{I_i, K_{i}\sigma}}_{i=1,\dots,N}, \set{\tuple{c, m_c\sigma}}_{c\in \mathbb{C}}, \set{E_i\sigma \rhd t_i\sigma}_{i=1, \dots, n}}$, \br
where  substitutions are applied to lists elementwise.

\begin{df}\label{def:execution}
An \emph{execution} $E_{PS}=\set{C_i\sigma}_{i=1,\dots,m}$ is an instance of a symbolic execution
 $\set{C_i}_{i=1,\dots,m}$ (where $C_i=\tuple{PS_i, \mathcal{K}_i, \mathcal{Q}_i, \mathcal{S}_i}$)
such that all terms of $C_i\sigma$ are ground and $\mathcal{S}_m$ is satisfied by $\sigma$.
\end{df}

Now we describe the transitions of Table~\ref{tab:conftransitions}.
Transition \ref{tr:isends} expresses the possibility of intruder $I$ controlling channel $\channel{f}{a}$ to impersonate $f$ and send to $a$ some message compliant with the expected by $a$ pattern $r$,
if the current knowledge of $I$ allows it. 
An intruder can also intercept messages sent on the channel that he controls (Transition \ref{tr:iintercepts}).
A message sent by an agent on the channel free from intruders is put to the end of the queue of this channel (Transition~\ref{tr:asends}).
Transition \ref{tr:areads} represents the reading of a message from the queue of the channel.

Let us explain where the constraint $\set{\enc{s}{k}} \rhd \enc{r}{k}$ comes from in the last transition.
Agent $a$ expects to read a message from the channel compatible with the pattern $r$. 
The first (possibly not yet instantiated) message  in the queue is $s$. 
Thus, $r$ and $s$ must be unifiable (modulo considered equational theory),
and even equivalent when we consider ground instances of the symbolic executions.
Since we will be interested only in concrete executions, but not symbolic,
we can use this constraint to express equivalence between $r$ and $s$ (Lemma~\ref{lemma:equalVSconst}).
\begin{lemma}\label{lemma:equalVSconst}
For terms $t_1$,  $t_2$ and substitution $\sigma$,
$\norm{t_1\sigma} =  \norm{t_2\sigma}$ is true
iff $\sigma$ is a model of  $\set{\enc{t_1}{k}} \rhd {\enc{t_2}{k}}$ for any term $k$, 
i.e.  $\norm{t_1\sigma} =  \norm{t_2\sigma} $ iff $ \norm{\enc{t_1}{k}\sigma}\in\der{\set{\norm{\enc{t_2}{k}\sigma}}}$. 
\end{lemma}

\paragraph{Offline communication}
At some point
the current knowledge of all local intruders can be shared to derive a secret  
which probably they cannot deduce separately.
In some cases these offline interactions are time-consuming and may be detected. Therefore 
we consider  reasonable that in the intruder strategy modelling  they take place after the protocol is over.

\paragraph{Coordinated attack problem}

Now we can formally state the problem.

\noindent\textit{Input:}
A finite set of agents $A$, a protocol session $PS = \set{\tuple{a_i, l_i}}_{i=1,\dots,k}$, a set of intruders $\mathbb{I}=\set{I_i}_{i=1,\dots,N}$ each 
with  initial knowledge  $K^0_{I_i}$, an intruder layout  $\iota$
and  some sensitive data given as a finite set of ground terms $S$. %
\noindent\textit{Output:}
$s\in S$ and an execution 
 $E_{PS}$
  of protocol session $PS$ with its last configuration
$\tuple{PS, \mathcal{K}, \mathcal{Q}, \mathcal{S}}$  such that
$s\in\der{\bigcup_{\tuple{K_I,I}\in \mathcal{K}} K_I}$.

\subsubsection{Solving the problem}
\label{ssec:mintruders-solving}

We proceeds as follows:
\begin{enumerate}
 \item Guess a sensitive datum $s$ from $S$. 
 \item Guess a symbolic execution $E^S_{PS}$ of some length $\leq \sum_{\tuple{a, l} \in PS }\limits\operatorname{length}(l) < \infty$.
 \item For the last configuration $\tuple{PS, \set{\tuple{K_I,I}}_{I\in\mathbb{I}}, \mathcal{Q}, \mathcal{S}}$ of $E^S_{PS}$, 
       if constraint system $\mathcal{S}\cup\set{\bigcup_{\tuple{K_I,I}\in \mathcal{K}} K_I \rhd s}$ is satisfiable with some $\sigma$, 
       then the protocol session is insecure and we return $E_{PS} = E^S_{PS}\sigma$. 
\end{enumerate}

We will show in Sections~\ref{sect:resolutionofcs} and \ref{sect:complexity} that the satisfiability of constraint systems is in $NP$ in the case of DY+ACI deduction theory.

\subsection{Attack exploiting XML format of messages}
\label{ssec:xmlinj}

Here we show how to model  (using our formalism) attacks based on an XML-representation of messages.
A different technique to handle this kind of attacks was presented in \cite{ChevalierLR07}.

We consider an e-shop that accepts e-cheques,
and we suppose that it is presented by a Web Service
using SOAP protocol for exchanging messages.

It consists of two services: 
\begin{itemize}
 \item the first exposes the list of goods for sale with their prices and process the orders by accepting payment, 
 \item the second is a delivery service; it receives information from the first one about successfully paid orders, and sends the ordered goods to the buyer.
\end{itemize}

\begin{figure}
\includegraphics[width=\columnwidth]{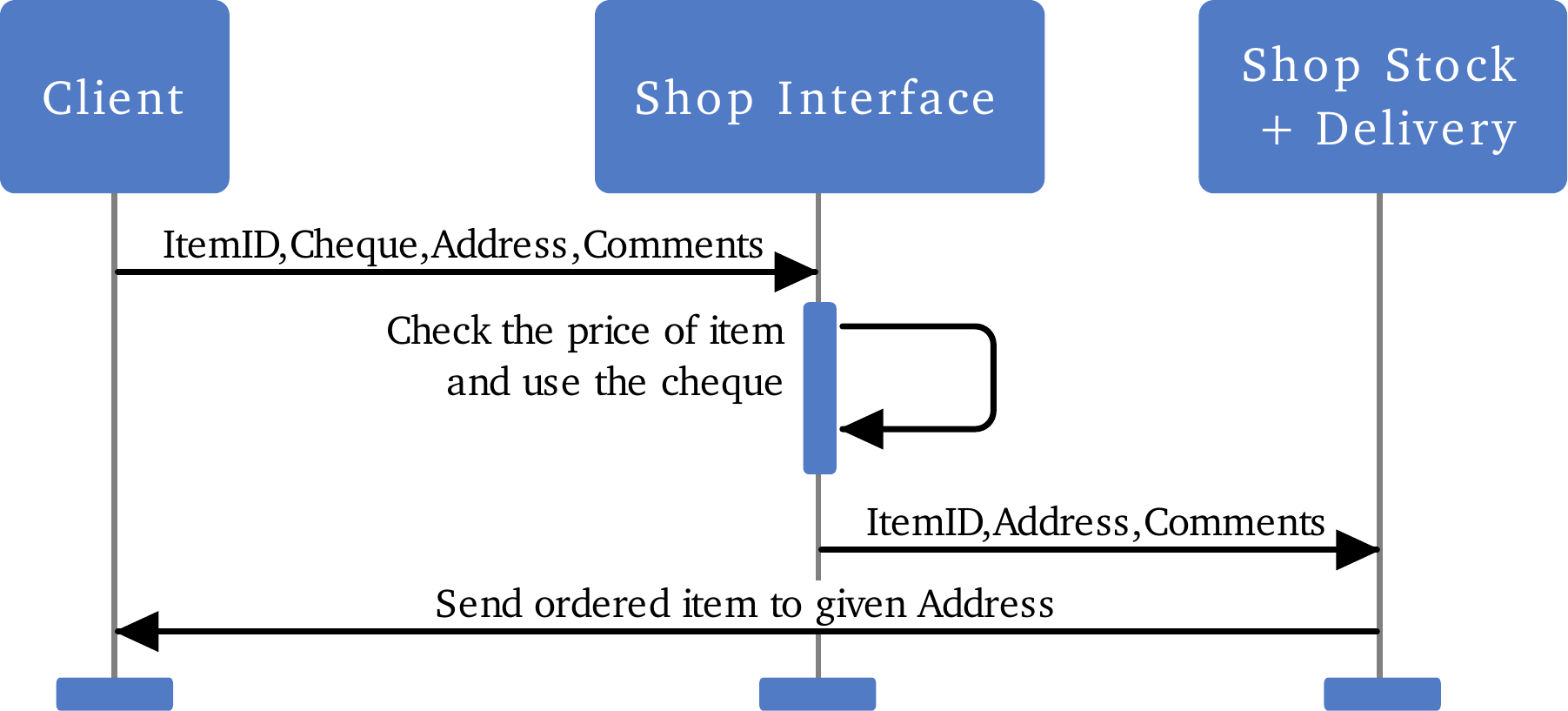}	
 \caption{Ordering item scenario}\label{fig:exInj}
\end{figure}

A simple scenario for ordering item is shown in \figurename{}~\ref{fig:exInj}.
First, a client sends an order using e-shop interface 
that consists of an item identifier, e-cheque, delivery address and some comments. 
Then, the first service of the e-shop checks whether the price of the ordered item corresponds to 
the received cheque. If it does, the service consumes the cheque 
and 
resends the order to the stock/delivery service (without the used e-cheque).
Stock and delivery service prepare a parcel with ordered item and send it to given address. 
The comment is automatically printed on the parcel to give some information to the postman about,
for example, delivery time or access instructions.

Suppose, Alice has an e-cheque for 5\officialeuro. She selected a simple pen (with ItemID simple) to buy, 
but she liked very much a 
more expensive 
gilded one (with ItemID gilded). 
Can we help Alice to get what she wants for  what she has?

Let us formalize the behaviour of scenario players (terms, normalization function and deduction system are defined as in \S~\ref{subs:def2} except that we will write $(t_1 \opaci \dots \opaci t_n)$ instead of $\aci{\lst{t_1,\dots,t_n}}$).
Identifiers starting from a capital letter are considered as variables; numbers and identifier starting from lower-case letter are considered as constants.
We model a delivery of item with some $ItemID$ to address $Address$ with comments $Comments$ by the following message:
$\sig{(ItemID\opaci Address\opaci Comments)}{\priv{k_s}}$ --- a message signed by e-shop, where $k_s$ its public key, such that no one can produce this message except the shop.
We abstract away from the procedure of checking price of the item and will suppose, that Shop Interface expects 5\officialeuro{} e-cheque for 
Item ``$simple$''.  For simplicity we assume only two items.

We will use notation for sending and receiving as in \S~\ref{ssec:severalintruders}.

For Shop Interface we have:
\begin{align*} 
&\rcv{Client}{(simple \opaci cheque5 \opaci IAddr\opaci IComm)}; \\
&\snd{Delivery}{(simple \opaci IAddr \opaci IComm)}.
\end{align*}
For Shop Stock/Delivery we have:
\begin{align*} 
&\rcv{Interface}{(DItemID\opaci DAddr \opaci DComm)}; \\
&\snd{Client}{\sig{(DItemID\opaci DAddr \opaci DComm)}{\priv{k_s}}}.
\end{align*}

Alice initially has:

\begin{centering}
\begin{tabular}{l c l}
 $simple,gilded$: & & identifiers of items;\\
 $cheque5$: & & an e-cheque for 5\officialeuro;\\
 $addr$: & & her address;\\
 $cmnts$: & & residence digital code;\\
 $k_s$:  & & a public key of the shop.
\end{tabular}

\end{centering}

Now we build a mixed constraint system (derivation constraints and equations) to know, whether Alice can do what she wants\LONG{:}\SHORT{\ (see \figurename{}~\ref{fig:exCS}).}
\LONG{%
\begin{empheq}[left=\left\lbrace, right=\right\rbrace]{align}
 & {\set{gilded,simple,cheque5,addr,cmnts,k_s} \  \rhd}  \LONG{ \nonumber \\}
  &   \LONG{{\hphantom{{\sig{gilded adi D}{\priv{k_s}}}}}}{ {(simple \opaci cheque5 \opaci IAddr \opaci IComm)}} \label{XML:constr1} \\
  & {(simple \opaci IAddr \opaci IComm) =_{ACI}}  
      {(DItemID\opaci DAddr\opaci DComm)} \label{XML:constr3}\\
   &{\left\{gilded,simple,cheque5,addr,cmnts,k_s, \right.}   \nonumber\\
   &{{\sig{(DItemID\opaci DAddr \opaci DComm)}{\priv{k_s}}}\} \  \rhd } \LONG{\nonumber \\  }
   &   \LONG{\hphantom{{simple  cheque5 IAdi \opaci I}}}     {{\sig{(gilded\opaci addr\opaci DComm)}{\priv{k_s}}}} \label{XML:constrgilded}  
\end{empheq}
}%
\SHORT{%
\begin{figure*}
\begin{empheq}[left=\left\lbrace, right=\right\rbrace]{align}
 & \set{gilded,simple,cheque5,addr,cmnts,k_s} \  && \rhd   &  { {(simple \opaci cheque5 \opaci IAddr \opaci IComm)}} \label{XML:constr1} \\
  & (simple \opaci IAddr \opaci IComm) 		 && =_{ACI}  &     {(DItemID\opaci DAddr\opaci DComm)} \label{XML:constr3}\\
   &{\left\{gilded,simple,cheque5,addr,cmnts,k_s, \right.}   \nonumber\\
   & {\sig{(DItemID\opaci DAddr \opaci DComm)}{\priv{k_s}}}\} \  && \rhd    &      {{\sig{(gilded\opaci addr\opaci DComm)}{\priv{k_s}}}} \label{XML:constrgilded}  
\end{empheq}
\caption{A constraint system describing possible vulnerability in E-shop scenario }\label{fig:exCS}
\end{figure*}
}%

Constraint \eqref{XML:constr1} shows, that Alice can construct a message expected by the shop from a client.
Constraint \eqref{XML:constr3} represents a request from the first to the second service of the shop: left-hand side is a message sent by 
the interface service, and right-hand side is a message expected by stock/delivery subservice.
The last constraint shows, that from the received values Alice can build a message that models a delivery of item with ItemID gilded.

To solve it, we first get rid of syntactic equations by applying most general unifier; 
and then of equations modulo ACI ($t_1 =_{ACI} t_2$ is equivalent to $\norm{t_1} = \norm{t_2}$) by encoding them into deduction rules (as it was done in \S~\ref{ssec:mintruders-solving}).

Then, one of the solutions is:
\begin{align*}
 & IAddr   &  \mapsto &  addr   &  & IComm & \mapsto  & (gilded\opaci cmnts) \\
 & DItemID & \mapsto  &  gilded &  & DAddr & \mapsto  & addr \\
 &         &          &         &  & DComm & \mapsto  & (simple\opaci cmnts)  
\end{align*}

From this solution we see, that Alice can send a not well-formed comments (that presents two XML-nodes), 
and Delivery service parser can choose an entry with ID gilded.
An attack-request can look like this:

\lstset{language=XML}

\begin{lstlisting}
    <ItemID>simple</ItemID>
    <Cheque>cheque5</Cheque>
    <Address>addr</Address>
    <Comments>cmnts</Comments>
    <ItemID>gilded</ItemID>
\end{lstlisting}

The parser of the first service can return value of the first occurrence of \lstinline!ItemID!: \lstinline!<ItemID>simple</ItemID>!.
But the parser of the second one can return \lstinline!<ItemID>gilded</ItemID>!.

This attack is possible, if Alice constructs a request ``by hand'',
but a similar attack is probably feasible using XML-injection:
Alice when filling a request form enters instead of her comments the following string:
\begin{lstlisting}
   cmnts</Comments>
   <ItemID>gilded</ItemID><Comments>
\end{lstlisting}
and in the resulting request we get:
\begin{lstlisting}
    <ItemID>simple</ItemID>
    <Cheque>cheque5</Cheque>
    <Address>addr</Address>
    <Comments>cmnts</Comments>
    <ItemID>gilded</ItemID><Comments>
    </Comments>
\end{lstlisting}

This kind of XML-injection attacks was described in \cite{owaspv3}.

\section{Satisfiability of  general DY+ACI constraint systems}
\label{sect:resolutionofcs}

In Section~\ref{sec:motiv} we reduced the  problem of protocol  insecurity in presence of several intruders 
to solving  a system of deducibility constraints. 
In this section we present a decision procedure for a constraint system where  
Dolev-Yao deduction system is extended by an associative-commutative-idempotent symbol (DY+ACI).
We consider operators for pairing, symmetric and asymmetric encryptions, decryption, signature 
and an  ACI operator that will be used as a set constructor.

As for the proof structure, after  introducing the formal notations, the main steps to show the decidability are as follows:
\begin{enumerate}
 \item We present an algorithm for solving a ground derivability in DY+ACI model.
 \item We prove, that the normalization does not change satisfiability: either we normalize a model or a constraint system.
 \item We show existence of a conservative solution of satisfiable constraint system: a substitution $\sigma$ that sends a variable to an ACI-set of quasi-subterms of the constraint system instantiated with $\sigma$ 
  together with $\oppriv$-ed atoms of the constraint system;
 \item We give a bound on size of a conservative solution, and, as consequence, we obtain decidability.
\end{enumerate}

\subsection{Formal introduction to the problem}

\subsubsection{Terms and notions}\label{subs:def2}

\begin{df}\label{def:term}
\emph{Terms}  are defined according to the following grammar:
\LONG{
 \begin{align*} 
& term & ::= & \, variable \, | \, atom \,|\, \pair{term}{term}  \,|\, \\
& & &  \enc{term}{term} \,|\,  \opaci(tlist) \,|\,  \priv{Keys}  \,|\, \\
& &  &  \aenc{term}{Keys} \,|\, \sig{term}{\priv{Keys}} \\
& Keys  & ::= & \, variable \, | \, atom \\
& tlist & ::= & \,  \ term \, | \, term,\, tlist \, 
\end{align*}
}%
\SHORT{
 \begin{align*} 
\SHORT{&} term \SHORT{&} ::= \SHORT{&} \, variable \, | \, atom \,|\, \pair{term}{term}  \,|\, \SHORT{\\}
\SHORT{&} \SHORT{&} \SHORT{&}  \enc{term}{term} \,|\,  \opaci(tlist) \,|\,  \priv{Keys}  \,|\, \\
\SHORT{&} \SHORT{&}  \SHORT{&}  \aenc{term}{Keys} \,|\, \sig{term}{\priv{Keys}} \\
\SHORT{&} Keys  \SHORT{&} ::= \SHORT{&} \, variable \, | \, atom \\
\SHORT{&} tlist \SHORT{&} ::= \SHORT{&} \, \{ \ term \ \llbracket, \  term\rrbracket* \, \}
\end{align*}
}%
where
$atom \in \UniAt{A}$ and
$variable \in \UniVar{X}$.
 We denote $\Universe(\UniAt{A},\UniVar{X})$ the set of all terms over 
a set of atoms $\UniAt{A}$ and a set of variables $\UniVar{X}$. 
For short, we write $\Universe$ instead of $\Universe(\UniAt{A},\UniVar{X})$.

\end{df}

By $\sig{p}{\priv{a}}$ we mean a signature of message $p$ with private key $\priv{a}$ 
We do not assume that one can retrieve the  message itself from the signature.

Note that we do allow complex keys for symmetric encryption only.
As a consequence, we have to introduce a condition on substitution applications: 
substitution $\sigma$ cannot be applied to the term $t$, if after replacing the resulting entity is not a term 
(for example, we cannot apply $\sigma=\set{x\mapsto \pair{a}{b}}$ to the term $\aenc{a}{x}$).

We denote a term on $i$-th position of a list $L$ as $L[i]$. 
Then $t\in L$ is a shortcut for $\exists i: t = L[i]$. 
We also define two binary relations $\subseteq$ and $\approx$ on lists as follows:
$L_1\subseteq L_2$ if and only if  any $t\in L_1$ implies  $t\in L_2$;
$L_1 \approx L_2$ if and only if $L_1 \subseteq L_2$ and $L_2 \subseteq L_1$,
and naturally extend them if $L_1$ or $L_2$ is a set. 

\begin{df}
We consider symbol $\opaci$ to be 
	associative,
	commutative,
	idempotent
(shortly, $ACI$).
\end{df}

We will use $\opbin$ throughout the paper as a generalization of all binary operators: $\opbin\in\set{\openc,\opaenc,\oppair,\opsig}$.

\begin{df}

 For every term $t\in \Universe$ we define its root symbol by 
\[
 \rut{t} = \left\{ 
 \begin{array}{rl} 
 \opbin, & \mbox{ if } t= \bin{p}{q} \\
 \opaci, &\mbox{ if } t= \aci{L},\\
 \oppriv, &\mbox{ if } t= \priv{p},\\
 t, &\mbox{ if } t \in \UniVar{X}\cup \UniAt{A},\\
\end{array}
\right. 
\]

\end{df}

\begin{df}\label{def:elems}
	For any term $t\in\Universe$ we define its \emph{set of elements} by:%
\[
\elems{t} =
 \begin{cases}
 	\bigcup_{p\in L} \elems{p}\, & \mbox{if }  t=\aci{L};\\
  	\set{t} , & \mbox{otherwise. }\\
 \end{cases}
\]
We extend $\elems{}$ to sets of terms or lists of terms $T$ by $\elems{T} = \bigcup_{t\in T} \elems{t}$. 
\end{df}

\begin{example}\label{ex:term}
 For term $t=\aci{\lst{a, \aci{\lst{b,a,\pair{a}{b}}},\pair{\aci{\lst{b,b}}}{a}}}$
set of its elements is $\elems{t}=\set{a,b,\pair{\aci{\lst{b,b}}}{a},\pair{a}{b}}$.
\end{example}

\begin{df}
Let $\prec$ be a strict total order 
 on $\Universe$,
such that comparing can be done in polynomial time. %

\end{df}

\begin{df}
The cardinality of a  set  $P$ is denoted by $\card{P}$.
\end{df}

\begin{df} \label{def:norm}
The \emph{normal form} of a term $t$ (denoted by $\norm{t}$) is recursively defined by:
\begin{itemize}
	\item $\norm{t}=t$, if $t\in \UniVar{X}\cup\UniAt{A}$
	\item $\norm{\bin{t_1}{t_2}} = \bin{\norm{t_1}}{\norm{t_2}} $%
	\item $\norm{\priv{t}}=\priv{\norm{t}}$
	\item 
\SHORT{
	$\norm{\aci{L}}= 
	     \begin{cases}
		\aci{L'},& \mbox{if } \card{\norm{\elems{{L}}}}>1 \SHORT{\\ &} \mbox{ and }  L'\approx  \norm{\elems{{L}}}\\
		  & \mbox{ and for all } i<j,\  L'[i]\prec L'[j];\\
		t',& \mbox{if } \norm{\elems{{L}}}=\set{t'}
	      
	     \end{cases},
	$
}%
\LONG{
	$\norm{\aci{L}}= 
	     \begin{cases}
		\aci{L'},& \mbox{if } \card{\norm{\elems{{L}}}}>1  \mbox{ and }  L'\approx  \norm{\elems{{L}}}\\
		  & \mbox{ and for all } i<j,\  L'[i]\prec L'[j];\\
		t',& \mbox{if } \norm{\elems{{L}}}=\set{t'}
	     \end{cases},
	$
}%

\end{itemize}
where for set of terms $T$, $\norm{T} = \set{\norm{t}:t\in T}$.
	
\end{df}

We can show easily that two terms  are congruent modulo the ACI properties of $''.''$ iff they have the same normal form. 
Other properties are stated in Lemma~\ref{lemma:normprop}. 

\begin{example}
 Referring to Example~\ref{ex:term} for the value of term $t$, we have
$\norm{t}=\aci{\set{a,b,\pair{a}{b},\pair{b}{a}}}$.
\end{example}

\begin{df}
 Let $t$ be a term. We define a set of quasi-subterms $\sub{t}$ as follows:
\[
\sub{t} =
 \begin{cases}
 	\{t\}, & \mbox{if } t \in \UniVar{X} \cup \UniAt{A};\\
  	\{t\} \cup \sub{t_1}, & \mbox{if } t = \priv{t_1};\\
 	\{t\} \cup \sub{t_1} \cup \sub{t_2}, & \mbox{if } t = \bin{t_1}{t_2} \\
 	\{t\} \cup \bigcup_{p\in\elems{L}}\sub{p}, & \mbox{if } t = \aci{L}\\	
 \end{cases}
\]
If $T$ --- set of terms, then $\sub{T} = \bigcup_{t \in T}{\sub{t}}$. 
If $\ConstrSys{S}=\set{E_i\rhd t_i}_{i=1,\dots, n}$ is a constraint system, we define $\sub{\ConstrSys{S}} = \bigcup_{t \in \bigcup_{i=1}^n E_i\cup\set{t_i}}{\sub{t}}$.
\end{df}

\begin{example}
 Referring to Example~\ref{ex:term}, we have
\begin{align*}
\sub{t}=\{
\aci{\lst{a, \aci{\lst{b,a,\pair{a}{b}}},\pair{\aci{\lst{b,b}}}{a}}}, \\
a,b,\pair{a}{b},\pair{\aci{\lst{b,b}}}{a}, \aci{\lst{b,b}}
\}.
\end{align*}
\end{example}

\begin{df}\label{df:vars}
 Let $t$ be a term. We define $\vars{t}$ as set of all the variables in $t$:
\[
\vars{t} = \UniVar{X} \cap \subII{t}
\]
\end{df}

We define $\osubtermsII(t)$ as the set of subterms of $t$ 
and the  DAG-size of a term, as the number of its different subterms. 
The DAG-size gives the size of a natural representation of a term in the 
considered ACI theory. 

\begin{df}\label{df:subdag}
 Let $t$ be a term. We define $\subII{t}$ as follows:
\[
\subII{t} =
 \begin{cases}
    \{t\}, & \mbox{if } t \in \UniVar{X} \cup \UniAt{A};\\
    \{t\} \cup \subII{t_1}, & \mbox{if } t = \priv{t_1};\\
    \{t\} \cup \subII{t_1} \cup \subII{t_2}, & \mbox{if } t = \bin{t_1}{t_2}\\
    \{t\} \cup \bigcup_{p\in L}\subII{p}, & \mbox{if } t = \aci{L}.
 \end{cases}
\]
If $T$ is a set of terms, then $\subII{T} = \bigcup_{t \in T}{\subII{t}}$.
If $\ConstrSys{S}=\set{E_i\rhd t_i}_{i=1,\dots, n}$ is a constraint system, we define $\subII{\ConstrSys{S}} = \bigcup_{t \in \bigcup_{i=1}^n E_i\cup\set{t_i}}{\subII{t}}$.
\end{df}

\begin{example}
 Referring to Example~\ref{ex:term}, we have
\begin{align*}
\subII{t}=\{
\aci{\lst{a, \aci{\lst{b,a,\pair{a}{b}}},\pair{\aci{\lst{b,b}}}{a}}}, \\
\aci{\lst{b,a,\pair{a}{b}}}, \pair{\aci{\lst{b,b}}}{a}, \LONG{\\}
a,b,\pair{a}{b}, \aci{\lst{b,b}}
\}.
\end{align*}
\end{example}

\begin{df}\label{df:sizedag}
    We define a DAG-size $\oDAGsize$ of a term $t$ as
\( \DAGsize{t} = \card{\subII{t}}\),
for set of terms $T$,
\( \DAGsize{T} = \card{\subII{T}}\)
and for constraint system $\ConstrSys{S}$ as
\( \DAGsize{\ConstrSys{S}} = \card{\subII{\ConstrSys{S}}}\).
\end{df}
Remark, that for a constraint system such a definition does not polynomially approximate a number of bits needed to write it down\LONG{ (cf. Def.~\ref{def:DAGSys})}.

We define a Dolev-Yao deduction system modulo ACI equational theory  (denoted DY+ACI). 
It consists of composition rules and decomposition rules, depicted in Table~\ref{tab:DYACI} 
where $t_1, t_2, \dots, t_m \in \Universe$.

\begin{table}[ht]
\centering
\begin{tabular}{|l|l|}
\hline
Composition rules & Decomposition rules \\
\hline
 $ {t_1, t_2} \rightarrow \norm{\enc{t_1}{t_2}}$ & ${\enc{t_1}{t_2},  \norm{t_2}} \rightarrow \norm{t_1}$ \\
 $ {t_1, t_2} \rightarrow \norm{\aenc{t_1}{t_2}}$ &  ${\aenc{t_1}{t_2},  \norm{\priv{t_2}}} \rightarrow \norm{t_1}$\\
 $ {t_1, t_2} \rightarrow \norm{\pair{t_1}{t_2}}$ & $ {\pair{t_1}{t_2}} \rightarrow \norm{t_1}$\\
 $ {t_1, \priv{t_2}} \rightarrow \norm{\sig{t_1}{\priv{t_2}}}$ & ${\pair{t_1}{t_2}} \rightarrow \norm{t_2}$\\ 
 $ {t_1, \dots, t_m} \rightarrow \norm{\aci{{t_1, \dots, t_m}}}$ & $\  \aci{{t_1, \dots, t_m}} \rightarrow \norm{t_i}$ for all $i$ \\
\hline
\end{tabular}
\caption{DY+ACI deduction system rules}\label{tab:DYACI}
\end{table}

We suppose, hereinafter,  that for a constraint system $\ConstrSys{S}$,  $\sub{\ConstrSys{S}} \cap \UniAt{A} \neq \emptyset$. Otherwise, we can add one constraint $\set{a} \rhd a$ to $\ConstrSys{S}$ which will be satisfied by any substitution.
We denote $\set{\priv{t}:t\in T}$ for set of terms $T$ as $\priv{T}$.
We define $\vars{\ConstrSys{S}} = \bigcup_{i=1}^n \vars{E_i} \cup \vars{t_i}$.
We say that $\ConstrSys{S}$ is normalized, iff for all $t\in \sub{\ConstrSys{S}}$, $t$ is normalized. 

\begin{example}\label{ex:constrsys}
 We give a sample of general constraint system and its solution within DY+ACI deduction system.
\[
   \ConstrSys{S}=\left\{ 
 \begin{array}{l l l}
	    { \enc{x}{a},\pair{c}{a}} & \rhd &  b \\
	    { \aci{\lst{x,c}}} & \rhd & a 
 \end{array}
  \right\},
\]  
where $a,b,c \in \UniAt{A}$ and $x \in \UniVar{X}$.
 One of the eventual models within DY+ACI is $\sigma = \set{x\mapsto \enc{\pair{a}{b}}{c}}$.
\end{example}

\begin{df}\label{def:pairing}
 Let $T=\set{t_1,\dots,t_k}$ be a non-empty set of terms. Then we define $\pairing{T}$ as follows:
 \[	
  \pairing{T} = \norm{\aci{{t_1, \dots{}, t_k}}}
 \]
Remark: $\pairing{\set{t}}=\norm{t}$.
 \end{df}

\begin{df}
	We denote $\sub{\ConstrSys{S}}\setminus \UniVar{X}$ as $\subo{S,\UniVar{X}}$ or, for shorter notation, $\subo{S}$.
\end{df}

We introduce a transformation $\pairing{\ahreal{\cdot}}$ on ground terms that replaces recursively all binary root symbols such that they are different from
all the non-variable quasi-subterms of the constraint system instantiated with its model $\sigma$,  with ACI symbol $\cdot$. 
Later, we will show, that $\pahs$ is also a model of $\ConstrSys{S}$.

\begin{df}\label{def:ah}

 Let us have a constraint system $\ConstrSys{S}$ which is satisfiable with model $\sigma$.
 Let us fix some $\alpha \in ( \UniAt{A} \cap \sub{\ConstrSys{S}})$.
 For given $\ConstrSys{S}$ and $\sigma$ we define a function $\ahreal{\cdot}: \UniverseG \rightarrow \bool{\UniverseG}$  as follows:

\[
 \ahreal{t} = \left\{ 
 \begin{array}{rl} 
   \set{\alpha}, & \mbox{if } {t}  \in (\UniAt{A} \setminus \sub{\ConstrSys{S}});\\
   \set{a}, & \mbox{if } {t} = a \in (\UniAt{A}\cap \sub{\ConstrSys{S}});\\
   
	\set{\priv{\pairing{\ahreal{t_1}}}} , & \mbox{if } t = \priv{t_1};\\

   \set{\bin{\pairing{\ahreal{t_1}}}{\pairing{\ahreal{t_2}}}} , & \mbox{if } {t} = \bin{t_1}{t_2}\\
		&  \norm{t} \in \norm{\subo{S}\sigma}\\

   \ahreal{t_1} \cup\ahreal{t_2} , & \mbox{if } {t} = \bin{t_1}{t_2}   \\
				&  \wedge\  \norm{t} \notin \norm{\subo{S}\sigma} \\
   \bigcup_{p\in L}\ahreal{p},  & \mbox{if } {t}=\aci{L}.\\
  
 \end{array}
 \right. 
\]

\end{df}

Henceforward, we will omit parameters and write $\ah{\cdot}$ instead of $\ahreal{\cdot}$ for shorter notation.

\begin{df}
 We define the superposition of $\pairing{\cdot}$ and $\ah{\cdot}$ on a set of terms $T=\set{t_1,\dots,t_k}$ as follows: $\pah{T} = \set{\pah{t}\,|\ t\in T}$.
\end{df}

\begin{df}
 Let $\theta =\set{x_1 \mapsto t_1, \dots, x_k \mapsto t_k }$  be a substitution. We  define $\pairing{\ah{\theta}}$ the  substitution $\set{x_1 \mapsto \pairing{\ah{t_1}}, \dots, x_k \mapsto \pairing{\ah{t_k}}}$.

\end{df}
Note, that $\dom{\pairing{\ah{\theta}}} = \dom{\theta}$.

\begin{example}\label{ex:constrsysmodel}
 We refer to Example~\ref{ex:constrsys} and show, that $\pah{\sigma}$ is also a model of $\ConstrSys{S}$.
$\pah{\enc{\pair{a}{b}}{c}} = \pairing{\ah{\pair{a}{b}}\cup \set{c}} = \pairing{\set{a}\cup\set{b} \cup \set{c}} = \aci{\lst{a,b,c}}$ (we suppose that $a\prec b \prec c$).
One can see, that $\pahs=\set{x\mapsto \aci{\lst{a,b,c}}}$ is also a model of $\ConstrSys{S}$ within DY+ACI.
\end{example}

\subsubsection{General properties used in proof}

The two following lemmas state simple properties of derivability.

\begin{lemma}\label{lemma:dertrans}
 Let $A,B,C \subseteq \UniverseG$. Then if $A\subseteq \der{B}$ and $B\subseteq \der{C}$ then $A\subseteq \der{C}$.

\end{lemma}

\begin{lemma}\label{lemma:derext}
 Let $A,B,C,D \subseteq \UniverseG$. Then if $A\subseteq \der{B}$ and $C\subseteq \der{D}$ then $A\cup C \subseteq \der{B\cup D}$.

\end{lemma}

In Lemma~\ref{lemma:normprop} we list some auxiliary properties that will be used in main proof.

\begin{lemma}\label{lemma:normprop}
	The following statements are true:
	\begin{enumerate}
		\item	\label{pACI}         %
			For terms $t, t_1,t_2$, we have $\norm{\aci{t, t}} = \norm{t}$, $\norm{\aci{t_1, t_2}}=\norm{\aci{t_2, t_1}}$, $\norm{\aci{\aci{t_1, t_2},  t_3}} = \norm{\aci{t_1, \aci{t_2, t_3}}} = \norm{\aci{t_1, t_2, t_3}}$
		\item 	\label{pNormNorm}      %
			if $t$ and $t\sigma$ are terms, then  $\norm{t\sigma}=\norm{\norm{t\sigma}}=\norm{\norm{t}\sigma}=\norm{t\norm{\sigma}}=\norm{\norm{t}\norm{\sigma}}$

		\item	\label{pSubNorm}     %
			$ s\in\sub{\norm{t}} \implies s=\norm{s}$

		\item	\label{pSubNormHasProimage}           %
			$ \forall s\in\subII{\norm{t}} \exists s' \in \subII{t} \,:\, s=\norm{s'}$

		\item \label{pNormElems}       %
			$\norm{\elems{t}} = \elems{\norm{t}}$ 
		
		\item \label{pNormDotNorm}        %
			$\norm{\aci{\norm{t_1}, \dots{}, \norm{t_m}}}=\norm{\aci{t_1, \dots, t_m}}$; $\pairing{T} = \pairing{\norm{T}}$

		\item \label{pDotNormElems}       %
			$\elems{\norm{\aci{\norm{t_1}, \dots, \norm{t_m}}}}={\elems{\aci{\norm{t_1}, \dots, \norm{t_m}}}} =$  \\ $\bigcup_{i=1,\dots,m}\elems{\norm{t_i}}$

		\item \label{pAhList}%
			$\ah{t} = \bigcup_{p\in\elems{t}}\ah{p}$, %
		\item \label{pAhNorm}                     %
		        ${\ah{t}} = \ah{\norm{t}}$

		\item \label{pPahNorm}                %
			$\pah{t}=\pah{\norm{t}}=\norm{\pah{t}}=\norm{\pah{\norm{t}}}$
		\item \label{pPairingOfPairing}          %
			$\pairing{T_1\cup\dots\cup T_m} = \pairing{\set{\pairing{T_1},\dots,\pairing{T_m}}}$
		\item \label{pSubSubSub}            %
			$\sub{\sub{t}} = \sub{t}$           
		\item \label{pNormSub}	%
			$\sub{\norm{t}}\subseteq \norm{\sub{t}}$
 		\item	\label{pSepVar}%
			$\sub{t\sigma}\subseteq \sub{t}\sigma\cup\sub{\vars{t}\sigma}$
		\item	\label{pSepVarII}          %
			$\subII{t\sigma} = \subII{t}\sigma\cup\subII{\vars{t}\sigma}$

		\item	\label{pNormCard}       %
			$\card{\norm{T}}\leq \card{T}$, $\card{T\sigma}\leq \card{T}$ %
		\item	\label{pSizesComparision}         %
			$\elems{t}\subseteq \sub{t} \subseteq \subII{t}$
		\item	\label{pNormSize}              %
			For term $t$, $\DAGsize{\norm{t}}\leq \DAGsize{t}$; \\ %
			for set of terms $T$, $\DAGsize{\norm{T}}\leq \DAGsize{T}$; \\
			for constraint system $\ConstrSys{S}$, $\DAGsize{\norm{\ConstrSys{S}}}\leq \DAGsize{\ConstrSys{S}}$

		\item   \label{pSubACI}         %
			$\sub{\aci{\lst{t_1, \dots, t_l}}}\subseteq \set{\aci{\lst{t_1, \dots,t_l}}}\cup \sub{t_1}\cdots\cup \sub{t_l}$
		\item	\label{pNormSubSize} %
			$\forall s\in\subII{t} \ \DAGsize{\norm{t\sigma}} \geq \DAGsize{\norm{s\sigma}}$.
	\end{enumerate}
\LONG{
\begin{proof}
We will give proofs of several statements. 
Some other technical proofs are given in \ref{app:lemma|normprop} \hfil \par
 \begin{description}
  \item [Statement~\ref{pNormElems}:]
      This statement is trivial, if $t\neq \aci{L}$.
      Otherwise, let $t=\aci{t_1,\dots,t_n}$. 
      \begin{itemize}
	\item if $\norm{\elems{t}}=\set{p}$, where $p\neq \aci{L_p}$. Then $\norm{t}=p$ and then $\elems{\norm{t}}=\elems{p} = \set{p} = \norm{\elems{t}}$.
	\item if $\norm{\elems{t}}=\set{p_1,\dots,p_k}$, $k>1$, where $p_i\neq \aci{L_i}$ for all $i$. Then $\norm{t} = \aci{L}$, where $L \approx \set{p_1,\dots,p_k}$. 
	      That means, that $\elems{\norm{t}} = \bigcup_{p\in\set{p_1,\dots,p_k}}\elems{p} = \set{p_1,\dots,p_k}$.
      \end{itemize}
  \item [Statement~\ref{pNormDotNorm}:]
      The first part follows from the definition of normal form and Statement~\ref{pNormElems}. %
      The second one directly follows from the first.
  \item [Statement~\ref{pAhNorm}:]    
	By induction on $\DAGsize{t}$:
	  \begin{itemize}
	  \item $\DAGsize{t}=1$ is possible in the only case: $t = a \in\UniAt{A}$ and as $a = \norm{a}$, the equality is trivial.
	  \item Suppose, that for any $t: \DAGsize{t} < k$ ($k>1$), ${\ah{t}}=\ah{\norm{t}}$ holds.
	  \item Given a term  $t: \DAGsize{t}=k$, $k>1$. We need to prove that ${\ah{t}}=\ah{\norm{t}}$.
		  \begin{itemize}
		  \item if $t=\priv{t_1}$, then ${\ah{t}}=\set{\priv{\pah{t_1}}}=$ (by induction supposition) 
		  $=\set{\priv{\pah{\norm{t_1}}}}=\ah{\priv{\norm{t_1}}}= \ah{\norm{t}}$.
		  \item if $t=\bin{p}{q}$ and $\norm{t}\in \norm{\subo{\ConstrSys{S}}\sigma}$.
			Then $\ah{\norm{t}} = \ah{\bin{\norm{p}}{\norm{q}}}=\set{\bin{\pah{\norm{p}}}{\pah{\norm{q}}}} = $
			(by induction supposition) $=\set{\bin{\pairing{\norm{\ah{p}}}}{\pairing{\norm{\ah{q}}}}}=$ (by Statement~\ref{pNormDotNorm}) 
			$=\set{\bin{\pah{p}}{\pah{q}}}={\ah{\bin{p}{q}}}$.
		  \item if $t=\bin{p}{q}$ and $\norm{t}\notin \norm{\subo{\ConstrSys{S}}\sigma}$.
			Then ${\ah{t}} = {\ah{p}}\cup{\ah{q}} = $ (by induction) $=\ah{\norm{p}}\cup\ah{\norm{q}} = $
			(as $\norm{\bin{\norm{p}}{\norm{q}}}=\norm{t}\notin \norm{\subo{\ConstrSys{S}}\sigma}$) \\ $=\ah{\bin{\norm{p}}{\norm{q}}}=\ah{\norm{t}}$ 
		  \item if $t=\aci{L}$, where $L=\lst{t_1,\dots,t_m}$.
			Note first, that as $t=\aci{L}$, we have for all$s \in \elems{t}$, $\DAGsize{s} < \DAGsize{t}$.
			Then, by Statement~\ref{pAhList}, $\ah{t} = \bigcup_{p\in\elems{t}}\ah{p} = $ (by induction supposition) $=\bigcup_{p\in\elems{t}}\ah{\norm{p}}$.
			On the other part, $\ah{\norm{t}} = \bigcup_{p\in\elems{\norm{t}}}\ah{p} =$  (by Statement~\ref{pNormElems}) $=\bigcup_{p\in\norm{\elems{t}}}\ah{p} =
			\bigcup_{p\in\elems{t}}\ah{\norm{p}}=\ah{t}$.
		  \end{itemize}

	  \end{itemize} 
  \item [Statement~\ref{pPairingOfPairing}:] 
	From definition of  $\opairing$ and Statement~\ref{pNormElems}, we obtain that \\ $\elems{\pairing{T_i}} = \norm{\elems{T_i}}$. 
	Next $\pairing{\set{\pairing{T_1},\dots,\pairing{T_m}}}=\norm{\aci{L}}$ (here we use $\norm{\aci{L}}$ to capture two cases from definition of normalization at once), where $L\approx \norm{\elems{\set{\pairing{T_1},\dots,\pairing{T_m}}}} = $ \\
	$\norm{\bigcup_{i=1,\dots,m}\norm{\elems{T_i}}} =\norm{\bigcup_{i=1,\dots,m}{\elems{T_i}}}$, \\
	while  $\pairing{T_1\cup\dots\cup T_m} = \norm{\aci{L'}}$, where $L' \approx \norm{\bigcup_{i=1,\dots,m}\elems{T_i}}$. 
  \item [Statement~\ref{pNormSub}:]
      By induction on $\DAGsize{t}$.
      \begin{itemize}
	\item $\DAGsize{t}=1$. 
	      Then $t\in\UniAt{A}\cup  \UniVar{X}$. As $\sub{t}=\set{t}$ and $t=\norm{t}$, the statement holds.
	\item Suppose, that for any $t: \DAGsize{t} < k$ ($k>1$), the statement is true.
	\item Given a term  $t: \DAGsize{t}=k$, $k>1$. Let us consider all possible cases:
	      \begin{itemize}
		\item $t=\bin{t_1}{t_2}$.%
		      On the one hand, $\sub{t}=\set{t}\cup\sub{t_1}\cup\sub{t_2}$.  
		      On the other hand, $\norm{t}=\bin{\norm{t_1}}{\norm{t_2}}$ and then,
		      $\sub{\norm{t}} = \set{\norm{t}}\cup\sub{\norm{t_1}}\cup\sub{\norm{t_2}}$.
		      Then, as $\sub{\norm{t_i}}\subseteq \norm{\sub{t_i}}$, we have that 
		      $\sub{\norm{t}} \subseteq \norm{\sub{t}}$.
		\item $t=\priv{t_1}$. Proof is similar to one for the case above.
		\item $t=\aci{L}$. We have $\sub{t} = \set{t}\cup\bigcup_{p\in\elems{{L}}}\sub{p}$.
		      From Statement~\ref{pNormElems} we have $\elems{\norm{\aci{L}}}  = \norm{\elems{\aci{L}}}$,
		      and then, $\sub{\norm{\aci{L}}} = $ \br$\set{\norm{\aci{L}}} \cup \bigcup_{p\in\elems{\norm{\aci{L}}}}\sub{p} = $ \br$
		      \norm{\set{\aci{L}}} \cup \bigcup_{p\in{\elems{\aci{L}}}}\sub{\norm{p}} \subseteq$ (by supposition) \\
		      $\subseteq \norm{\set{\aci{L}}} \cup \bigcup_{p\in{\elems{\aci{L}}}}\norm{\sub{p}} = $ \br$
		      \norm{{\set{\aci{L}}} \cup \bigcup_{p\in{\elems{\aci{L}}}}{\sub{p}}} = \norm{\sub{t}}$.
	      \end{itemize} 
      \end{itemize}      
  \item [Statement~\ref{pSepVar}:]
      By induction on $\DAGsize{t}$
      \begin{itemize}
	\item  $\DAGsize{t}=1$. 
	      \begin{itemize}
	      \item $t\in\as$. As $t\sigma = t$ and $\vars{t} = \emptyset$, the statement becomes trivial.
	      \item $t\in \UniVar{X}$. Then $\sub{t}\sigma = t\sigma$, $\vars{t} = \set{t}$; 
		    We have $\sub{t\sigma} \subseteq \set{t\sigma}\cup\sub{t\sigma}$.
	      \end{itemize}
	\item Suppose, that for any $t: \DAGsize{t} < k$ ($k\geq 1$), the statement is true.
	\item Given a term  $t: \DAGsize{t}=k$, $k>1$. Let us consider all possible cases:
	      \begin{itemize}
		\item $t=\bin{t_1}{t_2}$. %
		      Then 
		      $t\sigma =  \bin{t_1\sigma}{t_2\sigma}$ and $\vars{t} = \vars{t_1}\cup\vars{t_2}$.
		      $\sub{t\sigma} = \set{t\sigma} \cup \sub{t_1\sigma} \cup $ \br $ \sub{t_2\sigma} \subseteq $ (as $\DAGsize{t_i} < k$)
		      $\subseteq \set{t\sigma} \cup \sub{t_1}\sigma\cup\sub{\vars{t_1}\sigma} \cup \sub{t_2}\sigma\cup\sub{\vars{t_2}\sigma}
		      = \set{t\sigma} \cup \sub{t_1}\sigma  \cup \sub{t_2}\sigma \cup \sub{(\vars{t_1}\cup \vars{t_2})\sigma}
		      = \sub{t}\sigma \cup \sub{\vars{t}\sigma}$.
		\item $t=\priv{t_1}$. Proof is similar to one for the case above.
		\item $t=\aci{\lst{t_1,\dots,t_m}}$. We have $t\sigma =  \aci{\lst{t_1\sigma,\dots,t_m\sigma}}$ and $\vars{t} = \bigcup_{i=1,\dots,m}\vars{t_i}$.
		      Then we have $\sub{t\sigma} = \set{t\sigma} \cup \bigcup_{p\in\elems{\lst{t_1\sigma,\dots,t_m\sigma}}}\sub{p} \subseteq $ (using Statement~\ref{pSizesComparision})
		      $\subseteq \set{t\sigma} \cup \bigcup_{p\in\bigcup_{i=1}^m\sub{t_i\sigma}}\sub{p} = $
		      (as $\sub{\sub{p}}=\sub{p}$)
		      $= \set{t\sigma} \cup \bigcup_{i=1,\dots,m}\sub{t_i\sigma} \subseteq $
		      (as $\DAGsize{t_i} < k$) \\
		      $\subseteq  \set{t\sigma} \cup \bigcup_{i=1,\dots,m}\left(\sub{t_i}\sigma\cup\sub{\vars{t_i}\sigma} \right )
		      = \set{t\sigma} \cup \bigcup_{i=1,\dots,m}\sub{t_i}\sigma  \cup \sub{\left( \bigcup_{i=1,\dots,m}\vars{t_i}\right)\sigma}  $ \br $ 
		      = \sub{t}\sigma \cup \subII{\vars{t}\sigma}$. 
	      \end{itemize}
      \end{itemize}
 \end{description}
\end{proof}
} %
\end{lemma}

\begin{lemma}\label{lemma:normalsigma}
Given a  constraint system $\ConstrSys{S}$ and its model $\sigma$. Then 
substitution $\pahs$ is normalized 
\LONG{
\begin{proof}
 For any $x\in\dom{\pahs}$, $x\pahs = \pah{x\sigma} = \norm{\pah{x\sigma}}$ (by Lemma~\ref{lemma:normprop}).
\end{proof}
}%
\end{lemma}

\begin{lemma}\label{lemma:DAGvsQuasi}
 For any normalized term $t$, $\sub{t} = \subII{t}$.
\LONG{
\begin{proof} By induction on $\DAGsize{t}$.
 \begin{itemize}
  \item  $\DAGsize{t}=1$. Then  $t \in\UniVar{X} \cup \UniAt{A}$, and thus, $\sub{t} = \subII{t} = \set{t}$.
  \item Suppose, that for any $t: \DAGsize{t} < k$ ($k>1$), $\sub{t} = \subII{t}$.
  \item Given a term  $t: \DAGsize{t}=k$, $k>1$. We need to show that $\sub{t} = \subII{t}$.
	\begin{itemize}
	  \item $t=\bin{t_1}{t_2}$. %
	  	Then 
		$\sub{\bin{t_1}{t_2}} = \set{t} \cup \sub{t_1} \cup \sub{t_2} = $ (as $\DAGsize{t_i} < k$)
		$= \set{t} \cup \subII{t_1} \cup \subII{t_2} = \subII{t}$
	  \item $t=\priv{t_1}$. Then $\sub{\priv{t_1}} = \set{t}\cup\sub{t_1} = 
		\set{t}\cup\subII{t_1} = \subII{t}$
	  \item $t=\aci{L}$. As $t$ is normalized, we have that for all $p\in L$, $p\neq \aci{L_p}$. %
		Then $\elems{L} \approx L$. Thus, we have $\sub{t} = \set{t}\cup \bigcup_{p\in\elems{L}}\sub{p}
		= \set{t}\cup \bigcup_{p\in L}\sub{p} = \set{t}\cup \bigcup_{p\in L}\subII{p}=\subII{t}$.
	\end{itemize}

 \end{itemize}

\end{proof}
}%
\end{lemma}

In Proposition~\ref{prop:pairing} we remark, that ACI-set 
of normalized terms has the same deductive expressiveness as that set of normalized terms itself.

\begin{prop}\label{prop:pairing}
Let $T$ be a set of terms $T=\set{t_1,\dots, t_k}$. Then $\pairing{T}\in \der{\norm{T}}$ and  $\norm{T} \subseteq \der{\set{\pairing{T}}}$.

\end{prop}

In Proposition \ref{prop:normalSys}  we state that a  constraint system and its normal form have the same models. 
In Proposition \ref{prop:normalSigma} we show the equivalence, for a constraint system,  
between the existence of a model and the existence of a  normalized model. 
As a consequence we will need  only to consider  normalized constraints and models in the sequel.

\begin{prop}\label{prop:normalSys}
	The substitution $\sigma$ is a model of constraint system $\ConstrSys{S}$ if and only if $\sigma$ is a model of $\norm{\ConstrSys{S}}$.
\begin{proof}
	By definition, 
	$\sigma$ is a model of $\ConstrSys{S}=\set{E_i \rhd t_i}_{i=1, \dots, n}$, 
	iff $\forall i\in \set{1,\dots,n}$, $\norm{t_i\sigma} \in \der{\norm{E_i \sigma}}$. 
	But by Lemma~\ref{lemma:normprop} we have that  $\norm{t_i\sigma}=\norm{\norm{t_i}\sigma}$ and $\norm{E_i \sigma}=\norm{\norm{E_i} \sigma}$. 
	Thus, $\sigma$ is a model of ${\ConstrSys{S}}$ if and only if $\sigma$ is a model of $\norm{\ConstrSys{S}}$.
\end{proof}

\end{prop}

\begin{prop}\label{prop:normalSigma}
	The substitution $\sigma$ is a model of constraint system $\ConstrSys{S}$ if and only if $\norm{\sigma}$ is a model of $\ConstrSys{S}$.
\begin{proof}
	Proof is similar to one of Proposition~\ref{prop:normalSys}.
\end{proof}

\end{prop}

\subsection{Ground case of DY+ACI}\label{subs:groundcase}
In Algorithm~\ref{alg:solving} we need to check whether a ground substitution $\sigma$ satisfies a constraint system $\ConstrSys{S}$. For this, we have to check 
the derivability of a ground term from a set of ground terms. In this subsection we  present such an algorithm.

First, for the ground case we consider an  equivalent to DY+ACI deduction system DY+ACI' obtained from the first by replacing a set of rules 
 $$\forall i  \  \aci{{t_1, \dots, t_m}} \rightarrow \norm{t_i}$$
with   
 $$ \forall s\in\elems{t} \ {t} \rightarrow \norm{s}, \mbox{ if }  t=\aci{L}. $$

Now, we show an equivalence of the two deduction systems.
 
\begin{lemma}
	$t\in\xder{DY+ACI}{E} \iff t\in\xder{DY+ACI'}{E}$
\begin{proof}[Proof sketch] %

        We show that every   rule of one deduction system can be simulated by a combination 
of rules from the other. It is sufficient to show it for non common rules. 

        The DY+ACI' rules
	$\forall s\in\elems{t} \ {t} \rightarrow \norm{s}, \mbox{ if }  t=\aci{L}$ 
	are modeled by successive application of rules
	$\forall i  \  \aci{{t_1, \dots, t_m}} \rightarrow \norm{t_i}$.
	The converse simulation of $ \aci{{t_1, \dots, t_m}} \rightarrow \norm{t_i}$ by DY+ACI'
is based on getting all the normalized elements of $t_i$ 
	and, if $\card{\norm{\elems{t_i}}} \geq 2$ then reconstructing $\norm{t_i}$ by rule ${p_1,\dots, p_l} \rightarrow \norm{\aci{p_1, \dots, p_l}}$,
	where $p_1, \dots, p_l$ are $\norm{\elems{t_i}}$.
\end{proof}

\end{lemma}

\begin{algorithm}[H]
  \caption{Verifying derivability of term}
  \label{alg:ground}
  \KwIn{A normalized ground constraint $E \rhd t$}
  \KwOut{$t\in\xder{DY+ACI}{E}$}
  \BlankLine
Let  $S:=\sub{E}\cup\sub{t}\setminus E$\;
Let $D:=E$\;
\While{true}{ 
	\eIf{ exists DY rule $l\rightarrow r$, such that $l \subseteq D$ and $r\in S$}{  \nllabel{step:DY}
		$S:=S\setminus \set{r}$\;
		$D:=D\cup\set{r}$\;
}{
	\eIf{ exists $s\in S : \elems{s} \subseteq D$}{  \nllabel{step:setcompos}
		$S:=S\setminus\set{s}$\;
		$D:=D\cup\set{s}$\;
}
{
	\eIf{ exists $s\in D :  \elems{s} \nsubseteq D$}{ \nllabel{step:setdecompos}
		$S:=S\setminus \elems{s}$\;
		$D:=D\cup\elems{s}$\;
}{
		\Return{ $t\in D$}\; \nllabel{step:ret}
}}}
}
\end{algorithm}

\begin{lemma}\label{lemma:propalgo}
	For Algorithm~\ref{alg:ground} the following statements are true:
	\begin{itemize}
		\item for any step\footnote{Consider two sequential assignments as one step}, $D\cup S =\sub{E\cup\set{t}}$ and $D\cap S=\emptyset$;
		\item it terminates;
		\item for any step, $D\subseteq \xder{DY+ACI}{E}$.
	\end{itemize}

\end{lemma}

The following lemmas will be used to prove correctness of the algorithm. 

\begin{lemma}\label{lemma:strokerules} \
\begin{itemize}
	\item For any decomposition rule $l\rightarrow r$ of DY+ACI', if  $l$ is normalized, then $r$ is a quasi-subterm of $l$.
	\item For any composition rule $l\rightarrow r$ of DY+ACI' except $\set{t_1, \dots, t_m} \rightarrow \norm{\aci{t_1, \dots, t_m}}$, if $l$ is normalized, then $l\subseteq \sub{r}$.
\end{itemize}
\end{lemma}

\begin{lemma}\label{lemma:walkout}
	After the execution of Step~\ref{step:ret} of Algorithm~\ref{alg:ground}, if $l\rightarrow r$ is a DY+ACI' rule, such that $l\subseteq D$ and $r\notin D$, then $l\rightarrow r$ is a composition rule and $r\notin \sub{E\cup\set{t}}$.
\begin{proof}
	Suppose, $l\rightarrow r$ is a decomposition.   By Lemma~\ref{lemma:strokerules} we have that $r\in\sub{l}$ and thus, $r\in\sub{D}\subseteq D \cup S$. Then  $r\notin D$ implies $r\in S$, and then, Step~\ref{step:ret} must be skipped, as branch \ref{step:DY} or \ref{step:setdecompos} should have been visited. %
	
	Thus, $l\rightarrow r$ is  a composition.
	 As algorithm reached  Step~\ref{step:ret}, that means $r\notin S$ (otherwise one of three branches must be visited and this step would be skipped). As $r\notin S$ and $r\notin D$, we have $r\notin S\cup D = \sub{E\cup\set{t}}$.
\end{proof}

\end{lemma}

\begin{lemma}\label{lemma:composelems}
	Given a set of normalized terms $S$ such that for any $s\in S$, $\elems{s}\subseteq S$.
	Then for any DY+ACI' composition rule $l\rightarrow r$ such that $l\subseteq S$ we have $\elems{r}\subseteq S\cup\set{r}$.
\begin{proof}
	All cases of composition rules except $ {t_1, \dots, t_m} \rightarrow \norm{\aci{t_1, \dots, t_m}}$ are trivial, as for them $\elems{r}=\set{r}$.
	For this case, as $\elems{t_i}\subseteq S$ for all $i$, then (by Lemma~\ref{lemma:normprop}, Statement \ref{pDotNormElems}) $\elems{\norm{\aci{t_1, \dots, t_m}}} =  $ \\ $ \elems{\aci{t_1, \dots, t_m}} =\bigcup_{i=1}^m\elems{t_i}\subseteq S$.
\end{proof}

\end{lemma}

\begin{prop}
 Algorithm~\ref{alg:ground} is correct.
\begin{proof}
	If algorithm returns true, then, by Lemma~\ref{lemma:propalgo}, $t\in\xder{DY+ACI'}{E}$.
	
	Show, that output is correct, if algorithm returns false. Note, that we consider values of $D$ and $S$ that they have after finishing the algorithm. Suppose that output is false ($t\notin D$), but $t\in \xder{DY+ACI'}{E}$. Then there exists minimal by length derivation $\set{E_i}_{i=0,\dots,n}$ where $n\geq 1$, $D=E_0$ (as $D\subseteq \xder{DY+ACI'}{E}$ and $t\notin D$) and $t\in E_n$ and $E_{i+1}\setminus E_i \neq \emptyset$ and $E_i \rightarrow_{l_i\rightarrow r_i} E_{i+1}$ for all $i=0,\dots,n-1$.
	Then, applying Lemma~\ref{lemma:walkout} we have $l_0 \rightarrow r_0$ is a composition, and $r_0 \notin \sub{E\cup\set{t}}$.
	
	Let $m$ be the smallest index such that there exists $s \in S=\sub{E\cup\set{t}}\setminus D$ and $s\in E_m$.
	
	Let $k$ be the minimal integer, such that $l_k\rightarrow r_k$ is a decomposition.
	
	Show, $k\leq m$. Suppose the opposite, then $s$ is built by a chain of composition rules from $D$. If $l_{m-1}\rightarrow r_{m-1}$ (where $r_{m-1} =s $) is
	\begin{itemize}
	 \item a rule in form of  $ \set{t_1, \dots, t_c} \rightarrow \norm{\aci{t_1, \dots, t_c}}$, then $\elems{s} \neq \set{s}$ (otherwise it contradicts to minimality of the derivation) and from Lemma~\ref{lemma:composelems}, $\elems{s}\subseteq E_{m-1}$ ($m\neq 1$, otherwise this step would be executed in the algorithm). As $s\in S$, then $\elems{s}\subseteq \sub{s}\subseteq \sub{E\cup\set{t}}$.
	If $\elems{s}\subseteq D$ then we got contradiction with with the fact, that this step would be executed in the algorithm. If there exists $e\in\elems{s}$ and $e\notin D$ (that means, $e\in S$), then we get a contradiction with the minimality of $m$, as $e\in S$ was deduced before.
	 \item any other composition rule, then by Lemma~\ref{lemma:strokerules}, $l_{m-1} \subseteq \sub{s}$, and thus, $l_{m-1} \subseteq D\cup S$. Similarly to the previous case, $m\neq 1$ and we get a contradiction with either minimality of $m$, or with the fact, that the algorithm would have to add $s$ into $D$.%
	\end{itemize}
      
	Note, that this also shows, that decomposition rule is present in derivation.
	
	Show, $l_k\nsubseteq D$. Suppose the opposite. Then by Lemma~\ref{lemma:strokerules}, we have $r_k\subseteq D$ what contradicts to $E_{k+1}\setminus E_k \neq \emptyset$.
	Thus, at least one element from $l_k$ is not from $D$.
	Let us consider all possible decomposition rules $l_k\rightarrow r_k$:
	\begin{itemize}
		\item $\set{\pair{t_1}{t_2}} \rightarrow \norm{t_1}$. We know, that $\pair{t_1}{t_2}$ is not in $D$, thus, it was built by composition. As $E_i$ are normalized, the only possible way to build by composition $\pair{t_1}{t_2}$ from normalized terms is $\set{t_1,t_2}\rightarrow \pair{t_1}{t_2}$ (other ways, like $\pair{t_1}{t_2}, \pair{t_1}{t_2} \rightarrow \norm{\aci{\lst{\pair{t_1}{t_2}, \pair{t_1}{t_2}}}}$ would contradict the minimality of the derivation). 
		Thus, $t_1$ was derived before (or was in $D$), i.e. $t_1 \in E_k$. 
		That contradicts to $E_{k+1}\setminus E_k \neq \emptyset$.
		
		\item  $\set{\pair{t_1}{t_2}} \rightarrow \norm{t_2}$. Similar case.
		
		\item $\set{\enc{t_1}{t_2},  \norm{t_2}} \rightarrow \norm{t_1}$. The case where $\enc{t_1}{t_2}\notin D$ has similar explanations as two cases above. Thus, $\enc{t_1}{t_2}\in D$. That means, $t_2\in\sub{E\cup\set{t}}$ and 
		 $t_2\notin D$, i.e. $t_2\in S$. 
		This means, $t_2$ was derived before and $t_2\in S$, what contradicts to  $k\leq m$.

		\item $\set{\aenc{t_1}{t_2},  \norm{\priv{t_2}}} \rightarrow \norm{t_1}$ is a similar case to previous one. Note, that if $\priv{t_2}$ is not in $D$, that it must be obtained by decomposition.
		
		\item ${t} \rightarrow \norm{s}$, where $s\in\elems{t}$ and $t=\aci{L}$. By Lemma~\ref{lemma:composelems}, $\elems{t}\subseteq E_k$, that contradicts minimality of derivation ($E_{k+1}\setminus E_k \neq \emptyset$).
	\end{itemize}

\end{proof}
\end{prop}

\subsection{Existence of conservative solutions}

In this subsection we will show that for any satisfiable constraint system, there exist a model in special form (so called conservative solution). 
Roughly speaking, a model in this form can be defined per each variable by set of quasi-subterms of the constraint system and set of atoms (also from the constraint system) that must be prived.
This will bound a search space for the model (see \S~\ref{subs:bounds}).

First, we show, that on quasi-subterms of constraint system instantiated with its model, the transformation $\pah{\cdot}$ will be a homomorphism modulo normalization. %
\begin{prop}\label{prop:subt}
Given a normalized constraint system $\ConstrSys{S}$ and its  normalized model $\sigma$.
For all $t \in \sub{\ConstrSys{S}}$, $\norm{t\pahs} = \norm{\pairing{\ah{t\sigma}}} $.
 \begin{proof}
  We will prove it by induction on $\card{\subII{t}}$, where $t$ is normalized. 
\begin{itemize}
 \item Let $\card{\subII{t}} = 1$. Then:
	\begin{itemize}
 		\item either $t \in \UniAt{A}$.  In this case $t \in (\UniAt{A} \cap \sub{\ConstrSys{S}})$, and as $t \mu = t $ for any substitution $\mu$, then  $\pairing{\ah{t\sigma}} = \pairing{\ah{t}}  = \pairing{\set{t}}= t$ and $t\pahs = t$. Thus, $t\pahs = \pairing{\ah{t\sigma}} $.

		\item or $t \in \UniVar{X}$. As $\sigma$ is a model and $t\in \sub{\ConstrSys{S}}$, we have $t \in \dom{\sigma}$, and, by definition, $t \in \dom{\pahs}$. Then, by definition of $\pahs$, $t\pahs = \pairing{\ah{t\sigma}}$.
	\end{itemize}

 \item Assume that for some $k \geq 1$ if $\card{\subII{t}} \leq k$, then $\norm{t\pahs} = \norm{\pairing{\ah{t\sigma}}} $.
 \item Show, that for any $t$ such that $\card{\subII{t}} \geq k+1$, where $t=\bin{p}{q}$ 
      or $t=\priv{q}$ 
 or $t=\aci{t_1, \dots, t_m}$,
 but $\card{\subII{p}} \leq k$, $\card{\subII{q}} \leq k$ and $\card{\subII{t_i}}\leq k$, for all $i\in\set{1,\dots,m}$,
 statement  $\norm{t\pahs} = \norm{\pairing{\ah{t\sigma}}} $ is still true. 
 We have: 
	\begin{itemize}
 		\item either $t = \bin{p}{q}$. As $t = \bin{p}{q} \in \sub{\ConstrSys{S}} \Rightarrow p \in \sub{\ConstrSys{S}}$ and $q \in \sub{\ConstrSys{S}}$. As $\card{\subII{p}} < \card{\subII{t}}$  and from the induction assumption, we have $\norm{p\pahs} = \norm{\pairing{\ah{p\sigma}}}$. The same holds for $q$.

		Again, since $\bin{p}{q}\sigma \in \subo{S}\sigma$ (as $\bin{p}{q} \notin \UniVar{X}$ and $t\in \sub{\ConstrSys{S}}$) we have that 
		$\norm{\pah{\bin{p}{q}\sigma}} =  $ \br $ 
		\norm{\pah{\bin{p\sigma}{q\sigma}}}   =  \norm{\pah{\norm{\bin{p\sigma}{q\sigma}}}}  =   $ \br $ 
		\norm{\pairing{\ah{\bin{\norm{p\sigma}}{\norm{q\sigma}}}}}  =   $ \br $ 
		\norm{\pairing{\set{\bin{\pairing{\ah{\norm{p\sigma}}}}{\pairing{\ah{\norm{q\sigma}}}}}}} =  $ \br $ 
		\norm{\pairing{\set{\bin{\norm{\pairing{\ah{p\sigma}}}}{\norm{\pairing{\ah{q\sigma}}}}}}} =  $ \br $ 
		\norm{  \bin{\norm{\pairing{\ah{p\sigma}}}}{\norm{\pairing{\ah{q\sigma}}}}} = $ \br $ 
		\norm{\bin{\norm{p\pahs}}{\norm{q\pahs}}} =  $ \br $ 
		\norm{\bin{p\pahs}{q\pahs}} =  $ \br $ 
		\norm{\bin{p }{q }\pahs}=
		\norm{t\pahs}$.

		\item or $t=\aci{t_1, \dots, t_m}$. As $t$ is normalized, 
		it implies that for all $i\in\set{1,\dots,m}$, $t_i$ are not in form of $\aci{L_{i}}$ and then $t_i \in \sub{\ConstrSys{S}}$, %
		and thus, we have $t_i\in\sub{\ConstrSys{S}} \wedge \norm{\pah{t_i\sigma}} = \norm{t_i\pahs}$. 
		 $\pah{t\sigma}=
		 \pah{\aci{t_1\sigma,\dots,t_m\sigma}}=
		 \pairing{\ah{t_1\sigma} \cup\dots\cup \ah{t_m\sigma}}= $
		 (by Statement~\ref{pPairingOfPairing} of Lemma~\ref{lemma:normprop}) \br
		 $=
		 \pairing{\set{\pah{t_1\sigma},\dots,\pah{t_m\sigma}}}=  $ \br $ 
		 \pairing{\set{\norm{t_1\pahs},\dots,\norm{t_m\pahs}}}= $ \br $ 
		 \norm{\aci{\norm{t_1\pahs}, \dots, \norm{t_m\pahs}}}= $ \br $ 
		 \norm{\aci{t_1\pahs, \dots, t_m\pahs}}=  $ \br $ 
		 \norm{(\aci{t_1, \dots,t_m})\pahs} = \norm{t\pahs} $

		\item or $t=\priv{q}$. Then $q\in\sub{\ConstrSys{S}}$.
		
		$\pah{t\sigma}=\pairing{\set{\priv{\pah{q\sigma}}}} = \norm{\priv{\pah{q\sigma}}}= $ \br $ \norm{\priv{q\pahs}}=\norm{\priv{q}\pahs}=\norm{t\pahs}$.
		
	\end{itemize}
\end{itemize}

Thus, the proposition is proven.

 \end{proof}

\end{prop}

Now we show, that relation of derivability between a term and a set of terms is stable with regard to transformation $\pah{\cdot}$. %

\begin{lemma}\label{lemma:validstep}
Given a normalized constraint system $\ConstrSys{S}$ and its normalized model $\sigma$.
For any DY+ACI rule \ ${l_1,\dots,l_k}\rightarrow r$, \br $\pairing{\ah{r}} \in \der{\set{\pairing{\ah{l_1}}, \dots, \pairing{\ah{l_k}}} }$.

\begin{proof}[Proof idea] 
We proceed by considering all possible deduction rules. To give an idea, we show a proof for only one rule (see full proof in \ref{app:lemma|validstep}):
  	${\aenc{t_1}{t_2},\norm{\priv{t_2}}} \rightarrow \norm{t_1}$. Here we have to show that $\pah{\norm{t_1}}$ is derivable from $\set{\pah{\aenc{t_1}{t_2}}, \pah{\norm{\priv{t_2}}}}$.
	Consider two cases:
  	\begin{itemize}
		\item $\exists u\in \subo{S}$ such that $\norm{\aenc{t_1}{t_2}}=\norm{u\sigma}$. Then \br $\pah{\aenc{t_1}{t_2}}  = \aenc{\pah{t_1}}{\pah{t_2}}$, \br and then 
		$\pah{\norm{t_1}}=\pah{t_1} \in   $ \br $ \der{\set{\aenc{\pah{t_1}}{\pah{t_2}},\norm{\priv{\pah{t_2}}}}}$. \br
		On the other hand, $\pah{\norm{\priv{t_2}}} = \pah{\priv{t_2}} =  $ \br $ \pairing{\set{\priv{\pah{t_2}}}}=\norm{\priv{\pah{t_2}}}$.
		
		\item $\nexists u\in \sub{S}$ such that $\norm{\aenc{t_1}{t_2}}=\norm{u\sigma}$. 
		 Then \br $\pah{\aenc{t_1}{t_2}} = \pairing{\ah{t_1}\cup \ah{t_2}}$. Using Proposition~\ref{prop:pairing}, we have $\norm{\ah{t_1}\cup \ah{t_2}} \subseteq \der{\set{\pah{\aenc{t_1}{t_2}}}}$, thus (by Lemma~\ref{lemma:normprop}) $\norm{\ah{t_1}}\subseteq \der{\set{\pah{\aenc{t_1}{t_2}}}}$. And then, by Proposition~\ref{prop:pairing} we have that $\pah{t_1} \in \der{\norm{\ah{t_1}}}$. Therefore, by Lemma~\ref{lemma:dertrans},   $\pah{\norm{t_1}}=\pah{t_1} \in \der{\pah{\aenc{t_1}{t_2}}}$.
	\end{itemize}
\end{proof}

\end{lemma}

Using Proposition~\ref{prop:subt} and Lemma~\ref{lemma:validstep} we will show, that transformation $\pah{\cdot}$ preserves the property of substitution to be a model. %

\begin{theorem}\label{theorem:solution}
Given a normalized constraint system $\ConstrSys{S}$ and its normalized model $\sigma$.
Then substitution $\pahs$ also satisfies $\ConstrSys{S}$.
 \begin{proof}
 Suppose, without loss of generality, $\ConstrSys{S}= \set{E_i \rhd t_i}_{i=1, \dots, n}$.
  Let us take any constraint $(E \rhd t) \in \ConstrSys{S}$. As $\sigma$ is a model of $\ConstrSys{S}$,  there exists a derivation $D=\set{A_0, \dots, A_{k}}$  such that $A_0=\norm{E\sigma}$ and $\norm{t\sigma} \in A_{k}$. 

By Lemma~\ref{lemma:validstep} and Lemma~\ref{lemma:derext} we can easily prove that if $k>0$, $\pah{A_{j}} \subseteq \der{\pah{A_{j-1}}}, \  j=1,\dots,k$. 
Then, applying transitivity of $\der{\cdot}$ (Lemma~\ref{lemma:dertrans}) $k$ times, we have that $\pah{A_k} \subseteq \der{\pah{A_0}}$. In the case where $k=0$, the statement $\pah{A_k} \subseteq \der{\pah{A_0}}$ is also true.

 Using Proposition~\ref{prop:subt} we get $\pah{A_0} =\pah{E\sigma} = \norm{E\pahs}$, as $E \subseteq \sub{\ConstrSys{S}}$. The same for $t$: $\pah{t\sigma} = \norm{t\pahs}$, and as $\norm{t\sigma} \in A_k$, we have $\norm{t\pahs} \in \pah{A_k}$. 
Thus, we have that $\norm{t\pahs} \in \pah{A_k} \subseteq \der{\pah{A_0}} = \der{\norm{E\pahs}} $, that means $\pahs$ satisfies any constraint of $\ConstrSys{S}$. 

 \end{proof}
\end{theorem}

From now till the end of subsection we will study a very useful property of $\pahs$.
Proposition~\ref{prop:mainprop} and its corollary show, 
that if constraint system has a normalized model ($\sigma$) which sends different variables to different values, 
then there exists another normalized model ($\pahs$)  that  sends any variable of its domain
to an ACI-set of some non-variable quasi-subterms of constraint system instantiated by itself
and some private keys built with atoms of  the constraint system.

\begin{lemma}\label{lemma:goodSub}
	If $\norm{u\sigma}=\enc{p}{q}$, $\sigma$ is normalized, $u=\norm{u}$,  $u\notin \UniVar{X}$ and $x\sigma\neq y\sigma, x\neq y$,
	then there exists $s\in\sub{u}$ such that $s=\enc{p'}{q'}$ and $\norm{s\sigma}=\enc{p}{q}$. 
	The similar is true in the case of  $\norm{u\sigma}=\pair{p}{q}$, $\norm{u\sigma}=\aenc{p}{q}$, $\norm{u\sigma}=\sig{p}{q}$ and for $\norm{u\sigma}=\priv{p}$.
\LONG{
\begin{proof}
	As $u=\norm{u}$ and $\norm{u\sigma}=\enc{p}{q}$, we have:
	\begin{itemize}
		\item $u$ not in form of $\aci{L}$. Then, as $u\notin\UniVar{X}$ and $\norm{u\sigma}=\enc{p}{q}$, we have $u=\enc{p'}{q'}$ (where $\norm{p'\sigma}=p$ and $\norm{q'\sigma}=q$). 
		Then we can choose $s=\enc{p'}{q'}=u\in\sub{u}$.

		\item $u=\aci{t_1,\dots,t_m}$, $m>1$, as $u=\norm{u}$. Then, for all $i$,  $t_i$ is either a variable, or $\enc{p'_i}{q'_i}$. 
		But, as  $x\sigma\neq y\sigma, x\neq y$  and as $\sigma$ is normalized, we can claim, that $\set{t_1,\dots,t_m}$ contains at most one variable. 
		Then, as $m>1$, there exists $i$ such that $t_i = \enc{p'_i}{q'_i}$. Then by definition of normalization function, and from $\norm{u\sigma}=\enc{p}{q}$ we have, that 
		$\norm{\elems{u\sigma}} = \set{\enc{p}{q}}$ and as $t_i\sigma$ is an element of $u\sigma$, we have $\norm{\enc{p'_i}{q'_i}\sigma} = \enc{p}{q}$. 
		Thus, we can choose $s=t_i$, as $t_i \in \sub{u}$ and $t_i= \enc{p'_i}{q'_i}$.
	\end{itemize}
	
	The other cases  ($\oppair$, $\oppriv$, etc...) can be proved similarly.

\end{proof}
}
\end{lemma}

\begin{prop}\label{prop:mainprop}
 Given a normalized constraint system $\ConstrSys{S}$ 
 and its normalized model $\sigma$ such that for all $x,y\in\dom{\sigma}$, $x\neq y \implies x\sigma\neq y\sigma$. 
 Then  for all $x\in \dom{\pahs}$ there exist $k \in \mathbb{N}$ and $s_1,\dots, s_k \in \subo{S}\cup \priv{\sub{\ConstrSys{S}}\cap\UniAt{A}}$  such that $\rut{s_i} \neq \cdot$ 
and \br $ x\pahs = \pairing{\set{s_1\pahs, \dots, s_k\pahs}}$.

\begin{proof}
\raggedright
 By definition, $x\pahs = \pah{x\sigma}$. Let us take any $s\in \ah{x\sigma}$ (note, that $s$ is a ground term). Then, by definition of $\ah{\cdot}$ we have:
\begin{itemize}
 \item either $s \in \UniAt{A}$. Then, by definition of $\ah{\cdot}$, $s\in (\UniAt{A} \cap \sub{\ConstrSys{S}})$. Thus, $s\pahs = s$, $s \in  \subo{S}$, $s\neq \aci{L}$;
 
 \item or $s = \bin{\pah{t_1}}{\pah{t_2}}$ and there exists $u \in \subo{S}$ such that $\norm{u\sigma}=\norm{\bin{t_1}{t_2}}=\bin{\norm{t_1}}{\norm{t_2}}$. 
  As all conditions of Lemma~\ref{lemma:goodSub} are satisfied, then there exists $v\in\sub{u}$ such that $\norm{v\sigma}=\bin{\norm{t_1}}{\norm{t_2}}$ and 
  $v=\bin{p}{q}$ 
  and as $u\in\subo{S}$ then $v\in\subo{S}$.
  By Proposition~\ref{prop:subt}, $\norm{v\pahs}=\pah{v\sigma}=\pah{\norm{v\sigma}}=\pah{\bin{t_1}{t_2}}=\pairing{\set{\bin{\pah{t_1}}{\pah{t_2}}}} = \bin{\pah{t_1}}{\pah{t_2}} = s$. 
  That means that there exists  $v \in \subo{S}$ such that $v\neq \aci{L}$ and $s=\norm{v\pahs}$.

 \item or $s=\priv{\pah{t_1}}$. In this case, as $s$ is ground, $\pah{t_1}$ must be an atom, moreover, by definition of $\ah{\cdot}$, this atom is from $(\UniAt{A} \cap \sub{\ConstrSys{S}})$. 
	Therefore, $s=\priv{a}$, where $a\in \UniAt{A} \cap \sub{\ConstrSys{S}}$ (and of course, $s\neq  \aci{L}$).
\end{itemize}

Thus,  for all $s\in \ah{x\sigma}$, there exists $v \in   (\sub{\ConstrSys{S}})\cup \priv{\sub{\ConstrSys{S}}\cap\UniAt{A}} \setminus \UniVar{X}\,|\ s=\norm{v\pahs}$. 
Therefore, as  $x\pahs=\pah{x\sigma}$, we have that $x\pahs = \pairing{\set{\norm{s_1\pahs},\dots, \norm{s_k\pahs}}} =  \pairing{\set{s_1\pahs,\dots, s_k\pahs}}$, where   $s_1,\dots, s_k \in \subo{\ConstrSys{S}} \cup \priv{\sub{\ConstrSys{S}}\cap\UniAt{A}}$ and $s_i\neq \aci{L}, \forall 1\leq i \leq k$.  That proves the proposition.
\end{proof}

\end{prop}

\begin{cor}\label{cor:existsgood}
	 Given normalized constraint system $\ConstrSys{S}$ and $\sigma'$ --- its normalized model,   such that $x\neq y \implies x\sigma' \neq y\sigma'$. Then there exists a normalized model $\sigma$ of  $\ConstrSys{S}$ such that for all $x\in \dom{\sigma}$ there exist $k \in \mathbb{N}$ and $s_1,\dots, s_k \in \subo{S}\cup \priv{\sub{\ConstrSys{S}}\cap\UniAt{A}}$ such that $x\sigma = \pairing{\set{s_1\sigma, \dots, s_k\sigma}}$ and $s_i \neq s_j$, if $i\neq j$; $s_i\neq\aci{L}, \forall i$.

\end{cor}
 Any normalized model with property shown in Corollary \ref{cor:existsgood} we will call \emph{conservative}.

\subsection{Bounds on conservative solutions}
\label{subs:bounds}

To get a decidability result, 
we first show an upper bound on size of conservative model 
and then, 
by reducing any satisfiable constraint system to one that have 
conservative model and showing that reduced one is smaller
(by size) than original one,
we obtain an existence of a model with bounded size for any 
satisfiable constraint system.

\begin{lemma}\label{lemma:subsigma}
	Given a normalized constraint system $\ConstrSys{S}$ and its conservative model $\sigma$. Then for all $x \in \vars{\ConstrSys{S}}$ we have $\sub{x\sigma}\subseteq \norm{\sub{\ConstrSys{S}}\sigma}\cup \priv{\sub{\ConstrSys{S}}\cap\UniAt{A}}$.
\begin{proof}
	Given a ground substitution $\sigma$, let us define a  strict total order on variables: $x\sqsubset y \iff (\DAGsize{x\sigma}<\DAGsize{y\sigma}) \vee (\DAGsize{x\sigma}=\DAGsize{y\sigma} \wedge x\prec y)$.
	
	By Proposition~\ref{prop:mainprop} for all $x$ $x\sigma = \pairing{\set{s^x_1\sigma, \dots, s^x_{k^x}\sigma}}$, where  $s^x_i \in (\sub{\ConstrSys{S}} \setminus \UniVar{X})\cup \priv{\sub{\ConstrSys{S}}\cap\UniAt{A}}$  and $s^x_i\neq \aci{L}$. 

Let us show that if $y\in\vars{s^x_i}$ for some $i$, then $y\sqsubset x$.
	Suppose, that $y\in\vars{s^x_i}$ and $x\sqsubset y$. 
	Then $\DAGsize{x\sigma}  = \DAGsize{\pairing{\set{s^x_1\sigma, \dots, s^x_{k^x}\sigma}}} = 
	\DAGsize{\norm{\aci{s^x_1\sigma,\dots, s^x_{k^x}\sigma}}} \geq$ 
	(by Lemma~\ref{lemma:normprop}) 
	$ \geq \DAGsize{\norm{s^x_i\sigma}}> $ \br $ \DAGsize{\norm{y\sigma}}$,  
	because we know that $s^x_i=\bin{p}{q}$ or $s^x_i=\priv{p}$ and $y\in\vars{s^x_i}$ 
	(for example, in first case, $\DAGsize{\norm{s^x_i\sigma}} =  $ \br $ \DAGsize{\bin{\norm{p\sigma}}{\norm{q\sigma}}}=
	  1+\DAGsize{\set{\norm{p\sigma},\norm{q\sigma}}}$ and   \br  since $y\in\vars{\set{p,q}}$, 
	  using Statement~\ref{pNormSubSize} of Lemma~\ref{lemma:normprop}, 
	  we get   \br   $\DAGsize{\norm{s^x_i\sigma}} \geq 1+ \DAGsize{\norm{y\sigma}}$)
	And as $\DAGsize{\norm{y\sigma}} = \DAGsize{y\sigma}$ 
	That means, $y\sqsubset x$. Contradiction.
	
	Now we show by induction main property of this lemma.
	\begin{itemize}
		\item let $x=\min_{\sqsubset}(\vars{\ConstrSys{S}})$. \br 
		Then $ x\sigma = \pairing{\set{s^x_1\sigma, \dots, s^x_{k^x}\sigma}}= \norm{\aci{s^x_1\sigma, \dots, s^x_{k^x}\sigma}}$ and 
		 all $s^x_i$ \br are ground  (as there does not exists $y \sqsubset x$). 
		Then $x\sigma=\norm{\aci{s^x_1, \dots{}, s^x_{k^x}}}$.
 We have that $\sub{x\sigma}=\set{\norm{\aci{s^x_1, \dots, s^x_{k^x}}}}\cup\sub{s^x_1}\cup\dots\cup\sub{s^x_{k^x}} \subseteq \norm{\sub{\ConstrSys{S}}\sigma}\cup\priv{\UniAt{A}\cap\sub{\ConstrSys{S}}}$, 
as for any $s\in\sub{s^x_i}$, $s\in\UniverseG$ and $s\in\sub{\ConstrSys{S}}$ or $s=\priv{a}$ or $s=a$, where $a\in \sub{\ConstrSys{S}}\cap \UniAt{A}$, therefore  $s=\norm{s}=s\sigma\in\norm{\sub{\ConstrSys{S}}\sigma}\cup \priv{\sub{\ConstrSys{S}}\cap \UniAt{A}}$ 
and 
$\norm{\aci{s^x_1, \dots, s^x_{k^x}}} = x \sigma \in \norm{\sub{\ConstrSys{S}}\sigma}$.

		\item
			Suppose, that for all $z\sqsubset y$ we have \br  $\sub{z\sigma} \subseteq \norm{\sub{\ConstrSys{S}\sigma}}\cup \priv{\sub{\ConstrSys{S}}\cap\UniAt{A}}$.
		
		\item
			Show, that $\sub{y\sigma} \subseteq \sub{\ConstrSys{S}\sigma}\cup \priv{\sub{\ConstrSys{S}}\cap\UniAt{A}}$.
			We know that $ y\sigma = \pairing{\set{s^y_1\sigma, \dots, s^y_{k^y}\sigma}}= \norm{\aci{s^y_1\sigma,\dots, s^y_{k^y}\sigma}}$ 
			and for any  $z\in \vars{s^y_i}$, $z\sqsubset y$.
			Then we have $\sub{y\sigma} = $ \br $ \set{y\sigma}\cup\sub{\norm{s^y_1\sigma}}\cup\dots\cup\sub{\norm{s^y_{k^y}\sigma}}$. We know that $y\sigma\in\norm{\sub{\ConstrSys{S}}\sigma}$. Let us show that $\sub{\norm{s^y_i\sigma}}\subseteq\norm{\sub{\ConstrSys{S}}\sigma}\cup\priv{\sub{\ConstrSys{S}}\cap\UniAt{A}}$.
			By Lemma~\ref{lemma:normprop} 
			we have $\sub{\norm{s^y_i\sigma}}\subseteq  $ \br $ \norm{\sub{s^y_i\sigma}}\subseteq \norm{\sub{s^y_i}\sigma\cup\sub{\vars{s^y_i}\sigma}}=  $ \br $ 
			\norm{\sub{s^y_i}\sigma}\cup\sub{\vars{s^y_i}\sigma}$. We can see that $\norm{\sub{s^y_i}\sigma}\subseteq\norm{\sub{\ConstrSys{S}}\sigma}\cup\priv{\sub{\ConstrSys{S}}\cap\UniAt{A}}$  (as $s^y_i\in\sub{\ConstrSys{S}}\cup  $ \br $ \priv{\sub{\ConstrSys{S}}\cap\UniAt{A}}$); and by induction supposition and by statement proved above we have 
			$\sub{\vars{s^y_i}\sigma}\subseteq \norm{\sub{\ConstrSys{S}}\sigma}\cup $ \br $ \priv{\sub{\ConstrSys{S}}\cap\UniAt{A}}$. \br
			Thus, $\sub{y\sigma} \subseteq \norm{\sub{\ConstrSys{S}}\sigma}\cup\priv{\sub{\ConstrSys{S}}\cap\UniAt{A}}$.

	\end{itemize}

\end{proof}

\end{lemma}

\begin{prop}\label{prop:limit}
	For normalized constraint system $\ConstrSys{S}$ that have conservative  model $\sigma$, 
	for any $x\in \vars{\ConstrSys{S}}$ we have $\DAGsize{x\sigma}\leq 2\times\DAGsize{\ConstrSys{S}}$.
\begin{proof}
	As $\card{\norm{\subII{\ConstrSys{S}}\sigma}} \leq \card{\subII{\ConstrSys{S}}\sigma} \leq \card{\subII{\ConstrSys{S}}}=\DAGsize{\ConstrSys{S}}$, we have (using the fact that $\sigma$ is normalized and  Lemma~\ref{lemma:subsigma}) 
	that  $\card{\subII{x\sigma}} = \card{\sub{x\sigma}}\leq \card{\norm{\sub{\ConstrSys{S}}\sigma}\cup\priv{\UniAt{A}\cap\sub{\ConstrSys{S}}}} \leq \card{\norm{\sub{\ConstrSys{S}}\sigma}}+\card{\priv{\UniAt{A}\cap\sub{\ConstrSys{S}}}} \leq  \DAGsize{\ConstrSys{S}} + \card{\UniAt{A}\cap\sub{\ConstrSys{S}}} \leq 2 \times \DAGsize{\ConstrSys{S}}$;
thus, $\DAGsize{x\sigma}\leq 2 \times\DAGsize{\ConstrSys{S}}$.
\end{proof}

\end{prop}

From this proposition and Corollary~\ref{cor:existsgood} we obtain an existence of bounded model 
for a normalized constraint system 
that have a model sending different variables to different values.
We will reduce an arbitrary constraint system to already studied case. 
The target properties are stated in Proposition~\ref{prop:generalLimit} and Corollary~\ref{cor:polynom}.

\begin{lemma}\label{lemma:sizesubst}
Given any constraint system $\ConstrSys{S}$ and any substitution $\theta$ such that $\dom{\theta}=\vars{\ConstrSys{S}}$ and $\dom{\theta}\theta \subseteq \dom{\theta}$.	Then $\DAGsize{\ConstrSys{S\theta}}\leq\DAGsize{\ConstrSys{S}}$.
\LONG{
\begin{proof}
	From Lemma~\ref{lemma:normprop} we obtain $\DAGsize{\ConstrSys{S}\theta}= \card{\subII{\ConstrSys{S}\theta}} 
	= \card{\subII{\ConstrSys{S}}\theta\cup\subII{\vars{\ConstrSys{S}}\theta}}$, but $\vars{\ConstrSys{S}}\theta \subseteq  \dom{\theta}=\vars{\ConstrSys{S}}$ ($\vars{\ConstrSys{S}\sigma}$ consists only of variables), and then  $\subII{\vars{\ConstrSys{S}}\theta} = \vars{\ConstrSys{S}}\theta$. As  $\vars{\ConstrSys{S}} \subseteq\subII{ \ConstrSys{S}}$, we have  $\subII{\ConstrSys{S}}\theta\cup\subII{\vars{\ConstrSys{S}}\theta}=\subII{\ConstrSys{S}}\theta$. \br
	Thus, $\DAGsize{\ConstrSys{S}\theta}= \card{\subII{\ConstrSys{S}}\theta}\leq \card{\subII{\ConstrSys{S}}} = \DAGsize{\ConstrSys{S}}$.
\end{proof}
}%
\end{lemma}

\begin{df}
	Let $\sigma$ and $\delta$ be substitutions. Then $\sigma[\delta]$ is a substitution such that $\dom{\sigma[\delta]} = \dom{\delta}$ and for all $x\in\dom{\sigma[\delta]}$, $x\sigma[\delta] = (x\delta)\sigma$.
\end{df}

\begin{lemma}\label{lemma:substcomp}
	Let $\theta$ and $\sigma$ be substitutions such that $\dom{\theta}\theta=\dom{\sigma}$, $\dom{\sigma}\subseteq \dom{\theta}$ and $\sigma$ is ground. Then, for any term $t$, $(t\theta)\sigma= t\sigma[\theta]$.
\LONG{
\begin{proof}

When apply $\theta$ to $t$, every variable $x$ of $t$ such that $x\in\dom{\theta}$ is replaced by $x\theta$; then we apply $\sigma$ to $t\theta$:  every variable $y$ of $t\theta$ is replaced by $y\sigma$, thus, every variable $x$ from $\dom{\theta}$ will be replaced to $(x\theta)\sigma$ (as $\dom{\theta}\theta=\dom{\sigma}$); and no other variables will be replaced (as $\dom{\sigma}\subseteq \dom{\theta}$). Thus, we can see that it is the same as in definition of $\sigma[\theta]$.
\end{proof}
}%
\end{lemma}

\begin{prop} \label{prop:generalLimit}
  Given any satisfiable constraint system $\ConstrSys{S}$. Then there exists a model $\sigma$ of $\ConstrSys{S}$ such that for any $x\in\dom{\sigma}$, $\DAGsize{x\sigma}\leq 2\times\DAGsize{\ConstrSys{\norm{S}}}$ %
\begin{proof}[Proof idea]
Given a normalized model $\sigma'$ of $\ConstrSys{S}$ we build a substitution $\theta$ that maps different variables whose $\sigma'$-instnatces are the same to one.
In this way we obtain a new constraint system and its normalized model on which we can apply Corollary~\ref{cor:existsgood} and get its conservative model $\sigma''$,
and by applying Proposition~\ref{prop:limit} we get a bound on size for this model. 
On the other part, we use Lemma~\ref{lemma:substcomp} to show that $\sigma''[\theta]$ is a model of $\norm{\ConstrSys{S}}$.
And then, using obtained bound and Lemma~\ref{lemma:sizesubst} show existence of a model with stated property.
The detailed proof is given in \ref{app:prop|generalLimit}

\end{proof}

\end{prop}

\begin{cor}\label{cor:polynom}
Constraint system  $\ConstrSys{S}$ is satisfiable if and only if there exists  a normalized model of $\ConstrSys{S}$ defined on $\vars{\ConstrSys{S}}$ 
which maps a variable to a ground term in $\Universe(\UniVar{A}\cap \sub{\ConstrSys{\norm{S}}}, \emptyset)$  with size not greater than double $\DAGsize{\ConstrSys{S}}$.

\end{cor}

Using this result, we propose an algorithm of satisfiability of constraint system (Algorithm~\ref{alg:solving}).

\begin{algorithm}[H]
  \caption{Solving constraint system}
  \label{alg:solving}
  \SetKw{Guess}{Guess}
  \SetKw{Normalize}{Normalize}
  \KwIn{A constraint system  $\ConstrSys{S} = \set{E_i\rhd t_i}_{i=1,\dots,n}$}
  \KwOut{Model $\sigma$, if exists; otherwise $\bot$}
  \BlankLine
\Guess for every variable of $\ConstrSys{S}$ a value of ground normalized substitution $\sigma$ with size not greater than $2\times\DAGsize{\ConstrSys{{S}}}$\;
\eIf{ $\sigma$ satisfies $E_i\rhd t_i$ for all $i=1,\dots,n$}{
		\Return{ $\sigma$} }{\Return{ $\bot$ }}
\end{algorithm}

\begin{prop}
	Algorithm~\ref{alg:solving} is correct.
\begin{proof}
Let $\sigma$ be an output of Algorithm~\ref{alg:solving}. Then $\sigma$ is a ground substitution and $\sigma$ satisfies all constraints from $\ConstrSys{S'}$ and therefore, satisfies all constraints from $\ConstrSys{S}$ . This means, $\sigma$ is a model of $\ConstrSys{S}$.
\end{proof}

\end{prop}

\begin{prop}
	Algorithm~\ref{alg:solving} is complete.
\begin{proof}
Suppose, $\ConstrSys{S}$ is satisfiable.  Then, by Corollary~\ref{cor:polynom}, there exists a guess of value of ground substitution on every element of $\vars{\ConstrSys{S}}$  with size not greater than  $2\times\DAGsize{\ConstrSys{S}}$ which represents a model $\sigma$ of $\ConstrSys{S}$. Thus, algorithm~\ref{alg:solving} will return this $\sigma$.

\end{proof}

\end{prop}

\section{Complexity analysis}
\label{sect:complexity}

In this section we present complexity classes of proposed algorithms. 
First, we expose what we use as a representation of constraint systems to justify the selected measure of algorithms inputs.
Then, we notice that normalization algorithm is polynomial in time.
After that we will show the polynomial complexity of the ground derivability algorithm.
And as a consequence of the results given before, we obtain that the proposed algorithm for solving general constraint system within DY+ACI model is in $NP$.

To reason about complexity, we have to define a size of its input. For terms and set of terms, we will use $\DAGsize{\cdot}$ + $\card{\edges{\cdot}}$, 
where  $\edges{\cdot}$ is a set of edges of DAG-representation of its argument.
For system of constraints $\ConstrSys{S} = \set{E_i\rhd t_i}_{i=1,\dots,n}$ we will use $n\times\DAGsize{\ConstrSys{S}}$+ $\card{\edges{\css}}$ .
\LONG{
 The justification is given below.

\begin{df}\label{def:DAGSys}
	\emph{DAG-representation} of a constraint system $\ConstrSys{S} =  $ \br $ \set{E_i\rhd t_i}_{i=1,\dots,n}$ is a tagged graph with labeled edges $\mathbb{G=\tuple{V, E, \otaag}}$ ($\mathbb{V}$ is a set of vertices and $\mathbb{E}$ is a set of edges; $\otaag$ is a tagging function defined on $\mathbb{V}$) such that:
\begin{itemize}
 \item there exists a bijection $f: \mathbb{V} \mapsto \subII{\ConstrSys{S}}$; 
 \item $\forall v\in \mathbb{V}$  $\taag{v}=\tuple{s,m}$, where
		\begin{itemize}
			\item $s = \rut{f(v)}$;
			\item  $m$ is $2n$-bit integer, where $m[2i-1]=1 \iff f(v) \in E_i$ and $m[2i]=1 \iff f(v) = t_i$. 
		\end{itemize}

 \item  $v_1\xrightarrow{1}v_2 \in \mathbb{E} \iff \exists p\in\Universe: 
	  ( \exists \opbin: f(v_1)=\bin{f(v_2)}{p} )  
	  \vee f(v_1)=\priv{f(v_2)}
	$;
 \item  $v_1\xrightarrow{2}v_2 \in \mathbb{E} \iff \exists p\in\Universe: 
	    \exists \opbin: f(v_1)=\bin{p}{f(v_2)} 
	$;
 \item  $v_1\xrightarrow{i}v_2 \in \mathbb{E} \iff  
	 f(v_1) = \aci{L} \wedge L[i] = f(v_2)
	$;
\end{itemize}

\end{df}

\begin{example}
A constraint system
\[
\ConstrSys{S} = 
\left \{ 
\begin{array}{rl} 
 \set{\enc{a}{x}, \pair{b}{\enc{a}{a}}, c } & \rhd a\\
 \set{\priv{b}, c }& \rhd y \\
 \set{\enc{\sig{a}{\priv{c}}}{y}, \aenc{x}{b} }& \rhd \pair{\enc{a}{x}}{c}
\end{array}
\right. 
\]
will be represented as shown\footnote{Label ``1'' (resp.``2'') of an edge is represented by a left (resp. right) side of its source node} in \figurename{}~$\ref{fig:representationExample}$. Nodes of this graph represent an element from $\subII{\ConstrSys{S}}$ by indicating its root symbol (first part of its tag) and pointers to the children.

\begin{figure}
\includegraphics[width=\columnwidth]{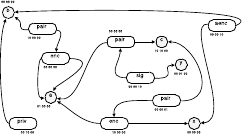}	
 \caption{DAG-representation of constraint system $\ConstrSys{S}$}\label{fig:representationExample}
\end{figure}
\end{example}

Remark that this representation can be refined, as we know that RHS of a constraint is exactly one term. That is why we could tag  a node not with $2n$ bits but with $n+\lceil \log(n+1) \rceil$ bits (concerning the second component of the tagging function).

The shown representation can be written in not more than $P(n\times \card{\mathbb{V}(\css)} + \card{\edges{\css}})$ bits of space,
where $\mathbb{V}(\cdot)$ is a set of edges in the DAG-representation, and $P$ is some polynomial with non-negative coefficients.
As we have a bijection between $\mathbb{V}(\css)$ and $\subII{\css}$, 
we obtain $\card{\mathbb{V}(\css)} = \DAGsize{\css}$.
On the other hand, as
we are not interested in rigorous estimation of complexity, 
but work in a polynomial class, 
we will estimate complexity of algorithms by taking $n\times \DAGsize{\css}+\card{\edges{\css}}$ as the measure of constraint system $\ConstrSys{S}$.

The DAG-representation of a term $t$ has the similar structure as it was shown for constraint system except that it does not need the second part of a tagging function: 
we need only $\rut{f(v)}$ as a node's tag. 
The size of this representation will be polynomially bounded by  $\DAGsize{t}+\card{\edges{t}}$.
Thus we give the following definition:

\begin{df}
 The \emph{measure} of term $t$ is defined as: $\measure{t} = \DAGsize{t}+\card{\edges{t}}$.
 For a constraint system $\css= \set{E_i\rhd t_i}_{i=1,\dots,n}$, its measure: $\measure{\css}= n\times \DAGsize{\css}+\card{\edges{\css}}$.
\end{df}

Note that for the normalized terms and constraint systems, number of edges in their DAG-representation are polynomially limited w.r.t. the number of vertices:
\begin{lemma}\label{lemma:edgesVSvertices}
 For any normalized term $t$, $\card{\edges{t}} < (\DAGsize{t})^2$. 
 For any normalized constraint system $\css$,  $\card{\edges{\css}} < (\DAGsize{\css})^2$.
\begin{proof}
  Since the term (resp. constraint system) is normalized, we cannot have more than two edge between two nodes.
It evidently holds for binary and unary nodes;
for $\opaci$-nodes it holds because of normalization: if a $\opaci$-node has two edges to one child, the term is not normalized (one of these edges should have been removed).
Therefore, as the graph is directed and acyclic, with as maximum two edges between two nodes, 
we have not more than $ \DAGsize{x} \times (\DAGsize{x} -1)$ edges  (where $x$ is a term $t$ or constraint system $\css$).
\end{proof}

\end{lemma}

}%

\subsection{Satisfiability of a general DY+ACI constraint systems is in 
  \texorpdfstring{$NP$}{%
    NP
  }%
}

\begin{lemma}\label{lemma:normComplexity} 
Given a term $t$. Normalization can be done in polynomial time on $\measure{t}$.
The same holds for a constraint system \css: normalization can be done in polynomial time on $\measure{\css}$.
\begin{proof}[Proof idea (for the case of terms)]
 The algorithm of term normalization works bottom-up by flattening nested ACI-sets, 
sorting children of ACI-set nodes, 
merging duplicated nodes while removing unnecessary duplicating edges and 
removing nodes without incoming edges (except the root-node of $t$).
\end{proof}

\end{lemma}

\begin{prop}\label{prop:alg|np}
	The general constraint system within DY+ACI satisfiability problem, that Algorithm \ref{alg:solving} solves, is in $NP$.
\begin{proof}
\allowdisplaybreaks
	Algorithm \ref{alg:solving} returns a proof  for the decision problem if it exists. We have to show, that the verification of this proof takes a polynomial time with regard to the input problem measure.
To do this, we will normalize $\ConstrSys{S}\sigma$ and then apply algorithm of checking ground derivability. Using the fact that $\DAGsize{x\sigma} \leq  2\times \DAGsize{\ConstrSys{S}}$ and polynomial complexity of the normalization and the ground derivability, we can overapproximate the execution time with polynomial on $\measure{\css}$. 
The details of the proof are given in \ref{app:prop|alg|np}.
\end{proof}

\end{prop}

On the other hand, we can reuse a technique presented in \cite{RusinowitchT03} 
to show that the satisfiability of a constraint system is an NP-hard problem.
The authors encoded 3-SAT problem into an insecurity problem of a single-session sequential protocol. 
Because the steps of the protocol are linearly ordered, the finding of an attack
is reduced to the satisfiability problem of a single constraint system.

\begin{theorem}\label{theo:dyaci-NPcomplete}
Satisfiability of general  DY+ACI constraint systems  is $NP$-complete.
\end{theorem}

\subsection{Ground derivability in DY+ACI is in 
  \texorpdfstring{$P$}{P}
}

\begin{prop}\label{prop:groundcomplexity}
 Algorithm \ref{alg:ground} has a polynomial complexity on \br $\DAGsize{E \cup \set{t}}$. 
\begin{proof} 
We will give a very coarse estimate. 

First remark, that in any step of algorithm, $\card{S}$ and $\card{D}$ don't exceed $\card{\sub{E\cup\set{t}}}$.

Building $\sub{E}\cup\sub{t}$ takes linear time on $\DAGsize{E \cup \set{t}}$.
Building $S$ will take not more than $O\left(\card{E}\times \card{\sub{E}\cup\sub{t}}\right)$, that is, not more than $O\left(\left(\DAGsize{E \cup \set{t}}\right)^2\right)$.

 The main loop has at most $\card{\sub{E \cup \set{t}}} - \card{E}$ steps.
 Searching for DY rule with left-hand side in $D$ and right-hand side in $S$ is not greater that $O(\card{S}\times \card{D}^2)$ and thus, not greater that $O((\DAGsize{E\cup\set{t}})^3)$. 
The next \texttt{if}  can be performed in $O(\card{S}\times \card{D}\times(\DAGsize{E\cup\set{t}}))$ steps and  the last \texttt{if} can be also done for cubic time. The check done in \texttt{return} statement is linear. 
And finally, thanks to the Statement~\ref{pSizesComparision} of Lemma~\ref{lemma:normprop}, we can easily justify the claimed complexity.
\end{proof}

\end{prop}

\section{Satisfiability of general DY constraint system} 
\label{sec:dy}

\begin{figure}
\center
\includegraphics[width=.8\columnwidth]{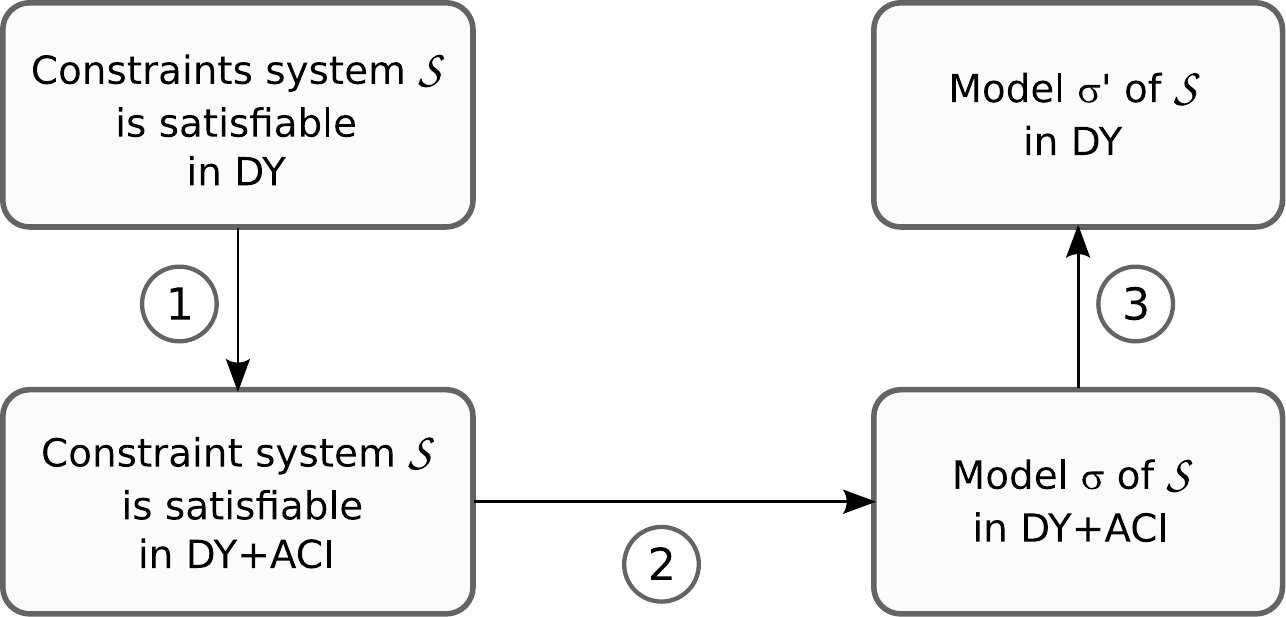}	
 \caption{ Proof Plan}\label{fig:DYvsACI}
\end{figure}

The previous result on constraint solving for  DY+ACI theory can be projected to the classical DY case. We cannot apply it directly, as in the resulting model we will probably have an ACI symbol. Thus, we need to prove the decidability of DY case. 
The scheme we follow to solve a constraint system within DY deduction system is shown in \figurename{}~\ref{fig:DYvsACI}.

First, we can show that if a constraint system is satisfiable within DY, then it is satisfiable within DY+ACI  (Proposition \ref{prop:Dy->Dyaci}).

Second, as we know, we can find a model of a given constraint system within DY+ACI.

Third, we will transform the  model obtained from previous step (which is in DY+ACI) in such a way, 
that the resulting substitution will be a model of initial constraint system within DY\LONG{ (Theorem~\ref{theorem:DY})}. 
The idea of  satisfactory transformation $\delta$ is simple: we replace any ACI list of terms with nested pairs: $\aci{\lst{t_1,\dots,t_n}}$ we replace with
$\pair{t_1}{\pair{...}{t_n}}$.
Note, that this transformation will have a linear complexity and the transformed model will have the DAG-size not more than  twice bigger than initial. 
This gives us a class of complexity, which is NP, for the problem of satisfiability of general constraint system within DY model.

\LONG{
\begin{df}
We define a replacement $\dlta{t}:\UniverseG\mapsto \UniverseG $ in the following way: 
\[
\dlta{t} = \left \{ 
\begin{array}{rl} 
  t, & \mbox{if } t \in \UniVar{X} \cup \UniAt{A};\\
  \bin{\dlta{p}}{\dlta{q}} , & \mbox{if } t = \bin{p}{q}, \\
  \priv{\dlta{p}} , & \mbox{if } t = \priv{p};\\
  \dlta{t_1} , & \mbox{if } t = \aci{t_1};\\
  \pair{\dlta{t_1}}{\dlta{\aci{t_2,\dots,t_m}}} , & \mbox{if } t = \aci{t_1,\dots,t_m}, m>1;\\
\end{array}
\right. 
\]
	
\end{df}
\begin{df}
	Given substitution $\sigma$. Then $\dlta{\sigma} = \set{x\rightarrow \dlta{x\sigma}}_{x\in \dom{\sigma}}$. For $T\subseteq \UniverseG$, $\dlta{T}=\set{\dlta{t}:t\in T} $.
\end{df}

Let us recall classical Dolev-Yao deduction system (DY) in Table~\ref{tab:DY}.
\begin{table}[ht]
\centering
\begin{tabular}{|l|l|}
\hline
Composition rules & Decomposition rules \\
\hline 
 $ {t_1, t_2} \rightarrow {\enc{t_1}{t_2}}$ & ${\enc{t_1}{t_2},  {t_2}} \rightarrow {t_1}$ \\
 $ {t_1, t_2} \rightarrow {\aenc{t_1}{t_2}}$ &  ${\aenc{t_1}{t_2},  {\priv{t_2}}} \rightarrow {t_1}$\\
 $ {t_1, t_2} \rightarrow {\pair{t_1}{t_2}}$ & $ {\pair{t_1}{t_2}} \rightarrow {t_1}$\\
 $ {t_1, \priv{t_2}} \rightarrow {\sig{t_1}{\priv{t_2}}}$ & ${\pair{t_1}{t_2}} \rightarrow {t_2}$\\ 
\hline
\end{tabular}
\caption{DY deduction system rules}\label{tab:DY}
\end{table}

\begin{df}
A constraint system $\ConstrSys{S}$ \emph{standard}, if for all $s\in \subII{\ConstrSys{S}}$ $\rut{s}\neq \cdot$.
The  definition is extended in natural way to  terms, sets of terms and substitutions.
\end{df}

We can redefine the notion of derivation for Dolev-Yao deduction system in a natural way, and denote it as $\oder_{DY}$.

\begin{lemma}
	Any standard constraint system is normalized.
\end{lemma}
 
\begin{lemma}\label{lemma:stdNorm}
	Let $t$ be a standard term, $\sigma$ be a normalized substitution. Then $t\sigma$ is normalized.
\end{lemma}

\begin{prop}\label{prop:Dy->Dyaci}
 If a standard constraint system $\ConstrSys{S}$ has a model $\sigma$ within DY deduction system, then $\ConstrSys{S}$ has a model within DY+ACI deduction system.
\begin{proof}
  It is enough to consider the same model $\sigma$ in DY+ACI. As $\ConstrSys{S}\sigma$ is normalized and as DY+ACI includes all the rules from DY, it is easy to show using the same derivation that proves $\sigma$ to be a model in DY, that $\sigma$ stays a model of $\ConstrSys{S}$ in DY+ACI. 
\end{proof}

\end{prop}

The goal of the following reasoning is to show that we can build a model of a constraint system within DY 
from a model of this constraint system within DY+ACI.

\begin{lemma}\label{lemma:validstepDY}
	For any DY+ACI rule $l_1,\dots,l_k\rightarrow r$, if $l_i$ are normalized for all $i=1,\dots,k$ then $\dlta{r} \in \derdy{\set{\dlta{l_1},\dots,\dlta{l_k}}}$.
\LONG{	
\begin{proof}
	Let us consider all possible rules: 

\begin{itemize}
 \item ${t_1,t_2} \rightarrow \norm{\pair{t_1}{t_2}}$
 
 	As $t_1$ and $t_2$ are normalized, then $\norm{\pair{t_1}{t_2}} = \pair{t_1}{t_2}$.
 	We can see, that $\dlta{\pair{t_1}{t_2}}=\pair{\dlta{t_1}}{\dlta{t_2}} \in \derdy{\set{\dlta{t_1},\dlta{t_2}}}$.
  
  \item ${t_1,t_2} \rightarrow \norm{\enc{t_1}{t_2}}$.  
	Proof of this case can be done by analogy of previous one.
  
  \item ${t_1,t_2} \rightarrow \norm{\aenc{t_1}{t_2}}$.  
	Proof of this case can be done by analogy of previous one.
  
  \item ${t_1,\priv{t_2}} \rightarrow \norm{\sig{t_1}{\priv{t_2}}}$.
 
 	As $t_1$ and $\priv{t_2}$ are normalized, then $\norm{\sig{t_1}{\priv{t_2}}} =  $ \br $ \sig{t_1}{\priv{t_2}}$.
 	We can see, that $\dlta{\sig{t_1}{\priv{t_2}}} =  $ \br $ \sig{\dlta{t_1}}{\dlta{\priv{t_2}}} =\sig{\dlta{t_1}}{\priv{\dlta{t_2}}} \in  $ \br $ \derdy{\set{\dlta{t_1},\priv{\dlta{t_2}}}}$, 
 	but $\priv{\dlta{t_2}}=\dlta{\priv{t_2}}$.
 	
  \item ${t_1, \dots, t_m} \rightarrow \norm{\aci{t_1, \dots, t_m}}$.  
  
  	The fact, that $\elems{\norm{\aci{t_1, \dots, t_m}}} = \bigcup_{i=1,\dots,m} \elems{t_i}$ follows from $t_i=\norm{t_i}$ (for all $i$) and Lemma~\ref{lemma:normprop}.
	
	We can (DY)-derive  from $\set{\dlta{t_i}}$ any term in $\dlta{\elems{t_i}}$, trivially, if $t_i\neq \aci{L}$ and 
	by applying rules $\pair{s_1}{s_2} \xrightarrow{DY} s_1$ and $\pair{s_1}{s_2} \xrightarrow{DY} s_2$  otherwise (proof by induction on size of $t_i$).
	
	One can observe, that $\dlta{t}$ is a pairing (composition of $\pair{\cdot}{\cdot}$ operator with itself) of $\dlta{\elems{t}}$ (by definition of $\dlta{\cdot}$ and normalization function). 
	And then, as $\dlta{t}$ is limited in size, we can  (DY)-derive $\dlta{t}$ from ${\dlta{\elems{t}}}$ by iterative use of rule $s_1,s_2 \xrightarrow{DY} {\pair{s_1}{s_2}}$, if needed. 
	
	Thus, first we can derive $\dlta{\elems{t_i}}$ for all $i$, and then rebuild (derive with composition rules) $\dlta{\norm{\aci{t_1, \dots, t_m}}}$.
	
  \item ${\enc{t_1}{t_2},\norm{t_2}} \rightarrow \norm{t_1}$. 

	As $\enc{t_1}{t_2}$ is normalized, then $t_1=\norm{t_1}$ and $t_2=\norm{t_2}$. Thus, $\dlta{t_1}\in \derdy{\set{\enc{\dlta{t_1}}{\dlta{t_2}},\dlta{t_2}}}$ and this is what we need, as $\dlta{\enc{t_1}{t_2}} = \enc{\dlta{t_1}}{\dlta{t_2}}$.
  
  \item ${\pair{t_1}{t_2}} \rightarrow \norm{t_1}$. Similar case.
  \item ${\pair{t_1}{t_2}} \rightarrow \norm{t_2}$. Similar case.

  \item ${\aenc{t_1}{t_2},\norm{\priv{t_2}}} \rightarrow \norm{t_1}$. Similar case. Note, that $\dlta{{\priv{t_2}}}=\priv{{\dlta{t_2}}}$

  \item ${\aci{t_1, \dots, t_m}} \rightarrow \norm{t_i}$. 
  
   	As said above, $\dlta{\elems{\aci{{t_1,\dots,t_m}}}} \subseteq \derdy{\set{\dlta{\aci{{t_1}, \dots, {t_m}}}}}$;
   	and as $\dlta{\elems{t_i}}\subseteq  \dlta{\elems{\aci{{t_1},\dots,{t_m}}}}$, we  can (DY)-derive (by composition rules) $\dlta{t_i}$ from $\dlta{\elems{t_i}}$.
\end{itemize}

\end{proof}
}%
\end{lemma}

\begin{prop}\label{prop:subtDY}
	Given a standard constraint system $\ConstrSys{S}$ and its normalized model $\sigma$ in DY+ACI. 
	Then, for any subterm of the system $t\in\subII{\ConstrSys{S}}$, we have $\dlta{t\sigma} = t\dlta{\sigma}$.
\begin{proof}
	The proof is done by induction as in Proposition~\ref{prop:subt}.
	
	\begin{itemize}
 \item Let $\DAGsize{t} = 1$. Then either $t \in \UniAt{A}$ or $t \in \UniVar{X}$. Both are trivial cases.
 \item Assume that for some $k \geq 1$ if $\DAGsize{t} \leq k$, then $\dlta{t\sigma} = t\dlta{\sigma}$.
 \item Show, that for $t$ such that $\DAGsize{t} \geq k+1$, where $t=\bin{p}{q}$ 
 or $t=\priv{p}$ and  $\DAGsize{p} \leq k$ and $\DAGsize{q} \leq k$, statement  $\dlta{t\sigma} = t\dlta{\sigma} $ is still true. We have: 
 	\begin{itemize}
 		\item either $t = \bin{p}{q}$. 
 		   As $\dlta{\bin{p}{q}\sigma} =\dlta{\bin{p\sigma}{q\sigma}}=  $ \br $ \bin{\dlta{p\sigma}}{\dlta{q\sigma}} = \bin{p\dlta{\sigma}}{q\dlta{\sigma}}=\bin{p}{q}\dlta{\sigma}$.

		\item or $t = \priv{p}$. In this case the proof can be done by analogy with previous one.
		
	\end{itemize}
	Remark: as $\ConstrSys{S}$ is standard, $t\neq \aci{L}$.
\end{itemize}

\end{proof}

\end{prop}

\begin{theorem} \label{theorem:DY}
	Given a standard constraint system $\ConstrSys{S} = \set{E_i \rhd t_i}_{i=1, \dots, n}$ and its normalized model $\sigma$ in DY+ACI. 
	Then $\dlta{\sigma}$ is a model in DY of $\ConstrSys{S}$.
\begin{proof}
	Let $E \rhd t$ be any element of $\ConstrSys{S}$. As $\sigma$ is a model of $\ConstrSys{S}$, then $\norm{t\sigma} \in \der{\norm{E\sigma}}$. As $\sigma$ is normalized and $\ConstrSys{S}$ is standard, using Lemma~\ref{lemma:stdNorm} we have $\norm{t\sigma} = t\sigma$ and $\norm{E\sigma}=E\sigma$.
	Then, $t\sigma \in \der{{E\sigma}}$. That means, there exists a DY+ACI derivation $D=\set{A_0, \dots, A_{k}}$  such that $A_0={E\sigma}$ and ${t\sigma} \in A_{k}$.

	By Lemma~\ref{lemma:validstepDY} and Lemma~\ref{lemma:derext} (which also works for DY case) we can easily prove that if $k>0$, $\dlta{A_{j}} \subseteq \derdy{\dlta{A_{j-1}}}, \  j=1,\dots,k$. Note, that $\dlta{A}$ is a  set of  standard terms (and thus, normalized) for any set of terms $A$.
	Then, applying transitivity of $\derdy{\cdot}$ (Lemma~\ref{lemma:dertrans} for DY) $k$ times, we have that $\dlta{A_k} \subseteq \derdy{\dlta{A_0}}$. In the case where $k=0$, the statement $\dlta{A_k} \subseteq \derdy{\dlta{A_0}}$ is also true.

	 Using Proposition~\ref{prop:subtDY} we have that $\dlta{A_0} =\dlta{E\sigma} = {E\dlta{\sigma}}$, as $E \subseteq \sub{\ConstrSys{S}}$. The same for $t$: $\dlta{t\sigma} = {t\dlta{\sigma}}$, and as ${t\sigma} \in A_k$, we have ${t\dlta{\sigma}} \in \dlta{A_k}$. 
  
Thus, we have that ${t\dlta{\sigma}} \in \dlta{A_k} \subseteq \derdy{\dlta{A_0}} = \derdy{E\dlta{\sigma}} $, that means $\dlta{\sigma}$  DY-satisfies any constraint of $\ConstrSys{S}$. 
	
\end{proof}

\end{theorem}

We present an example illustrating the theorem.
}

\begin{example}\label{ex:DYconstrsys}
 Let us consider a standard constraint system similar to one in Example~\ref{ex:constrsys}.
\[
   \ConstrSys{S}=\left\{ 
 \begin{array}{l l l}
	    { \enc{x}{a},\pair{c}{a}} & \rhd &  b \\
	    { \pair{x}{c}} & \rhd & a 
 \end{array}
  \right\},
\]  
Using Algorithm~\ref{alg:solving}, we can get a model of $\ConstrSys{S}$ within DY+ACI, let's say, as in Example~\ref{ex:constrsysmodel},
$\sigma = \set{x\mapsto \aci{\lst{a,b,c}}}$.

Then, by applying transformation $\dlta{\cdot}$, we will get $\sigma'=\dlta{\sigma} =  $ \br $  \set{x\mapsto \pair{a}{\pair{b}{c}}}$.
We can see, that $\sigma'$ is also a model of $\ConstrSys{S}$ within DY\LONG{ (as it was proven in Theorem~\ref{theorem:DY})}.

\end{example}

\LONG{
\begin{cor}[of Theorem~\ref{theorem:DY} and Proposition~\ref{prop:Dy->Dyaci}]
 A standard constraint system $\ConstrSys{S}$ is satisfiable within DY iff it is satisfiable within DY+ACI.
\end{cor}

\begin{cor}
 Satisfiability of constraint system within DY is in $NP$.
\end{cor}
}

\section{Conclusions} 
\label{sec:concl}

In this work we presented a decision algorithm of satisfiability  of general constraint system 
within Dolev-Yao deduction system 
as well as one extended with ACI symbol that can be used to represent sets of terms.
The complexity class of the algorithm was proved to be in $NP$-complete.

We have given also two applications of the presented result: 
protocol insecurity with non-communicating intruders
and discovering XML-based attacks.

\appendix
  \section*{APPENDIX}

\section{General constraints for subterm theories}
\label{sec:undec}

{

\renewcommand{\sub}[1]{\oBASEsubterms\left(#1\right)}
\begin{itemize}
 \item composition rules: for all public functional symbols $f$,  $x_1,\dots, x_k \rightarrow f(x_1,\dots, x_k)$
 \item decomposition rules: $t_1,\dots,t_m \rightarrow s$, where $s$ is a subterm of $t_i$ for some $i$.
\end{itemize}

We show that the satisfiability of constraint system within subterm deduction system is undecidable in general.
More precisely: 
\LONG{\begin{description}}
\SHORT{\begin{description}[\IEEEsetlabelwidth{\bf Question}]}
\item[{\bf Instance}:] a subterm deduction system D, a constraint system C.
\item[{\bf Question}:] is C satisfiable ? 
\end{description}
To show this, we reduce the halting problem of a Deterministic Turing Machine (TM) $M$
that works on a single tape. 
We consider the tape  alphabet 
${\Gamma} =\{ {0}, {1}, {\flat} \}$,
and $\flat$  is the  blank symbol.
The states of the TM $M$ are in a finite set 
${Q} =\{ {q}_1,{q}_2, \ldots ,{q}_n \}$. W.l.o.g. we can assume that 
${q}_1$ (resp. ${q}_n$) is the unique initial (resp. accepting) state. 

In order to represent  Turing machine configuration as   terms  we shall 
introduce a set of variables ${\call X}$ and an alphabet $\call F$ 
\[ {\call F} :=  \{ 0,1,\flat , \bot \} \cup Q,   \]
where ${\call F} \setminus \set{\bot} $ are public functional symbols.

The TM configuration with tape $ \bot~abcde~\bot$, (where $\bot$ 
is an endmarker), with 
symbol $d$ under the head, and state $q$  will be represented by the following term
of $ q(c(b(a(\bot), d(e(\bot), x)$ where $x\in {\call X}$ and $a,b,c,d,e \in  \{ 0,1,\flat \}$.

The composition rules we consider for the TM are
$u \rightarrow f(u)$ for each $f\in \set{0,1,\flat}$ and  $u,v,w \rightarrow q(u,v,w)$ for each $q\in Q$.
For each TM transition of $M$ we will introduce some decomposition deduction 
rule that can be applied  on a term representation $q(u,v,q'(u',v',x'))$ iff 
the  transition can be applied to a configuration represented by  $q(u,v,\_)$ 
and generate a configuration  represented by  $q'(u',v',\_)$.
\medskip
For each TM instruction of type:
\noindent{\it ``In state $q$ reading $a$ go to state $q'$ and write $b$''},
 we define the following rule for $a,b \in \{ 0,1,\flat \}$:

$$q(u,a(v),q'(u,b(v),x)) \rightarrow q'(u,b(v),x) $$

For each instruction of type:
\noindent{\it ``In state $q$ reading $a$ go to state $q'$ and move right''}, 
we define the following  rules for $a \in \{ 0,1,\flat \}$ :
$$q(u,a(v),q'(a(u),v,x)) \rightarrow q'(a(u),v,x)$$

A  rule is  for extending the tape on the right when needed:

$$q(u,\bot,q'(\flat(u),\bot,x)) \rightarrow q'(\flat(u),\bot,x)$$

For each instruction of type:
\noindent{\it ``In state $q$ reading $a$ go to state $q'$ and move left''}, 
we define the following  rules for $a \in \{ 0,1,\flat \} $ 
:

$$q(a(u),v,q'(u,a(v),x)) \rightarrow  q'(u,a(v),x)$$

A rule is  for extending the tape on the left when needed:
$$q(\bot,a(v),q'(\bot,\flat(a(v)),x)) \rightarrow q'(\bot,\flat(a(v)),x)$$

The resulting deduction system $D_{M}$  is obviously a subterm deduction system.

Let us consider a constraint  $\ConstrSys{S}$ to be solved modulo  $D_{M}$: 
\[
\set{q_1(\bot,\bot,x)} \rhd q_n(y,z,w)
\]

This constraint is satisfiable iff there is a sequence of transitions of $M$ 
from a configuration with  initial state $q_1$ and empty tape 
to a configuration with an accepting state.
Hence the constraint solving problem is undecidable. 

Let us recall the definition of some  properties of constraint systems.  
These two properties are natural  for modeling standard  security protocols: 

\SHORT{ \begin{description}[\IEEEsetlabelwidth{variable origination:}]}
\LONG{ \begin{description}}
 	\item[variable origination:] $\forall i,\, \forall x\in \vars{E_i}\  \exists j<i ~~ x\in \vars{t_j}$,
 	\item[monotonicity:]  $j<i \implies E_j \subseteq E_i $.
 \end{description}

Note that $\set{\set{q_1(\bot,\bot,x)}  \rhd q_n(y,z,w)}$ is obviously monotonic.

As a consequence, satisfiability of monotonic constraint systems (but without variable origination) is undecidable. 
Here is another constraint system, where  variable origination is satisfied, but monotony is not. It can be used for 
reducing the halting problem again:
\[
	\set{
		\set{\bot} \rhd x,\
		\set{q_1(\bot,\bot,x)}  \rhd q_n(y,z,w)
	}
\]

As a consequence, satisfiability of  constraint systems with variable origination (but without monotonicity) is undecidable. 

\medskip

We should note by contrast (see ~\cite{Baudet05}), that constraint solving in subterm 
convergent theories is decidable if the constraint system $\ConstrSys{S}=\set{E_i\rhd t_i}_{i=1,\dots,n}$ 
satisfies both variable origination and monotonicity.

}

\section{Proofs}
\label{sec:proofs}

{

\subsection{Proofs of several statements of Lemma~\ref{lemma:normprop}}
\label{app:lemma|normprop}
\begin{description}
  \item [Statement~\ref{pACI}:]
      Follows from the definition of the normalization function and Definition~\ref{def:elems}.
\DRAFT{
  \item [Statement~\ref{pNormNorm}:]
???
}
  \item [Statement~\ref{pSubNorm}:]
      By induction on $\DAGsize{s}$. Let us fix $t$.
      \begin{itemize}
       \item $\DAGsize{s} = \DAGsize{\norm{t}}$. 
	     Then $s= \norm{t}$ and $\norm{s}= \norm{\norm{t}}$, 
	     and from Statement~\ref{pNormNorm} (by taking empty $\sigma$) we have $\norm{t}=\norm{\norm{t}}$, 
	     and thus $\norm{s}=s$.
       \item Suppose, that for some $k$, for any $s\in\sub{\norm{t}}$, such that $\DAGsize{s} > k$, $s=\norm{s}$.
       \item Consider case, where  $s\in\sub{\norm{t}}$ and $\DAGsize{s} = k$. 
	     Then, by definition of $\sub{\cdot}$, $s$ is in 
	     \begin{itemize}
	      \item $\priv{s}\in \sub{\norm{t}}$. By induction supposition we have $\norm{\priv{s}} = \priv{s}$, and as $\norm{\priv{s}}=\priv{\norm{s}}$, we have $s=\norm{s}$.
	      \item $\bin{s}{p} \in \sub{\norm{t}}$. By induction we have $\norm{\bin{s}{p}} = \bin{s}{p}$, and as $\norm{\bin{s}{p}}=\bin{\norm{s}}{\norm{p}}$, we have $s=\norm{s}$.
	      \item $\bin{p}{s} \in \sub{\norm{t}}$. The similar case.
	      \item $s \in \elems{\aci{L}}$,$\aci{L}\in \sub{\norm{t}}$. 
		    As $\DAGsize{\aci{L}}>k$, we have $\aci{L}=\norm{\aci{L}}$, 
		    that means (from Definition~\ref{def:norm}), that $L$ is a list of normalized non-ACI-set terms, and as $\elems{{L}} \approx L$, we have that $s$ is normalized.
	     \end{itemize}

      \end{itemize}
  \item [Statement~\ref{pSubNormHasProimage}:] 
      Suppose the opposite and let us take $s\in \subII{\norm{t}}$ with maximal $\DAGsize{s}$ that does not satisfy the desired property.
      Note that the ``biggest'' term in $\subII{\norm{t}}$, i.e. $\norm{t}$, does satisfy the property, as we can choose $s'=t\in \subII{t}$. 
      By definition of $\subII{\cdot}$ if $s\in\subII{\norm{t}}$ and $s\neq \norm{t}$ then there exists $r \in\subII{\norm{t}}$ such that
      \begin{itemize}
       \item  $r = \bin{p}{s}$ or $r=\bin{s}{p}$ or $r=\priv{s}$. 
	     Without loss of generality we consider only the first case ($r = \bin{p}{s}$) as other ones are similar.
	     As $\DAGsize{r} > \DAGsize{s}$, there exists $r'\in\subII{t}$ such that $r=\norm{r'}$. 
	     By definition of $\norm{\cdot}$:
	     \begin{itemize}
	      \item either $r'= \bin{p'}{s'}$ and $\norm{p'} = p$ and $\norm{s'} = s$. As $s' \in \subII{r'}\subseteq \subII{t}$ the property is proved.
	      \item or $r'= \aci{L}$ and $\norm{\elems{L}} = \set{r}$. Since for all $q \in \elems{L}$,  $\rut{q}\neq \opaci$, then there exists $q \in \elems{L}$ such that 
	      $q = \bin{p'}{s'} $  and $\norm{p'} = p$ and $\norm{s'} = s$. Using Statement~\ref{pSizesComparision} we have $s' \in \subII{t}$.
	     \end{itemize}

       \item $r = \aci{L}$ and $s \in L$. Then, (since $\DAGsize{r} > \DAGsize{s}$) there exists  $r' \in \subII{t}$ such that $\norm{r'} = r$. %
	  Using Lemma~ref{lemma:DAGvsQuasi} and Statement~\ref{pSubNorm} we obtain $r$ --- normalized, and thus, $\rut{s} \neq \opaci$.
	  Then by definition of $\norm{\cdot}$ we have $r' = \aci{L'}$ and $L \approx \norm{\elems{L'}}$, and thus, $s\in\norm{\elems{L'}}$,
	  that is there exists $s' \in \elems{L'}$ such that $s = \norm{s'}$. Using again Statement~\ref{pSizesComparision} we have $s' \in \subII{t}$.
      \end{itemize}
  \item [Statement~\ref{pDotNormElems}:]    
      To get the first part of equality we first use Statement~\ref{pNormElems}: 
      $\elems{\norm{\aci{\norm{t_1}, \dots, \norm{t_m}}}} = \norm{\elems{\aci{\norm{t_1}, \dots, \norm{t_m}}}}$; then
      from Definition~\ref{def:elems} and Statement
      \ref{pNormElems} 
      we have that $\elems{{\aci{\norm{t_1}, \dots, \norm{t_m}}}}$ is a set of normalized terms.
      The second part directly follows from Definition~\ref{def:elems}.
  \item [Statement~\ref{pAhList}:]    
	By induction on $\DAGsize{t}$.
	  \begin{itemize}
	  \item $\DAGsize{t}=1$, implies $t = a \in\UniAt{A}$ and then $\elems{a}=\set{a}$, i.e.  the equality becomes trivial.
	  \item Suppose, that for any $t: \DAGsize{t} < k$ ($k>1$), $\ah{t} = \bigcup_{p\in\elems{t}}\ah{p}$ holds.
	  \item Given a term  $t: \DAGsize{t}=k$, $k>1$. We shold prove  $\ah{t} = \bigcup_{p\in\elems{t}}\ah{p}$.
		\begin{itemize}
		  \item $t=\priv{t_1}$ or  $t=\bin{p}{q}$. In both cases, $\elems{t}=\set{t}$, and thus, he equality is trivial.
		  \item $t=\aci{L}$. Note, that for all $s\in L$, $\DAGsize{s} <k$. Then, on the one hand, $\ah{\aci{L}}=\bigcup_{p\in L}\ah{p} = $ 
			(by induction supposition) $=\bigcup_{p\in L}\bigcup_{p'\in \elems{p}}\ah{p'}$.
			On the other hand, $\bigcup_{p\in\elems{\aci{L}}}\ah{p}=\bigcup_{p\in \bigcup_{p'\in L} \elems{p'}}\ah{p}
			=$ \\ $\bigcup_{p'\in L}\bigcup_{p\in\elems{p'}}\ah{p}$. Thus, $\ah{t} = \bigcup_{p\in\elems{t}}\ah{p}$.
		\end{itemize}
	  \end{itemize}
  \item [Statement~\ref{pPahNorm}:]    
	This follows from Statements \ref{pAhNorm}, \ref{pNormDotNorm}, Definition~\ref{def:pairing} and from  equality $\norm{\norm{t}}=\norm{t}$ (Statement~\ref{pNormNorm}).
  \item [Statement~\ref{pSubSubSub}:]
	$\sub{t} \subseteq \sub{\sub{t}}$ is trivial as $t\in\sub{t}$.
	Now we prove by induction on $\DAGsize{t}$ that $\sub{\sub{t}} \subseteq {\sub{t}}$
      \begin{itemize}
	\item $\DAGsize{t}=1$. 
	      Then $t\in\UniAt{A}\cup  \UniVar{X}$. As $\sub{t}=\set{t}$ the statement is trivial.
	\item Suppose, that for any $t: \DAGsize{t} < k$ ($k\geq 1$), the statement is true.
	\item Given a term  $t: \DAGsize{t}=k$, $k>1$. Let us consider all possible cases:
	      \begin{itemize}
		\item $t=\bin{t_1}{t_2}$. %
		      By definition $\sub{t}=\set{t}\cup\sub{t_1}\cup\sub{t_2}$.  
		      Then, $\sub{\sub{t}}=\sub{t} \cup $\br $ \sub{\sub{t_1}} \cup \sub{\sub{t_2}}$ and,
		      as $\DAGsize{t_i} < k$ for $i=1,2$, by using induction supposition we obtain the wanted property.
		\item $t=\priv{t_1}$. Proof is similar to one for the case above.
		\item $t=\aci{L}$. Then $\sub{t} = \set{t}\cup\bigcup_{p\in\elems{L}}\sub{p}$.
		      And $\sub{\sub{t}} = \sub{t} \cup \bigcup_{p\in\elems{L}}\sub{\sub{p}}$,
		      but since $\DAGsize{p} < k$ for such any $p$ we can apply induction supposition and get
		      $\bigcup_{p\in\elems{L}}\sub{\sub{p}} = \bigcup_{p\in\elems{{L}}}\sub{p} = \sub{t}\setminus \set{t}$.
		      Then \br $\sub{\sub{t}} = \sub{t} \cup (\sub{t}\setminus \set{t}) = \sub{t}$.
	      \end{itemize} 
      \end{itemize}      
  \item [Statement~\ref{pSepVarII}:]    
      By induction on $\DAGsize{t}$
      \begin{itemize}
	\item  $\DAGsize{t}=1$. 
	      \begin{itemize}
	      \item $t\in\UniAt{A}$. As $t\sigma = t$ and $\vars{t} = \emptyset$, the statement becomes trivial.
	      \item $t\in \UniVar{X}$. Then $\subII{t}\sigma = t\sigma$, $\vars{t} = \set{t}$; 
		    and as for any term p, $p\in\subII{p}$, we have $\subII{t\sigma} = \set{t\sigma}\cup\subII{t\sigma}$.
	      \end{itemize}
	\item Suppose, that for any $t: \DAGsize{t} < k$ ($k\geq 1$), the statement is true.
	\item Given a term  $t: \DAGsize{t}=k$, $k>1$. Let us consider all possible cases:
	      \begin{itemize}
		\item $t=\bin{t_1}{t_2}$. %
		      Then 
		      $t\sigma =  \bin{t_1\sigma}{t_2\sigma}$ and $\vars{t} = \vars{t_1}\cup\vars{t_2}$.
		      $\subII{t\sigma} = \set{t\sigma} \cup \subII{t_1\sigma} \cup \subII{t_2\sigma} = $ (as $\DAGsize{t_i} < k$)
		      $= \set{t\sigma} \cup \subII{t_1}\sigma\cup\subII{\vars{t_1}\sigma} \cup \subII{t_2}\sigma\cup\subII{\vars{t_2}\sigma}
		      = \set{t\sigma} \cup \subII{t_1}\sigma  \cup \subII{t_2}\sigma \cup \subII{(\vars{t_1}\cup \vars{t_2})\sigma}
		      = \subII{t}\sigma \cup \subII{\vars{t}\sigma}$.
		\item $t=\priv{t_1}$. Proof is similar to one for the case above.
		\item $t=\aci{\lst{t_1,\dots,t_m}}$. We have $t\sigma =  \aci{\lst{t_1\sigma,\dots,t_m\sigma}}$ and $\vars{t} = \bigcup_{i=1,\dots,m}\vars{t_i}$.
		      Then we have $\subII{t\sigma} = \set{t\sigma} \cup \bigcup_{i=1,\dots,m}\subII{t_i\sigma} = $ (as $\DAGsize{t_i} < k$) \br
		      $= \set{t\sigma} \cup \bigcup_{i=1,\dots,m}\left(\subII{t_i}\sigma\cup\subII{\vars{t_i}\sigma} \right )
		      = \set{t\sigma} \cup \bigcup_{i=1,\dots,m}\subII{t_i}\sigma  \cup \subII{\left( \bigcup_{i=1,\dots,m}\vars{t_i}\right)\sigma} =  $ \br $ 
		      \subII{t}\sigma \cup \subII{\vars{t}\sigma}$.
	      \end{itemize}
      \end{itemize}
  \item [Statement~\ref{pNormCard}:]    
  It follows from the fact that $f(t) = \norm{t}$ and $g_\sigma(t) = t\sigma$ are deterministic functions,
  and thus return at most one value for one given argument.  
  \item [Statement~\ref{pSizesComparision}:]    
  First we prove that $\elems{t} \subseteq \sub{t}$.
  We use induction on $\DAGsize{t}$. 
  \begin{itemize}
   \item If $\rut{t}\neq \opaci$, then $\elems{t} = \set{t} \subseteq \sub{t}$.
	 This case includes all $t$ such that $\DAGsize{t} = 1$.
         Thus we need to consider only $t=\aci{L}$.
   \item Suppose that for any $t: \DAGsize{t} < k$ ($k\geq 1$), the statement holds.
   \item If for some $t$ we have $\DAGsize{t}=k$, $k>1$, then 
	  $\elems{t} = \bigcup_{p\in L} \elems{p}$ and
	    $\sub{t} = \set{t} \bigcup_{p\in L} \sub{p}$.
	    And since  $\DAGsize{p} < k$ using the induction supposition we obtain the wanted statement.
  \end{itemize}
  Now we show that $\sub{t}\subseteq \subII{t}$.
  Again, applying proof by induction on on $\DAGsize{t}$ we have:
  \begin{itemize}
	\item If $\DAGsize{t} = 1$, then $\sub{t} = \subII{t} = \set{t}$.
	\item Suppose that for any $t: \DAGsize{t} < k$ ($k\geq 1$), the statement holds.
	\item If for some $t$ we have $\DAGsize{t}=k$, $k>1$, then 
	      \begin{itemize}
		\item $t=\bin{t_1}{t_2}$.%
		      Then 
		      $\sub{t} = \set{t}\cup \sub{t_1} \cup \sub{t_2}$ and 
		      $\subII{t} = \set{t}\cup \subII{t_1} \cup \subII{t_2}$, \br
		      where $\max\{\DAGsize{t_1}, \DAGsize{t_2}\} < k$. And then using induction supposition 
		      we can conclude for this case.
		\item $t=\priv{t_1}$. Proof is similar to one for the case above.
		\item $t=\aci{\lst{t_1,\dots,t_m}}$. 
		      Then we have \br $\sub{t} = \set{t} \cup \bigcup_{p\in\elems{\lst{t_1,\dots,t_m}}}\sub{p} \subseteq $ (using the already proved part of the property) \br
		      $ \subseteq  \set{t} \cup \bigcup_{p\in\sub{\set{t_1,\dots,t_m}}}\sub{p} = $ (as $\sub{\sub{t}}=\sub{t}$)
		      $= \set{t} \cup \bigcup_{p\in\set{t_1,\dots,t_m}}\sub{p} \subseteq $ (by induction supposition, as $\DAGsize{t_i} < k$ for all $i$)
		      $\subseteq \set{t} \cup \bigcup_{p\in\set{t_1,\dots,t_m}}\subII{p}  = \subII{t}$.
	      \end{itemize}
  \end{itemize}
  \item [Statement~\ref{pNormSize}:]    
  Using Statement~\ref{pSubNormHasProimage} and the fact that $\norm{\cdot}$ is a deterministic function we obtain
  $\forall p,q \in \subII{\norm{t}} p\neq q \, \exists p',q' \in \subII{t} : p=\norm{p'} \wedge q = \norm{q'} \wedge p\neq q$.
  And thus, $\card{\subII{\norm{t}}} \leq \card{\subII{{t}}}$.
  \item [Statement~\ref{pSubACI}:]    
  We have $\sub{\aci{\lst{t_1, \dots, t_l}}} = \set{\aci{\lst{t_1, \dots, t_l}}} \cup  $ \br $  \bigcup_{i=1}^l\sub{\elems{t_i}}$.
  Using Statement~\ref{pSizesComparision} and \ref{pSubSubSub} we have \br $\sub{\elems{t_i}} \subseteq \sub{\sub{t_i}} = \sub{t_i}$.
  Thus, \br
  $\sub{\aci{\lst{t_1, \dots, t_l}}} \subseteq \set{\aci{\lst{t_1, \dots,t_l}}}\cup \sub{t_1}\cdots\cup \sub{t_l}$.  
\DRAFT{
  \item [Statement~\ref{pNormSubSize}:]    
TODO
}

\end{description}

\subsection{Proof of Property~\ref{prop:alg|np}}
\label{app:prop|alg|np}
\LONG{	
        As was stated before, the measure of the problem input is $\measure{\css} = n\times\DAGsize{\css}+\card{\edges{\css}}$,
        where $\css=\set{E_i\rhd t_i}_{i=1,\dots,n}$.

        Algorithm \ref{alg:solving} returns a normalized proof $\sigma$ for decision problem if it exists. 
         Moreover, $\DAGsize{x\sigma}  \leq 2\times\DAGsize{\css}$ for any $x\in\vars{\css}$.

	First, we will normalize $\ConstrSys{S}\sigma$. 
	From Lemma~\ref{lemma:normComplexity} follows, 
	that we can do it for the time $T_{\norm{}} \leq P_{\norm{}}(\measure{\css\sigma})$, 
	where $P_{\norm{}}$ is some polynomial with non-negative coefficients  of some degree $m'' > 0$.

	From the Proposition~\ref{prop:groundcomplexity} 
 	we will know that check of derivability of a normalized ground term $g$ from set of normalized ground terms $G$ takes a polynomial time depending on $\DAGsize{G\cup \set{g}}$. That is,  there exists a polynomial $P_g$ with non-negative coefficients, such that number of operations (execution time) to verify the derivability ($g$ from $G$) will be limited by $P_g(\DAGsize{G\cup \set{g}})$. Then the execution time for checking a set of ground constraints $\set{G_i\rhd g_i}_{i=1,\dots,n}$ will be limited by $\sum_{i=1}^n P_g(\DAGsize{G_i\cup \set{g_i}})$.

	To show that the algorithm is in $NP$ we need to show, that execution time of check is polynomial limited by measure of algorithm's input, i.e. there exists a polynomial $P$, such that execution time does not exceed  $O(P(n\times \DAGsize{\ConstrSys{S}} + \card{\edges{\css}}))$ steps. 

	In our case, execution time $T$ of a check will be $T=T_{\norm{}} + T_g$, 
	where $T_g$ is a time needed for checking ground derivability of $\ConstrSys{S}\sigma$: 
	$T_g \leq  $ \br $ \sum_{i=1}^n P_g(\DAGsize{\norm{(E_i\cup\set{t_i})\sigma}})$. As $P_g$  is a polynomial, let us say, of degree $m' > 0$, with non-negative coefficients, we can use the fact, that for any positive integers ${x_1,\dots,x_k}$ we have $\sum_{i=1}^k P_g(x_i) \leq P_g(\sum_{i=1}^k x_i)$. Then we have $T_g\leq P_g(\sum_{i=1}^n \DAGsize{\norm{(E_i\cup\set{t_i})\sigma}})$ and by Statement~\ref{pNormSize}  of Lemma~\ref{lemma:normprop} we have $T_g\leq P_g(\sum_{i=1}^n \DAGsize{(E_i\cup\set{t_i})\sigma})$; using the same lemma, we have 
	\begin{align*} 
	\begin{split}
	T_g\leq P_g\left(\sum_{i=1}^n \left(\DAGsize{E_i\cup\set{t_i}} + \DAGsize{\bigcup_x x\sigma}\right)\right) \leq \\
	\leq P_g\left(\sum_{i=1}^n \left(\DAGsize{E_i} + \DAGsize{t_i} + \sum_x\DAGsize{x\sigma}\right)\right) \leq \\
	\leq P_g\left(\sum_{i=1}^n \left(2\DAGsize{\ConstrSys{S}}\right) +  n\times \sum_x (\DAGsize{x\sigma})\right) \leq \\
	\leq    P_g\left(2\times n\times \DAGsize{\ConstrSys{S}} + n\times \sum_x (2\times \DAGsize{\ConstrSys{S}})\right)\leq \\
	\leq P_g\left(2\times n\times \DAGsize{\ConstrSys{S}} + 2\times n\times (\DAGsize{\ConstrSys{S}})^2\right)\leq \\
	\leq P_g\left(4\times n\times (\DAGsize{\ConstrSys{S}})^2\right)\leq P_g\left(4\times (n\times \DAGsize{\ConstrSys{S}}+\card{\edges{\css}})^2\right) =\\
	= O\left(\left(\measure{\css}\right)^{2m'}\right).
	\end{split}
	\end{align*}

	On the other hand, let us consider $T_{\norm{}}$. \br
 	We have $T_{\norm{}} \leq P_{\norm{}}(n\times\DAGsize{\css\sigma}+\card{\edges{\css\sigma}})$.
 	One can see that the number of edges in DAG-representation of $\css\sigma$ (where every variable $x$ of \css is replaced by $x\sigma$)
 	will not exceed the number of edges in $\css$ plus the number of edges of all $x\sigma$: 
 	$\card{\edges{\css\sigma}}\leq \card{\edges{\css}} + \sum_{x\in\vars{\css}}\card{\edges{x\sigma}}$. 
 	And since $\sigma$ is normalized, we can use Lemma~\ref{lemma:edgesVSvertices}: 
 	$T_{\norm{}} \leq P_{\norm{}}(n\times\DAGsize{\css\sigma} + \card{\edges{\css}} + \sum_{x\in\vars{\css}}(\DAGsize{x\sigma})^2)$.

 	Then, using Lemma~\ref{lemma:normprop} (Statement~\ref{pSepVarII}) we obtain
	$\subII{\css\sigma} = \subII{\css}\sigma\cup\subII{\vars{\css}\sigma}$, and
	 thus,
	$\DAGsize{\css\sigma} \leq \card{\subII{\css}\sigma} + \sum_{x\in\vars{\css}}\DAGsize{x\sigma}$.
	From Statement~\ref{pNormCard} of Lemma~\ref{lemma:normprop} follows that $\card{\subII{\css}\sigma} \leq \DAGsize{\css}$.
	Since $\DAGsize{x\sigma} \leq 2\times\DAGsize{\css}$ and $\card{\vars{\css}} \leq \DAGsize{\css}$, we obtain
	$\DAGsize{\css\sigma} \leq \DAGsize{\css}+2\times(\DAGsize{\css})^2$. 	
	In the same way, $\sum_{x\in\vars{\css}}(\DAGsize{x\sigma})^2 \leq \DAGsize{\css}\times (2\times \DAGsize{\css})^2$.
	
	Therefore, 
	$T_{\norm{}} \leq P_{\norm{}}(n\times(\DAGsize{\css}+2\times(\DAGsize{\css})^2) + \card{\edges{\css}} + \DAGsize{\css}\times (2\times \DAGsize{\css})^2) = O\left( (\measure{\css})^{3m''} \right)$.

	Thus,  $T = O\left(\left(\measure{\css}\right)^{3m''} + \left(\measure{\css}\right)^{2m'}\right)$  that shows, that a test of a proof returned by the algorithm takes polynomial time what gives us a class of complexity.
}%

\subsection{Proof of Lemma~\ref{lemma:validstep}}
\label{app:lemma|validstep}
Let us consider all the cases of DY+ACI rules:
\begin{itemize}
 \item ${t_1,t_2} \rightarrow \norm{\pair{t_1}{t_2}}$ 
	We have two cases:
	\begin{itemize}
		\item $\exists u\in \subo{S}$ such that $\norm{\pair{t_1}{t_2}}=\norm{u\sigma}$. Then we have 
		$\pah{\norm{\pair{t_1}{t_2}}} = \pah{\pair{\norm{t_1}}{\norm{t_2}}} = $ \br $ 
		\pairing{\set{\pair{\pah{\norm{t_1}}}{\pah{\norm{t_2}}}}} =  $ \br $ 
		\norm{\pair{\pah{t_1}}{\pah{t_2}}}$ and then $\pah{\norm{\pair{t_1}{t_2}}} \in  $ \br $ \der{\set{\pah{t_1},\pah{t_2}}}$.
		
		\item $\nexists u\in \subo{S}$ such that $\norm{\pair{t_1}{t_2}}=\norm{u\sigma}$. 
		Then  (by definition, Lemma~\ref{lemma:normprop} and Proposition~\ref{prop:pairing}) $\pah{\norm{\pair{t_1}{t_2}}} =  $ \br $ 
		\pairing{\ah{\norm{t_1}}\cup \ah{\norm{t_2}}} \in \der{\norm{\ah{t_1}\cup \ah{t_2}}}$. 
		By Proposition~\ref{prop:pairing}, $\norm{\ah{t_1}} \subseteq \der{\set{\pah{t_1}}}$ and 
		$\norm{\ah{t_2}} \subseteq  $ \br $ \der{\set{\pah{t_2}}}$, then by Lemma~\ref{lemma:derext}, 
		$\norm{\ah{t_1}}\cup \norm{\ah{t_2}}  $ \br $  \subseteq \der{\set{\pah{t_1}}\cup \set{\pah{t_2}}}$. 
		Now, by applying Lemma~\ref{lemma:dertrans}, we have $\pah{\norm{\pair{t_1}{t_2}}} \in \der{\set{\pah{t_1}}\cup \set{\pah{t_2}}}$.
	\end{itemize}

	So, in this case $\pairing{\ah{r}} \in \der{\set{\pairing{\ah{l_1}}, \pairing{\ah{l_2}}} }$.
  \item ${t_1,t_2} \rightarrow \norm{\enc{t_1}{t_2}}$.  
	Proof of this case can be done by analogy of previous one.
  \item $\set{t_1,t_2} \rightarrow \norm{\aenc{t_1}{t_2}}$.  The same.
  
  \item ${t_1,\priv{t_2}} \rightarrow \norm{\sig{t_1}{\priv{t_2}}}$. 
  	\begin{itemize} 

		\item \raggedright $\exists u\in \subo{S}$ such that $\norm{\sig{t_1}{\priv{t_2}}}=\norm{u\sigma}$.  Then $\pah{\norm{\sig{t_1}{\priv{t_2}}}} = \pah{\sig{t_1}{\priv{t_2}}} = \pairing{\set{\sig{\pah{\norm{t_1}}}{\pah{\norm{\priv{t_2}}}}}} = \norm{\sig{\pah{t_1}}{\priv{\pah{t_2}}}}$ and then $\pah{\norm{\sig{t_1}{\priv{t_2}}}} \in \der{\set{\pah{t_1},\pah{\priv{t_2}}}}$ (as $\pah{\priv{t_2}}= \priv{\pah{t_2}}$).
		
		\item $\nexists u\in \subo{S}$ such that $\norm{\sig{t_1}{\priv{t_2}}}=\norm{u\sigma}$. This case can be proved in similar way as done for $\set{t_1,t_2} \rightarrow \norm{\pair{t_1}{t_2}}$.
	\end{itemize}
	
  \item ${t_1, \dots, t_m} \rightarrow \norm{\aci{{t_1}, \dots, {t_m}}}$.  
	On one hand, $\pah{\norm{\aci{{t_1}, \dots, {t_m}}}} = \pah{\aci{{t_1}, \dots, {t_m}}} = \pairing{\ah{t_1}\cup \dots\cup \ah{t_m}}\in $ \br $ 
	\der{\norm{\ah{t_1}\cup \dots\cup \ah{t_m}}}$.
	On the other hand, $\norm{\ah{t_i}}\subseteq $ \br $ \der{\set{\pah{t_i}}}$. And thus, by Lemma~\ref{lemma:derext}, $\pah{\norm{\aci{{t_1}, \dots, {t_m}}}} \in \der{\set{\pah{t_1},\dots,\pah{t_m}}}$.
	
  \item ${\enc{t_1}{t_2},\norm{t_2}} \rightarrow \norm{t_1}$. Here we have to show that $\pah{\norm{t_1}}$ is derivable from $\set{\pah{\enc{t_1}{t_2}}, \pah{\norm{t_2}}}$.
	Consider two cases:
  	\begin{itemize}
		\item $\exists u\in \subo{S}$ such that $\norm{\enc{t_1}{t_2}}=\norm{u\sigma}$. Then \br $\pah{\enc{t_1}{t_2}}  = \enc{\pah{t_1}}{\pah{t_2}}$, and 
		$\pah{\norm{t_1}}=  $ \br $  \pah{t_1} \in \der{\set{\enc{\pah{t_1}}{\pah{t_2}},\norm{\pah{\norm{t_2}}}}}$.
		
		\item $\nexists u\in \subo{S}$ such that $\norm{\enc{t_1}{t_2}}=\norm{u\sigma}$. Then \br $\pah{\enc{t_1}{t_2}} = \pairing{\ah{t_1}\cup \ah{t_2}}$. Using Proposition~\ref{prop:pairing}, we have $\norm{\ah{t_1}\cup \ah{t_2}} \subseteq \der{\set{\pah{\enc{t_1}{t_2}}}}$, thus \br (by Lemma~\ref{lemma:normprop}) $\norm{\ah{t_1}}\subseteq \der{\set{\pah{\enc{t_1}{t_2}}}}$. And then, by Proposition~\ref{prop:pairing} we have that $\pah{t_1} \in \der{\norm{\ah{t_1}}}$. Therefore, by Lemma~\ref{lemma:dertrans}, we have 
		\br  $\pah{\norm{t_1}}=\pah{t_1} \in \der{\pah{\enc{t_1}{t_2}}}$.
	\end{itemize}
	
  \item 
  	${\aenc{t_1}{t_2},\norm{\priv{t_2}}} \rightarrow \norm{t_1}$. Here we have to show that $\pah{\norm{t_1}}$ is derivable from $\set{\pah{\aenc{t_1}{t_2}}, \pah{\norm{\priv{t_2}}}}$.
	Consider two cases:
  	\begin{itemize}
		\item $\exists u\in \subo{S}$ such that $\norm{\aenc{t_1}{t_2}}=\norm{u\sigma}$. Then \br $\pah{\aenc{t_1}{t_2}}  = \aenc{\pah{t_1}}{\pah{t_2}}$, \br and then 
		$\pah{\norm{t_1}}=\pah{t_1} \in   $ \br $ \der{\set{\aenc{\pah{t_1}}{\pah{t_2}},\norm{\priv{\pah{t_2}}}}}$. \br
		On the other hand, $\pah{\norm{\priv{t_2}}} = \pah{\priv{t_2}} =  $ \br $ \pairing{\set{\priv{\pah{t_2}}}}=\norm{\priv{\pah{t_2}}}$.
		
		\item $\nexists u\in \sub{S}$ such that $\norm{\aenc{t_1}{t_2}}=\norm{u\sigma}$. 
		 Then \br $\pah{\aenc{t_1}{t_2}} = \pairing{\ah{t_1}\cup \ah{t_2}}$. Using Proposition~\ref{prop:pairing}, we have $\norm{\ah{t_1}\cup \ah{t_2}} \subseteq \der{\set{\pah{\aenc{t_1}{t_2}}}}$, thus (by Lemma~\ref{lemma:normprop}) $\norm{\ah{t_1}}\subseteq \der{\set{\pah{\aenc{t_1}{t_2}}}}$. And then, by Proposition~\ref{prop:pairing} we have that $\pah{t_1} \in \der{\norm{\ah{t_1}}}$. Therefore, by Lemma~\ref{lemma:dertrans},   $\pah{\norm{t_1}}=\pah{t_1} \in \der{\pah{\aenc{t_1}{t_2}}}$.
	\end{itemize}
	
  \item ${\pair{t_1}{t_2}} \rightarrow \norm{t_1}$. Here, as usual, we consider two cases:
	\begin{itemize}
		\item $\exists u\in  \subo{S}$ such that $\norm{\pair{t_1}{t_2}}=\norm{u\sigma}$. 
		Then \br $\pah{\pair{t_1}{t_2}} =  \pair{\pah{t_1}}{\pah{t_2}}$ and then \br
		$\pah{\norm{t_1}}=\norm{\pah{t_1}}\in \der{\set{\pah{\pair{t_1}{t_2}}}}$.
		
		\item $\nexists u\in  \subo{S}$ such that $\norm{\pair{t_1}{t_2}}=\norm{u\sigma}$. 
		Then \br $\pah{\pair{t_1}{t_2}} = \pairing{\ah{t_1}\cup \ah{t_2}}$. Then by  Proposition~\ref{prop:pairing}, 
		we have $\norm{\ah{t_1}\cup \ah{t_2}} \subseteq \der{\set{\pah{\pair{t_1}{t_2}}}}$, 
		thus \br  $\norm{\ah{t_1}}\subseteq \der{\set{\pah{\pair{t_1}{t_2}}}}$. 
		And then, by Proposition~\ref{prop:pairing} we have that $\pah{t_1} \in \der{\norm{\ah{t_1}}}$. 
		Therefore, by Lemma~\ref{lemma:dertrans}, $\pah{\norm{t_1}}=\pah{t_1} \in \der{\pah{\pair{t_1}{t_2}}}$.
	\end{itemize}

  \item ${\pair{t_1}{t_2}} \rightarrow \norm{t_2}$. Proof like above.
  
  \item ${\aci{{t_1}, \dots, {t_m}}} \rightarrow \norm{t_i}$. 
  We have $\pah{\aci{t_1, \dots, t_2}}=\pairing{\ah{t_1}\cup \dots \cup \ah{t_m}}$. 
  Then by  Proposition~\ref{prop:pairing}, $\norm{\ah{t_1}\cup \dots \cup \ah{t_m}} \subseteq $ \br $ \der{\pah{\aci{t_1, \dots, t_m}}}$; 
  thus $\norm{\ah{t_i}}\subseteq \der{\pah{\aci{t_1, \dots, t_m}}}$. 
  As $\pah{t_i} \in \der{\norm{\ah{t_i}}}$, 
  by Lemma~\ref{lemma:dertrans} we have $\pah{\norm{t_i}}=\pah{t_i} \in \der{\pah{\aci{t_1, \dots, t_2}}}$.

\end{itemize}
  As all possible cases satisfy lemma conditions, we proved the lemma.

\subsection{Proof of Property~\ref{prop:generalLimit}}
\label{app:prop|generalLimit}
\begin{proof}
	From proposition~\ref{prop:normalSys} and \ref{prop:normalSigma} we know that if $\sigma'$ is a model of $\ConstrSys{S}$ then $\norm{\sigma'}$ is a model of $\ConstrSys{S}$  and  $\norm{\sigma'}$ is a model of $\norm{\ConstrSys{S}}$.
	Then, there exists a substitution $\theta: \dom{\theta}=\dom{\norm{\sigma'}}, \, \dom{\theta}\theta \subseteq \dom{\theta}, \, \sigma''=\norm{\sigma'}|_{\dom{\theta}\theta}$  and $\sigma''$ is a model of $\norm{\ConstrSys{S}}\theta$ such that $x\sigma''\neq y\sigma''$, if $x\neq y$ (this is true because we can show how to build $\theta$ : given the $\norm{\sigma'}$ --- simply split $\dom{\norm{\sigma'}}$ into  the classes of equivalence modulo $\norm{\sigma'}$, i.e. $x\equiv y \iff x\norm{\sigma'}=y\norm{\sigma'}$; for every class choose one representative  $[x]_\equiv$, and then $x\theta = [x]_\equiv$). Note, that $\theta\sigma''=\sigma'$, that's why $\sigma''$ is a model of $\norm{\ConstrSys{S}}\theta$.
	
	Then, as $\sigma''$ is a model of $\norm{\ConstrSys{S}}\theta$, using 
	Proposition~\ref{prop:normalSys}, we can say that $\sigma''$ is a model of $\norm{\norm{\ConstrSys{S}}\theta}$. Moreover, $\sigma''$ is normalized and $x\sigma'' \neq y\sigma''$ for all $x,y\in\dom{\sigma''}$ such that $x\neq y$. Then, we can apply Corollary~\ref{cor:existsgood}, which gives us existence of conservative model $\delta$ of $\norm{\norm{\ConstrSys{S}}\theta}$.
	That is why we can apply  Proposition~\ref{prop:limit}: 
	for any $x\in \vars{\norm{\norm{\ConstrSys{S}}\theta}}$, $\DAGsize{x\delta}\leq 2\times\DAGsize{\norm{\norm{\ConstrSys{S}}\theta}}$.
	
	Note, that using Proposition~\ref{prop:normalSys},
	Lemma~\ref{lemma:substcomp} and definition of ``model'', we can easily show that $\delta[\theta]$ is a model of $\norm{S}$. Moreover, $\delta[\theta]$ is normalized.
	By definition of $\delta[\theta]$ we can say, that for all $x\in\dom{\delta[\theta]}$ there exists $y\in\dom{\theta}\theta$ such that $x\delta[\theta]=y\delta$; and as $y\in\UniVar{X}$ (by definition of $\theta$), then $\DAGsize{x\delta[\theta]}=\DAGsize{y\delta}\leq 2\times\DAGsize{\norm{\norm{\ConstrSys{S}}\theta}} \leq 2\times\DAGsize{\norm{\ConstrSys{S}}\theta}$. Applying Lemma~\ref{lemma:sizesubst}, we have $\DAGsize{x\delta[\theta]}\leq 2\times\DAGsize{\norm{\ConstrSys{S}}}$.
	
	Summing up, we have a normalized model $\sigma=\delta[\theta]$ of $\norm{\ConstrSys{S}}$ such that for all $x\in \dom{\sigma}$, $\DAGsize{x\sigma}\leq 2\times\DAGsize{\norm{\ConstrSys{S}}}$.
 
\end{proof}

}

\bibliographystyle{elsarticle-num}
\bibliography{constr_ACI}

\begin{thebibliography}{10}
\expandafter\ifx\csname url\endcsname\relax
  \def\url#1{\texttt{#1}}\fi
\expandafter\ifx\csname urlprefix\endcsname\relax\def\urlprefix{URL }\fi
\expandafter\ifx\csname href\endcsname\relax
  \def\href#1#2{#2} \def\path#1{#1}\fi

\bibitem{MillenShmatikov2001}
J.~Millen, V.~Shmatikov,
  \href{http://doi.acm.org/10.1145/501983.502007}{Constraint solving for
  bounded-process cryptographic protocol analysis}, in: Proceedings of the 8th
  ACM conference on Computer and Communications Security, CCS '01, ACM, New
  York, NY, USA, 2001, pp. 166--175.
\newblock \href {http://dx.doi.org/http://doi.acm.org/10.1145/501983.502007}
  {\path{doi:http://doi.acm.org/10.1145/501983.502007}}.
\newline\urlprefix\url{http://doi.acm.org/10.1145/501983.502007}

\bibitem{ofmc2005}
D.~Basin, S.~M\"odersheim, L.~Vigan\`o, Ofmc: A symbolic model checker for
  security protocols, International Journal of Information Security 4 (2005)
  181--208.

\bibitem{Turuani06}
M.~Turuani, {The CL-Atse Protocol Analyser}, in: Term Rewriting and
  Applications (RTA), 2006, pp. 277--286.

\bibitem{Cr2008Scyther}
C.~Cremers, The {S}cyther {T}ool: Verification, falsification, and analysis of
  security protocols, in: Computer Aided Verification, 20th International
  Conference, CAV 2008, Princeton, USA, Proc., Vol. 5123/2008 of Lecture Notes
  in Computer Science, Springer, 2008, pp. 414--418.
\newblock \href {http://dx.doi.org/10.1007/978-3-540-70545-1_38}
  {\path{doi:10.1007/978-3-540-70545-1_38}}.

\bibitem{BasinMV05}
D.~Basin, S.~M{\"o}dersheim, L.~Vigan{\`o}, Algebraic intruder deductions, in:
  Logic for Programming, Artificial Intelligence, and Reasoning (LPAR), 2005,
  pp. 549--564.

\bibitem{ChevRus-Combin2009}
Y.~Chevalier, M.~Rusinowitch,
  \href{http://www.sciencedirect.com/science/article/B6V1G-4XKBYW5-2/2/1191e69%
4e6322c89670136e608d98620}{Symbolic protocol analysis in the union of disjoint
  intruder theories: Combining decision procedures}, Theoretical Computer
  Science 411~(10) (2010) 1261 -- 1282, iCALP 2005 - Track C: Security and
  Cryptography Foundations.
\newblock \href {http://dx.doi.org/DOI: 10.1016/j.tcs.2009.10.022}
  {\path{doi:DOI: 10.1016/j.tcs.2009.10.022}}.
\newline\urlprefix\url{http://www.sciencedirect.com/science/article/B6V1G-4XKB%
YW5-2/2/1191e694e6322c89670136e608d98620}

\bibitem{CDL05-survey}
V.~Cortier, S.~Delaune, P.~Lafourcade,
  \href{http://www.lsv.ens-cachan.fr/Publis/PAPERS/PDF/surveyCDL.pdf}{A survey
  of algebraic properties used in cryptographic protocols}, Journal of Computer
  Security 14~(1) (2006) 1--43.
\newline\urlprefix\url{http://www.lsv.ens-cachan.fr/Publis/PAPERS/PDF/surveyCD%
L.pdf}

\bibitem{ChevalierKRT05xor}
Y.~Chevalier, R.~K{\"u}sters, M.~Rusinowitch, M.~Turuani, An np decision
  procedure for protocol insecurity with xor, Theor. Comput. Sci. 338~(1-3)
  (2005) 247--274.

\bibitem{DLLT-IC07}
S.~Delaune, P.~Lafourcade, D.~Lugiez, R.~Treinen,
  \href{http://www.lsv.ens-cachan.fr/Publis/PAPERS/PDF/DLLT-ic07.pdf}{Symbolic
  protocol analysis for monoidal equational theories}, Information and
  Computation 206~(2-4) (2008) 312--351.
\newblock \href {http://dx.doi.org/10.1016/j.ic.2007.07.005}
  {\path{doi:10.1016/j.ic.2007.07.005}}.
\newline\urlprefix\url{http://www.lsv.ens-cachan.fr/Publis/PAPERS/PDF/DLLT-ic0%
7.pdf}

\bibitem{Mazare05}
L.~Mazar{\'e}, {Satisfiability of Dolev-Yao Constraints}, Electronic Notes in
  Theoretical Computer Science 125~(1) (2005) 109--124.

\bibitem{MazareThese}
L.~Mazar{\'e},
  \href{http://www.lsv.ens-cachan.fr/Publis/PAPERS/PDF/these-mazare.pdf}{{Comp%
utational Soundness of Symbolic Models for Cryptographic Protocols}}, Ph.D.
  thesis, Institut National Polytechnique de Grenoble (October 2006).
\newline\urlprefix\url{http://www.lsv.ens-cachan.fr/Publis/PAPERS/PDF/these-ma%
zare.pdf}

\bibitem{Syverson00dolev-yaois}
P.~Syverson, C.~Meadows, I.~Cervesato, {Dolev-Yao} is no better than
  {Machiavelli}, in: First Workshop on Issues in the Theory of Security —
  WITS’00, 2000, pp. 87--92.

\bibitem{Baudet05}
M.~Baudet, Deciding security of protocols against off-line guessing attacks,
  in: ACM Conference on Computer and Communications Security, 2005, pp. 16--25.

\bibitem{RusinowitchT03}
M.~Rusinowitch, M.~Turuani, Protocol insecurity with a finite number of
  sessions, composed keys is np-complete, Theor. Comput. Sci. 1-3~(299) (2003)
  451--475.

\bibitem{ChevalierKRT03exp}
Y.~Chevalier, R.~K{\"u}sters, M.~Rusinowitch, M.~Turuani, Deciding the security
  of protocols with diffie-hellman exponentiation and products in exponents,
  in: FSTTCS, 2003, pp. 124--135.

\bibitem{Comon-Lundh03intruderdeductions}
H.~Comon-Lundh, V.~Shmatikov, Intruder deductions, constraint solving and
  insecurity decision in presence of exclusive or, IEEE Comp. Soc. Press, 2003,
  pp. 271--280.

\bibitem{Shmatikov04decidableanalysis}
V.~Shmatikov, Decidable analysis of cryptographic protocols with products and
  modular exponentiation, in: In Proc. 13th European Symposium on Programming
  (ESOP ’04), volume 2986 of LNCS, Springer-Verlag, 2004, pp. 355--369.

\bibitem{THESE-delaune06}
S.~Delaune,
  \href{http://www.lsv.ens-cachan.fr/Publis/PAPERS/PDF/these-delaune.pdf}{V{\'%
e}rification des protocoles cryptographiques et propri{\'e}t{\'e}s
  alg{\'e}briques}, Th{\`e}se de doctorat, Laboratoire Sp{\'e}cification et
  V{\'e}rification, ENS Cachan, France (Jun. 2006).
\newline\urlprefix\url{http://www.lsv.ens-cachan.fr/Publis/PAPERS/PDF/these-de%
laune.pdf}

\bibitem{Bursuc07associative-commutativededucibility}
S.~Bursuc, H.~Comon-lundh, S.~Delaune, Associative-commutative deducibility
  constraints, in: Proceedings of the 24th Annual Symposium on Theoretical
  Aspects of Computer Science (STACS’07), volume 4393 of Lecture Notes in
  Computer Science, Springer, 2007, pp. 634--645.

\bibitem{ChevalierLR07}
Y.~Chevalier, D.~Lugiez, M.~Rusinowitch, Towards an automatic analysis of web
  service security, in: FroCoS 2007, Liverpool, UK, September 10-12, Vol. 4720
  of Lecture Notes in Computer Science, Springer, 2007, pp. 133--147.

\bibitem{owaspv3}
O.~Foundation,
  \href{http://www.owasp.org/index.php/Testing_for_XML_Injection_(OWASP-DV-008%
)}{{OWASP-DV-008, OWASP Testing Guide, v3.0}},
  \url{http://www.owasp.org/index.php/Testing_for_XML_Injection_(OWASP-DV-008)}
  (2008).
\newline\urlprefix\url{http://www.owasp.org/index.php/Testing_for_XML_Injectio%
n_(OWASP-DV-008)}

\end{thebibliography}

\end{document}